\documentclass[12pt,preprint]{aastex}

\newcommand{\teff}{$T_{\rm eff}$}

\newcommand{\msun}{$M_\odot$}
\newcommand{\lsun}{$L_\odot$}

\newcommand{\logg}{$\log g$}

\newcommand{\kms}{km~s$^{-1}$}
\newcommand{\U}{$F336W$}
\newcommand{\B}{$F439W$}
\newcommand{\V}{$F555W$}
\newcommand{\I}{$F814W$}

\slugcomment{To appear in the ApJ in ... }

\shorttitle{Spectroscopy of warm stars in globular clusters}
\shortauthors{De Marco et al.}

\begin{document}


\title{A Spectroscopic Analysis of Blue Stragglers, Horizontal Branch and Turn-Off Stars 
    in Four Globular Clusters}


\author{
Orsola De Marco\altaffilmark{1}, 
Michael M. Shara\altaffilmark{1},
D. Zurek\altaffilmark{1}, 
John A. Ouellette\altaffilmark{1}, 
Thierry Lanz\altaffilmark{2,3,4}, 
Rex A. Saffer\altaffilmark{5,6} \& 
Jeremy F. Sepinsky\altaffilmark{5}}


\altaffiltext{1}{Astrophysics Department, American Museum of Natural History, New York, NY 10024}
\altaffiltext{2}{Department of Astronomy, University of Maryland, College Park, MD 20742, USA}
\altaffiltext{3}{NASA Goddard Space Flight Center, Code 681, Greenbelt, MD 20771, USA}
\altaffiltext{4}{Visiting fellow at Columbia University}
\altaffiltext{5}{Villanova University, UNIT-Central Services, 800 Lancaster Avenue, Villanova, PA 19085, USA}
\altaffiltext{6}{Department of Astronomy and Astrophysics, Villanova University, 800 Lancaster Avenue, Villanova, PA 19085}


\begin{abstract}
We present a spectroscopic analysis of
HST/STIS and FOS low- and intermediate-resolution 
spectroscopy of 55 stars in four globular clusters
(47~Tucanae, M~3, NGC~6752, and NGC~6397). Stars hotter than \teff = 5750~K
and with a signal-to-noise ratio larger than 15 were analyzed with non-Local Thermodynamic Equilibrium 
model atmospheres, and values for their effective temperatures and gravities were obtained.
Using photometric fluxes, we also obtained radii, luminosities and spectroscopic masses. 

Twenty-four stars in our sample are blue stragglers (BS). 
Their photometric colors and magnitudes place these BSs above and redward of the
clusters zero-age main sequence: this is consistent with the gravities 
we find for these stars, which are lower than zero-age main sequence gravities.
A comparison with stellar evolutionary tracks shows that almost all of our BSs are in the Hertzsprung gap.
This is contrary to theory, because of the short timescale expected for stars in this evolutionary phase.

Mean BS masses are 1.04~\msun\ for 14 non-variable stars, or 1.07~\msun\ counting 
all 24 BSs in our sample. For the non-variable stars the mean BS mass for individual 
clusters are 1.73, 1.01, 0.95 and 0.72~\msun\ for NGC~6397, NGC~6752, 47~Tuc and M~3, respectively.
Adding the variable stars (which improves the statistics but increases
the uncertainty) the mean masses become 1.27, 1.05, 0.99 and 0.99~\msun, respectively. 
Though there is considerable scatter, the BS spectroscopic masses correlate
with both effective temperature and brightness of the stars, as expected. The BSs in

The mean non-variable turn-off star mass (0.58~\msun) is significantly below the values determined for the BSs and
below the main sequence turn-off mass. The mean non-variable horizontal-branch (HB) star mass is higher than expected
(0.79~\msun). In particular,
several HB stars have masses well above the main sequence turn-off mass. Some of these
HB stars are suspected of actually being BSs, since most of them
reside at ambiguous locations on the CMD making them prone to misclassification. 

Values and limits to the stellar rotation rates ($v \sin i$)
are imposed by fitting weak metal lines, the Ca~{\sc ii} K line wings, or the
helium lines for the hotter stars. Five BSs with reasonably 
constrained rotations show average and median
$v \sin i$ values of 109 and 100~\kms, respectively, suggesting $v \sim 160$~\kms. 
At least some GC BSs are are very rapid rotators, but this information cannot yet
constrain their origin as stellar collision or binary mergers, because of the lack of 
clear theoretical predictions. Six extreme HB stars have rotation rates $v \sin i$ between
50 and 200~\kms, which are high for these stars and might indicate a binary origin.

De Marco et al. found that four BSs and two HB star 
in our sample have
Balmer jumps which are too large for the effective temperatures implied by the slopes of their
Paschen continua. Two additional HB stars are now identified in the current study as having the same feature.
For these stars, the presence of a disk of partly ionized material is suspected, although high stellar
rotation rates could also partly explain the data. 

\end{abstract}



\section{Introduction}
\label{sec:introduction}

Fifty years ago Sandage (1953) discovered a unique class of stars, the Blue Stragglers (BSs).
In the color-magnitude diagrams (CMD) of clusters, these stars appear brighter and bluer than the
main sequence turn-off.
They appear in the CMD of open and globular
clusters (GC), as well as in the field, and their brightnesses and colors are consistent with 
evolutionary models for stars more massive than the cluster (or population) turn-off. To explain
how stars more massive than the cluster's turn-off stars could still be present in GCs, or in old populations,
several scenarios have been proposed. After years of debate (for a review, see Bailyn 1995), the consensus
has settled on stellar collisions or binary mass transfer and mergers as the most likely events responsible for the formation
of BS stars. 

BS populations have a great diversity of properties. For instance, their specific frequencies and
mean brightnesses vary greatly. It 
is likely that these differences are due to differences in the clusters' dynamical histories. 
However, to date,
attempts to correlate BS population photometric properties with other cluster parameters
have failed to determine the formation scenarios most likely to apply (e.g., Piotto et al. 2004).
As a result, the information the BS population contains about the history of its cluster's dynamics 
remains inaccessible due to our ignorance of the BS formation channels.

Critical BS properties which have not been fully investigated due to the difficulty of the observations
are their masses and rotation rates.
Whether the BSs are the result of a collision or a binary merger, they are expected
to be more massive than the main sequence turn-off mass. However, models of the
two scenarios do not predict BSs which are systematically different, with the exception that
masses larger than 
twice the turn-off mass for the cluster are likely to be the result of 3-body collisions. 
The fact that BSs are more centrally-concentrated than subgiants,
and, by inference, than TO stars,
already suggested that BSs are more massive than the clusters' turn-off
mass (Nemec \& Harris 1987).
Spectroscopic evidence that BS masses are larger than the turn-off mass for the cluster was previously
obtained by Shara et al. (1997), who measured the mass
of one of the most luminous BS in the core of 47~Tuc, showing that it has twice the cluster's turn-off mass.
This determination was also confirmed by Ouellette (2000) in a distance-independent way,
by comparing stellar temperatures and gravities with evolutionary calculations.
Saffer et al. (2002) carried out a spectroscopic analysis on 6 BSs in NGC~6397, also
deriving masses higher than the turn-off mass for 5 out of 6 objects.
Masses larger than the cluster turn-off were also reported for 4 BSs by Gilliland et al. (1998), based on their pulsation
frequency measurements. However the distribution of the BS masses remains an open question since 
the samples analyzed were invariably small. Since mass predictions do not yet allow us to differentiate
between formation scenarios, one can hope that determining masses for a large enough sample will
reveal correlations with other stellar and cluster properties.

Rotation rates for stars formed either through collisions or binary mergers are also expected to be higher than
normal. Simulations of
collision mergers (e.g., Sills et al. 2002, 2005), show merger product to be rapidly rotating.
Simulations of main sequence binary mergers are lacking, but it is likely that these systems also
inherit considerable angular momentum. However, rotation rates measured for BSs in open 
cluster M67 appear to be
{\it lower} than those for comparable main-sequence stars (Mathys 1991). 
Shara et al. (1997) determined the stellar rotation for the 47~Tuc BS BSS19 to be
a high $v \sin i$ = 155~km~s$^{-1}$. 
Theoretically, rapidly-rotating BSs are expected to slow down:
Sills et al. (2005) argue that a magnetically-locked accretion disk or wind
can draw angular momentum from the star and slow it down. However, Matt \& Pudritz (2005) have recently
argued that this might not be the case. 
De Marco et al. (2004) suggested that low mass disks might be present around some BSs,
but Porter \& Townsend (2005) showed that the same data can also be interpreted by 
accounting for the distortion of the stellar shape due to rotation. 
It is clear that a great deal of work remains to be done both observationally and theoretically.

With all this in mind, we 
obtained Hubble Space Telescope (HST) spectroscopic observations of
a sample of BSs drawn from four different clusters with different central
concentration (47~Tucanae [47~Tuc; NGC~104], M~3 [NGC5272], NGC~6752 and NGC~6397),
to determine spectroscopic masses and rotation rates.

In Section~\ref{sec:observations} we describe our HST
observations. 
The details of the spectroscopic data reduction and extraction are given
in Section~\ref{sec:datareduction}. In
Section~\ref{sec:photometry} we present the stellar 
photometry, while in Section~\ref{sec:identification} we use the photometry to classify the
analyzed stars. In Section~\ref{sec:stellarparameters}
we derive the stellar parameters, followed by an assessment of 
the uncertainties in Section~\ref{sec:uncertainties}. Stellar masses 
and rotation rates are presented in Sections~\ref{sec:masses} and \ref{sec:rotation},
respectively. In Section~\ref{sec:disk} we describe the spectroscopic indicators
of circumstellar disks or rapid rotation, while in Section~\ref{sec:individualobjects} individual
stars are discussed. A comparison with stellar evolutionary
models is presented in
Section~\ref{sec:evolutionarytracks}, followed by our summary in Section~\ref{sec:conclusions}.

\section{Spectroscopic Observations}
\label{sec:observations}

\begin{figure}
\vspace{14cm}
\includegraphics{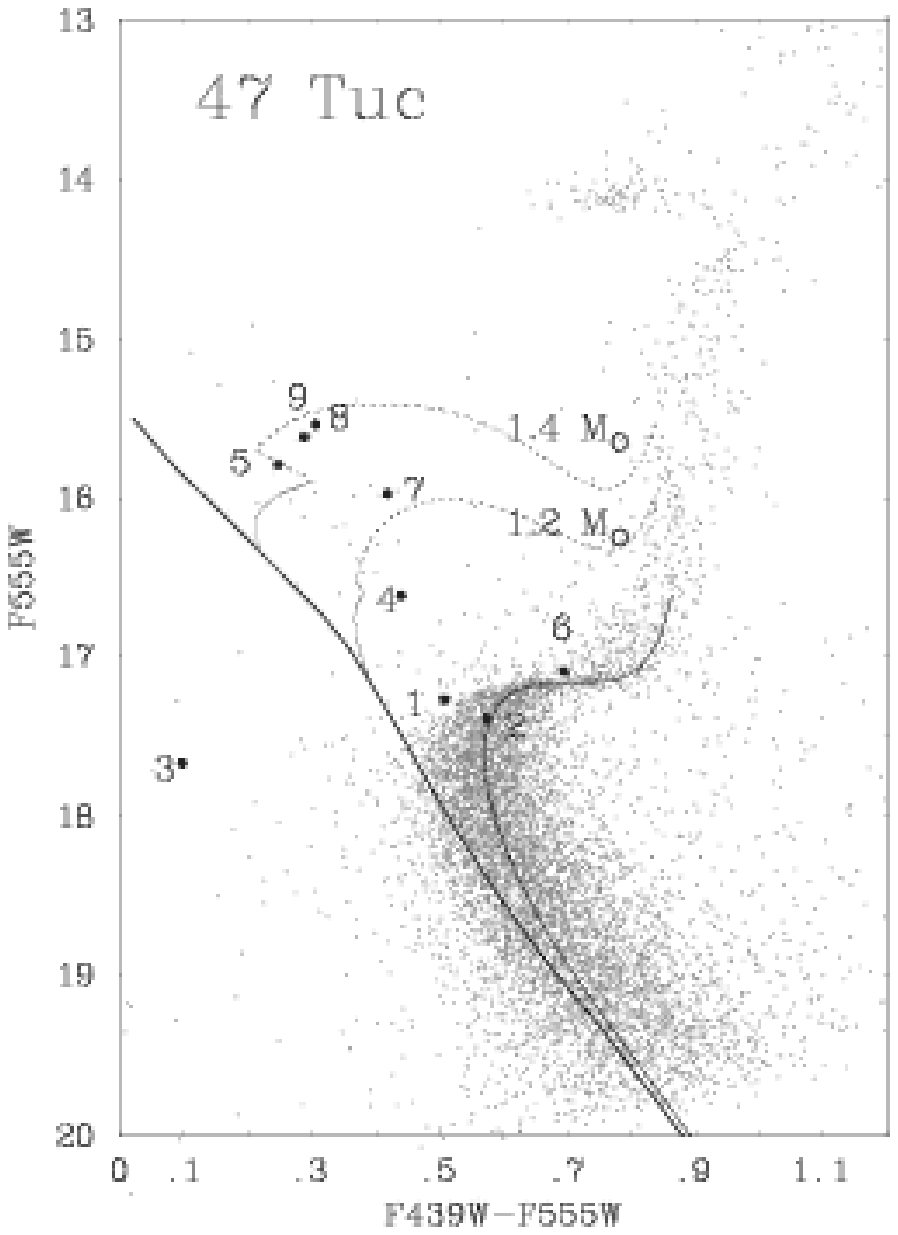}
\caption{Same-epoch (September 1, 1999) \B--\V\ vs \V\ color-magnitude diagrams for 47~Tuc, showing the 
positions of the stars analyzed in this paper (larger filled circles).
The ZAMS and a 12 Gyr isochrone (shifted by 13.4~mag in \V\ and 0.038~mag in \B--\V)
are plotted (thin and thick solid lines,
respectively), as are evolutionary tracks of different initial mass values
(dotted lines).
\label{fig:cmd_ngc104_bv}}
\end{figure}

\begin{figure}
\vspace{14cm}
\includegraphics{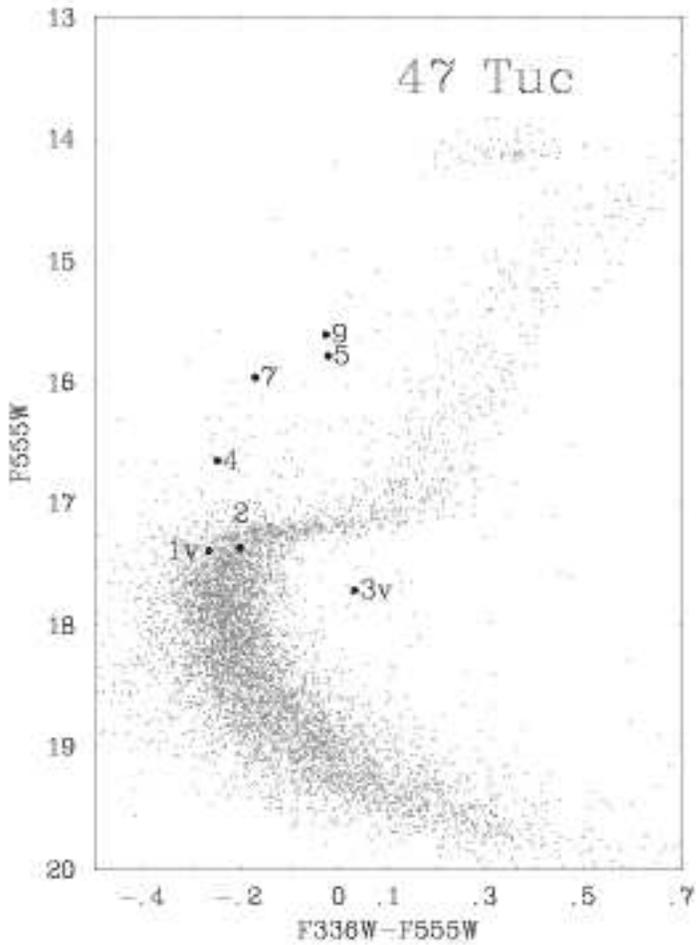}
\caption{Different-epoch \U--\V\ vs \V\ color-magnitude diagrams for 47~Tuc, showing the
positions of the non-variable stars analyzed in this paper (larger filled circles).
A ``v" indicates suspected variability detected because of color shifts between different CMDs. 
Stars 6 and 8 do not appear in this plot because they have
no \B\ values. 
\label{fig:cmd_ngc104_uv}}
\end{figure}

\begin{figure}
\vspace{17cm}
\includegraphics{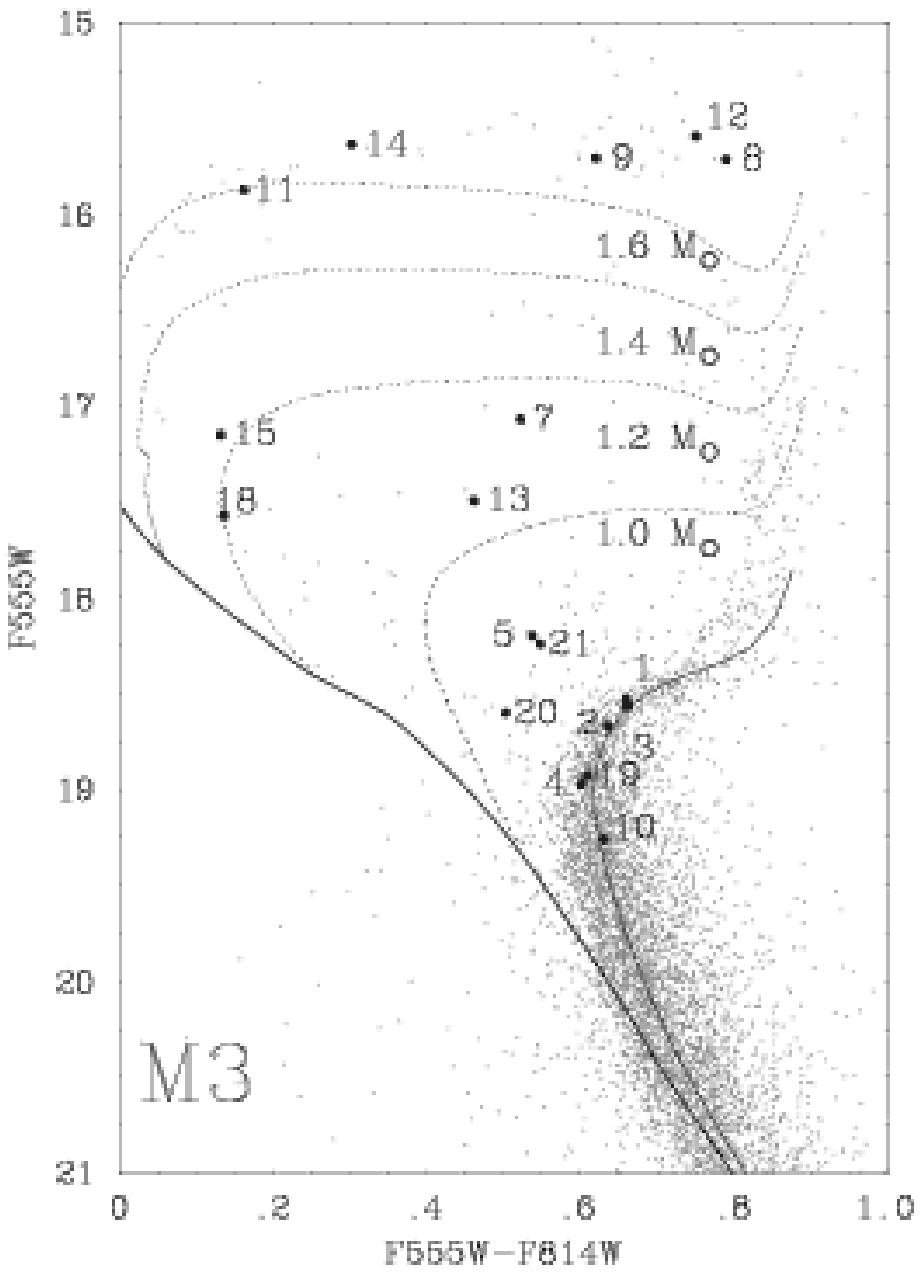}
\caption{Same-epoch (May 14, 2001) \V--\I\ vs \V\ color-magnitude
diagram for M~3, showing the
positions of the stars analyzed in this paper (larger filled circles).
The ZAMS and a 14 Gyr isochrone (shifted by 14.05~mag in \V\ and 0.040~mag in \V--\I) are plotted (thin and thick solid lines,
respectively), as are evolutionary tracks of different initial mass values
(dotted lines). For stars 6, 16 and 17 there is no same-epoch photometry. 
\label{fig:cmd_ngc5272_vi}}
\end{figure}

\begin{figure}
\vspace{17cm}
\includegraphics{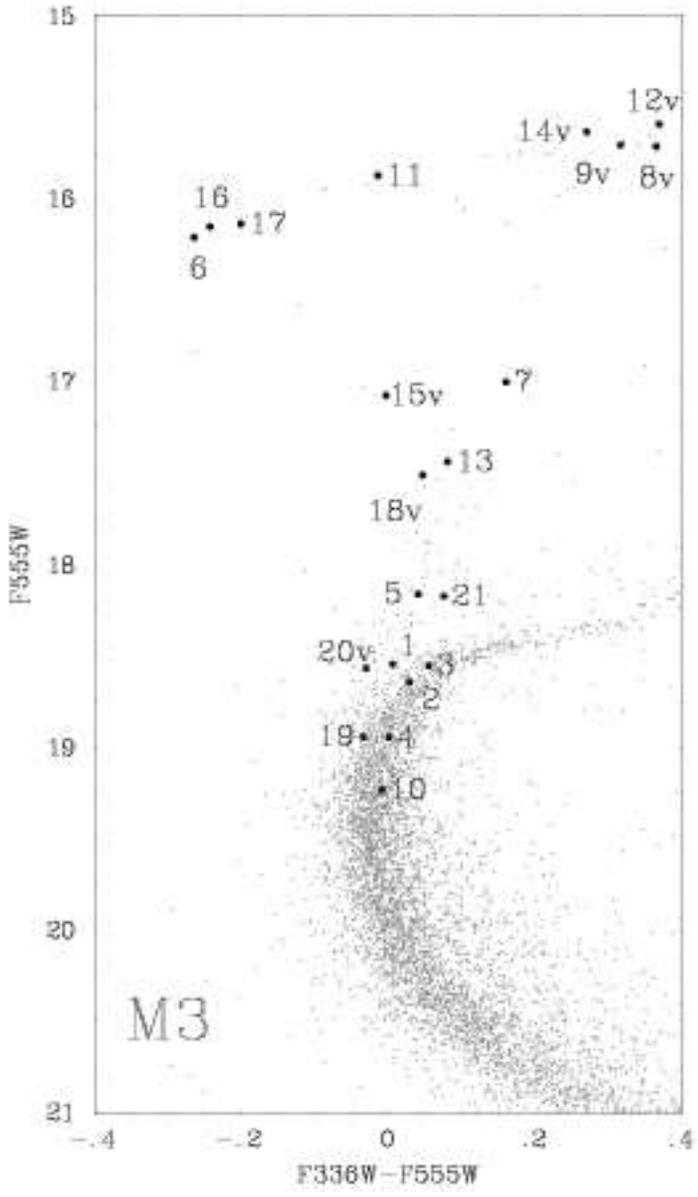}
\caption{Different-epoch \U--\V\ vs \V\ color-magnitude
diagram for M~3, showing the
positions of the stars analyzed in this paper (larger filled circles). Stars suspected of variability 
detected because of color shifts between different CMDs are
marked with a ``v".
\label{fig:cmd_ngc5272_uv}}
\end{figure}

\begin{figure}
\vspace{17cm}
\includegraphics{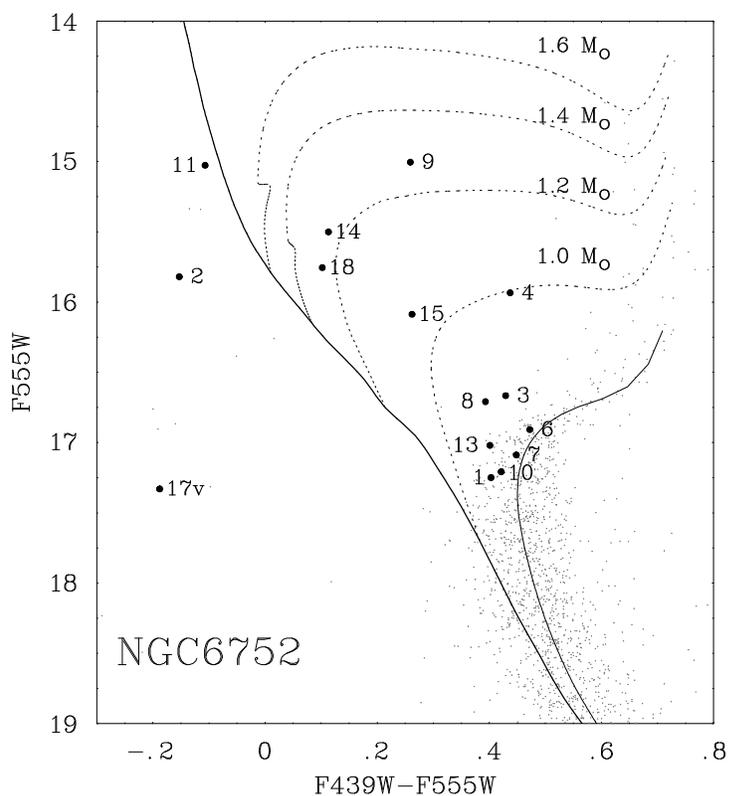}
\caption{Different-epoch \B--\V\ vs \V\ color-magnitude diagrams for NGC~6752, showing the
positions of the stars analyzed in this paper (larger filled circles).
The ZAMS and a 14 Gyr isochrone (shifted by 13.30~mag in \V\ and 0.040~mag in \B--\V) are plotted (thin and thick solid lines,
respectively), as are evolutionary tracks of different initial mass values
(dotted lines). Star 17 is suspected of variability (``v") on the grounds 
of color shifts between different CMDs
and its position on the different-epoch CMD is artificially shifted.
\label{fig:cmd_ngc6752_bv}}
\end{figure}

\begin{figure}
\vspace{11cm}
\includegraphics{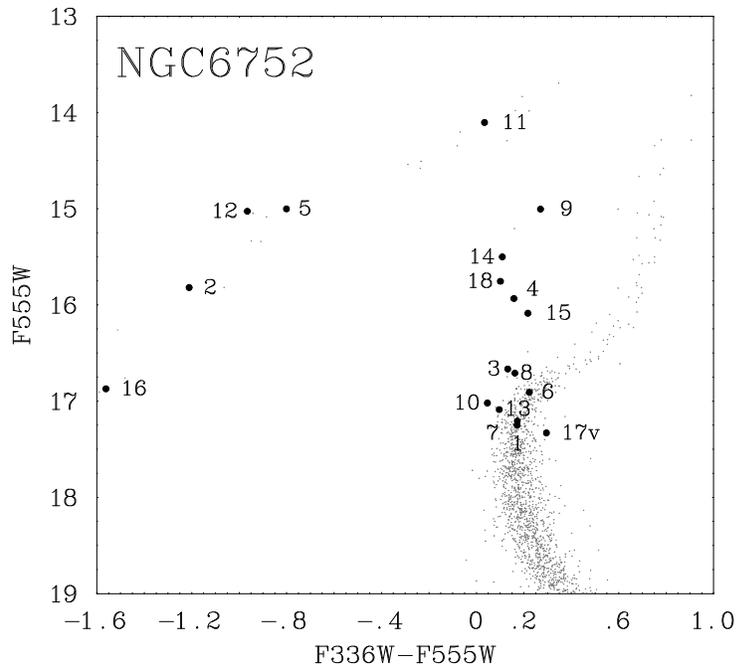}
\caption{Same-epoch (March 22, 2001) \U--\V\ vs \V\ color-magnitude diagrams for NGC~6752, showing the
positions of the stars analyzed in this paper (larger filled circles).
\label{fig:cmd_ngc6752_uv}}
\end{figure}

\begin{figure}
\vspace{17cm}
\includegraphics{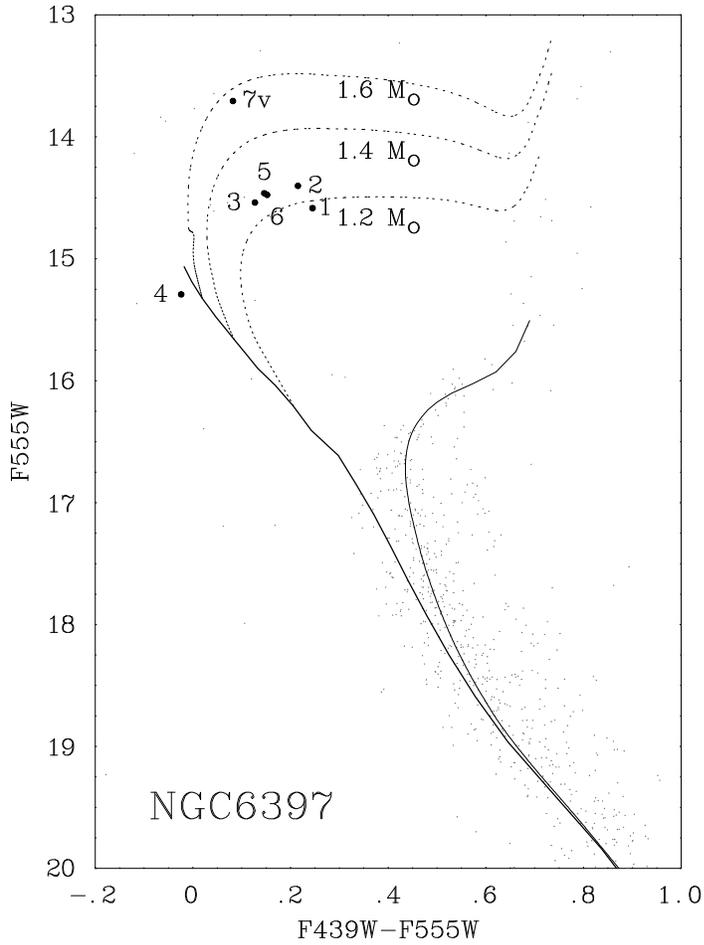}
\caption{Same epoch (March 6-7, 2001) \B--\V\ vs \V\ color-magnitude diagram for NGC~6397, showing the
positions of the stars analyzed in this paper (larger filled circles).
The ZAMS and a 14 Gyr isochrone (shifted by 12.80~mag in \V\ and 0.060~mag in \B--\V) are plotted (thin and thick solid lines,
respectively), as are evolutionary tracks of different initial mass values
(dotted lines).
\label{fig:cmd_ngc6397_bv}}
\end{figure}

\begin{figure}
\vspace{17cm}
\includegraphics{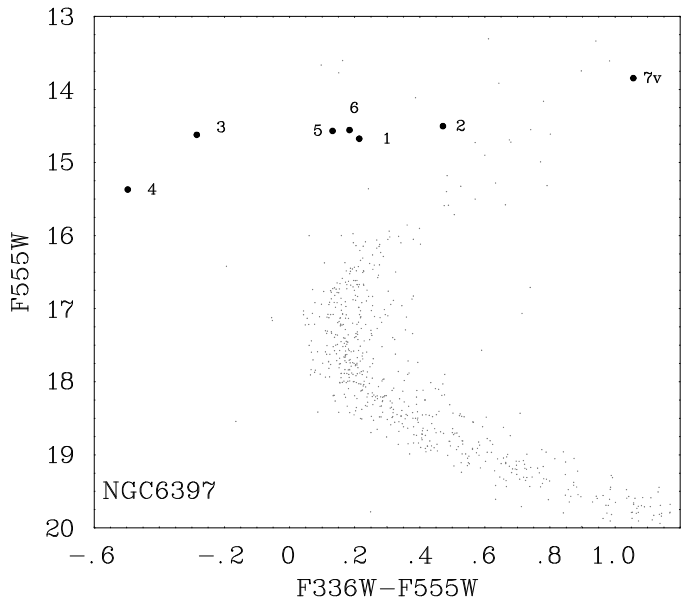}
\caption{Different-epoch \U--\V\ vs \V\ color-magnitude diagram for NGC~6397, showing the
positions of the stars analyzed in this paper (larger filled circles). Stars
1, 2, 5, 6 and 7 are suspected of variability because they acquire different colors depending on the
date in which the images were acquired.
\label{fig:cmd_ngc6397_uv}}
\end{figure}

In order to obtain a sizable sample of BSs, we selected three clusters, rich in BSs:
47~Tuc (NGC~104), M~3 (NGC~5272) and NGC~6752. We also re-analyzed data for the cluster NGC~6397, in which
6 BSs and one HB star were previously identified and analyzed by Saffer et al. (2002).
These clusters span a range in metallicity and central star density. Details of the clusters
are presented in Table~\ref{tab:clusterproperties}.
The CMDs of these clusters are presented in Figures~\ref{fig:cmd_ngc104_bv} to \ref{fig:cmd_ngc6397_uv},
where the BSs can be
clearly seen at positions brighter and bluer than the main sequence turn-off.

Spectra of stars in the GCs 47~Tuc, M~3 and NGC~6752 
were acquired through HST 
proposal GO-8226 (PI: Shara), 
using the Space Telescope Imaging Spectrograph (STIS) between April 1999
and May 2000, with a few frames acquired in March 2001.
Low (grating G430L) and intermediate (grating G430M) resolution spectroscopy 
was obtained using the 52$\times$0.5~arcsec$^2$ slit. This setup afforded
5.5-\AA\ resolution spectra in the range 3200-5600~\AA\ and 0.56-\AA\
resolution spectra in the range 3800-4100~\AA . The slit width was chosen to allow
some uncertainty in the target acquisition, in view of the fact that more than
one target was to be observed per slit. 
Several slit positions were used for each cluster. Each slit position was
chosen so as to maximize the number of BSs that fell in the slit.

HST/FOS (Faint Object Spectrograph) spectroscopy of 6 BSs and one HB star in 
NGC~6397 (proposal GO-6697; PI: De Marchi)
was acquired on October 5, 1996 using the G400H (wavelength range 3240--4822~\AA ) 
and G570H (wavelength range 4574--6872~\AA ) gratings with the 0.5-arcsec aperture.
This setup afforded 2.82 and 4.09-\AA\ resolution for the G400H and G570H gratings,
respectively (using the nominal 3.07 and 4.45 \AA\ per diode for the G400H and
G570H gratings, respectively, and the size of the 
post-COSTAR full width at half maximum [FWHM] of 0.92 diodes for the 0.5-arcsec
aperture). 

\begin{figure}
\plotone{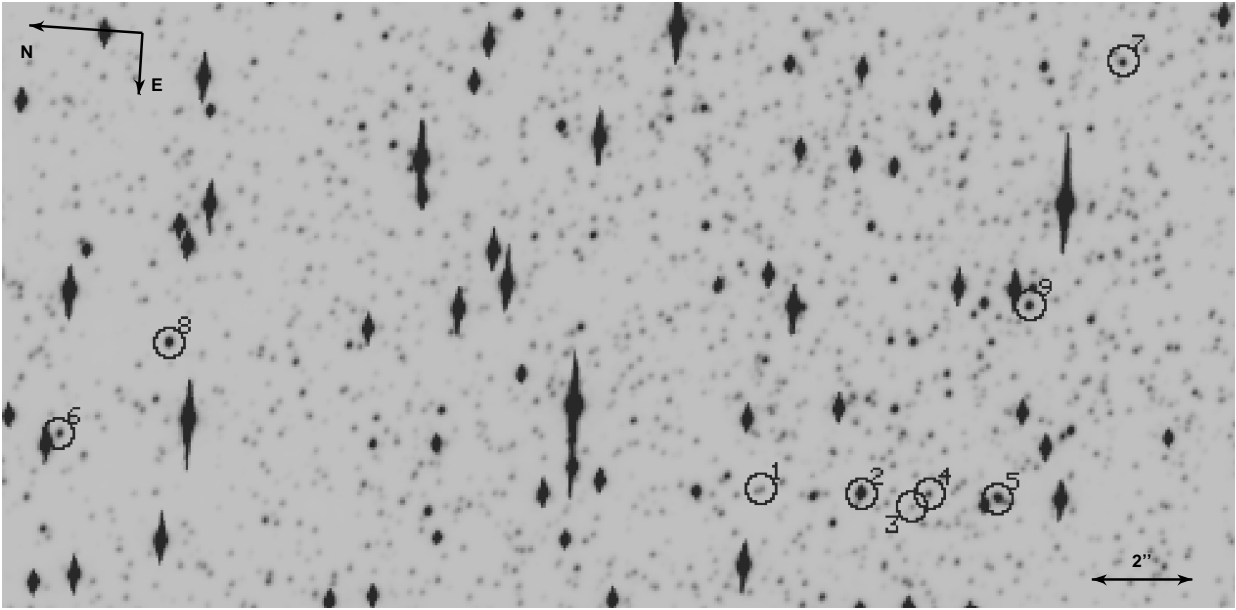}
\caption{An ACS/F435W image of 47~Tuc (j6ll01yiq). The analyzed objects are marked with
a circle and a numerical label, corresponding to the labels in Table~\ref{tab:results}.
N104-1 is a blend of two similar brightness stars, the brighter of which is a BS; N104-2
is heavily blended with a giant whose F555W magnitude is 14.57~mag. 
Blends are discussed in Sections~\ref{sec:identification} and \ref{ssec:errvar}.
\label{fig:47tuc_image}}
\end{figure}

\begin{figure}
\plotone{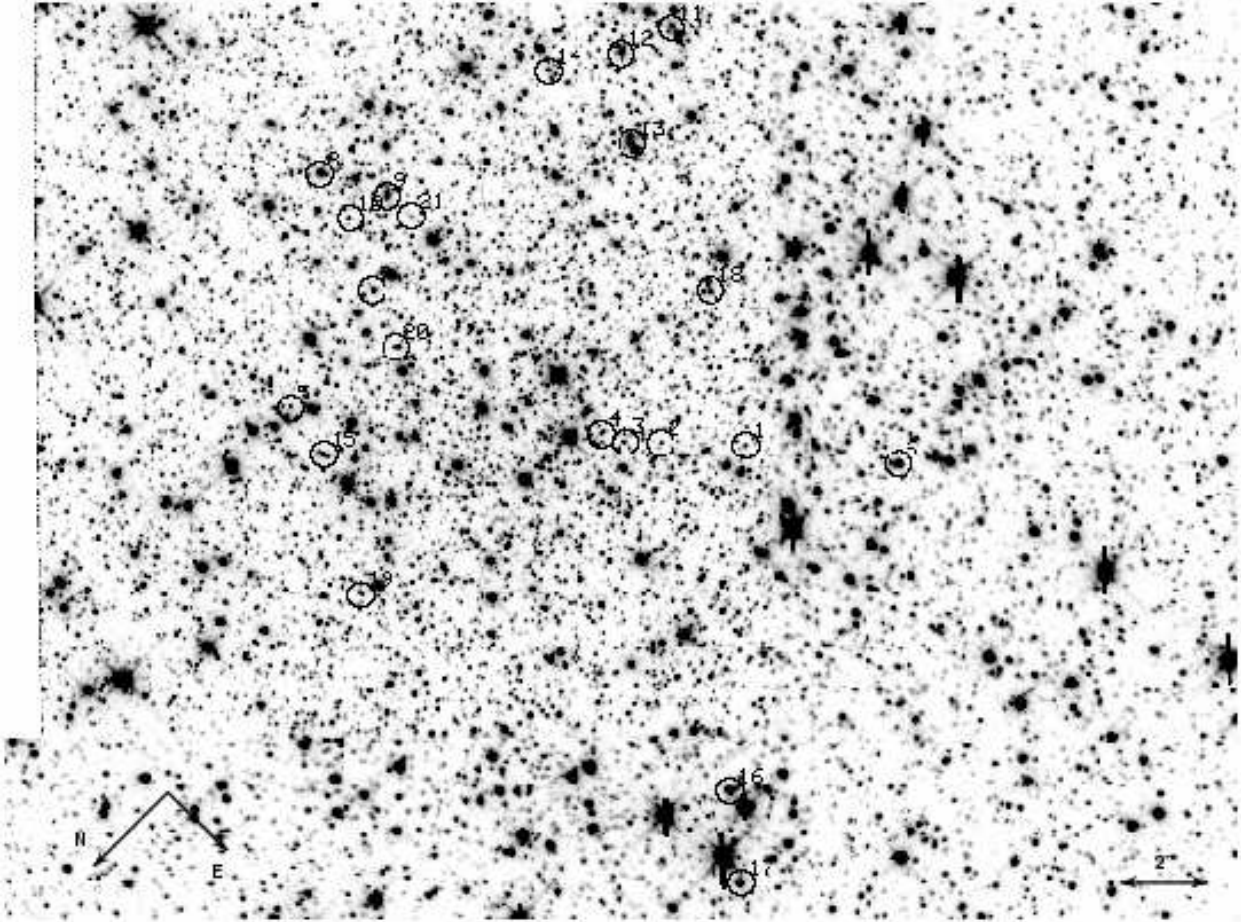}
\caption{A WFPC2/F555W image of M~3 (u2li010ct). The analyzed objects are marked with
a circle and a numerical label, corresponding to the labels in Table~\ref{tab:results}. 
Blends are discussed in Sections~\ref{sec:identification} and \ref{ssec:errvar}.
\label{fig:ngc6752_image}}
\end{figure}

\begin{figure}
\plotone{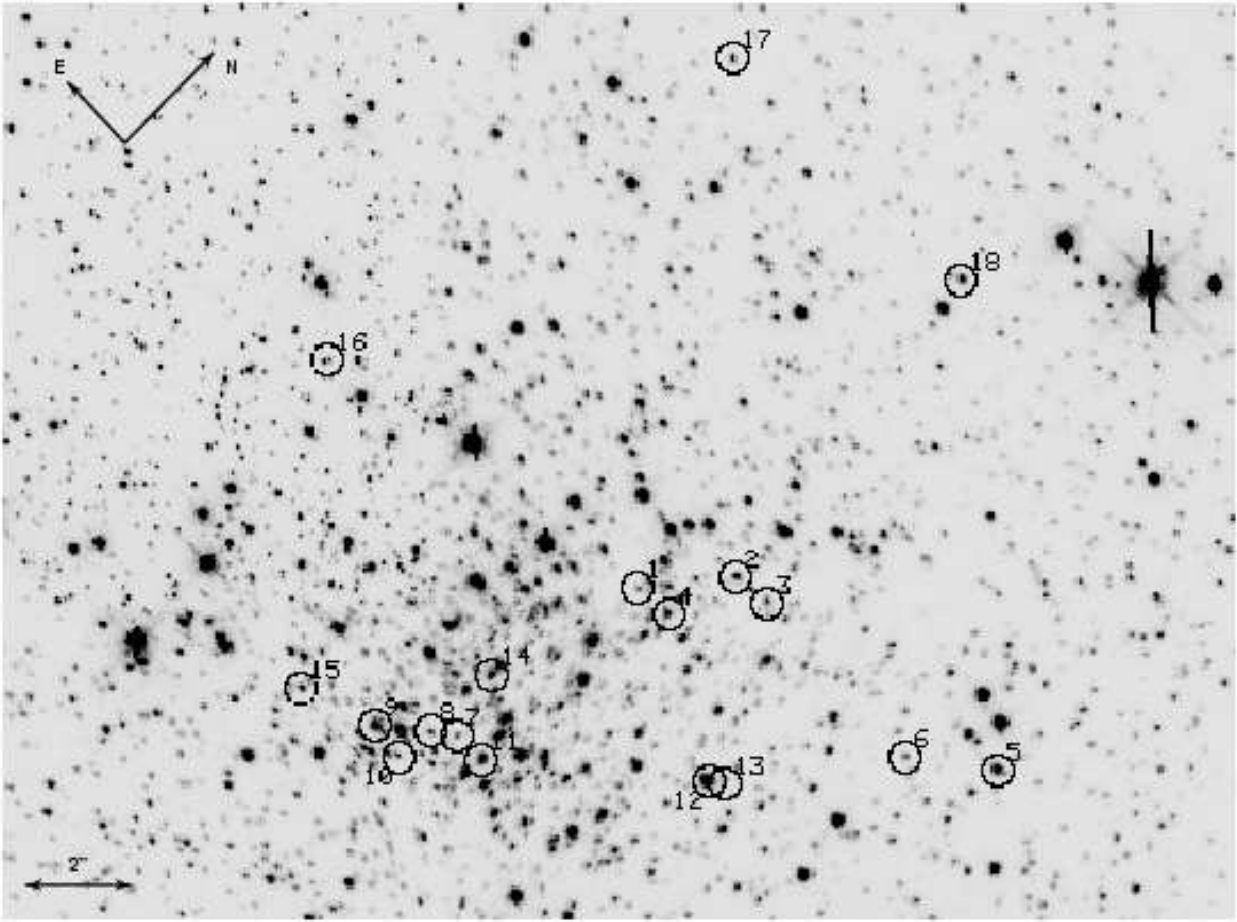}
\caption{A WFPC2/F555W image of NGC~6752 (u5b80301r). The analyzed objects are marked with
a circle and a numerical label, corresponding to the labels in Table.~\ref{tab:results}. 
Star N6752-16 is the brighter of the two that appear within the circle.
Blends are discussed in Sections~\ref{sec:identification} and \ref{ssec:errvar}.
\label{fig:m3_image}}
\end{figure}

\begin{figure}
\plotone{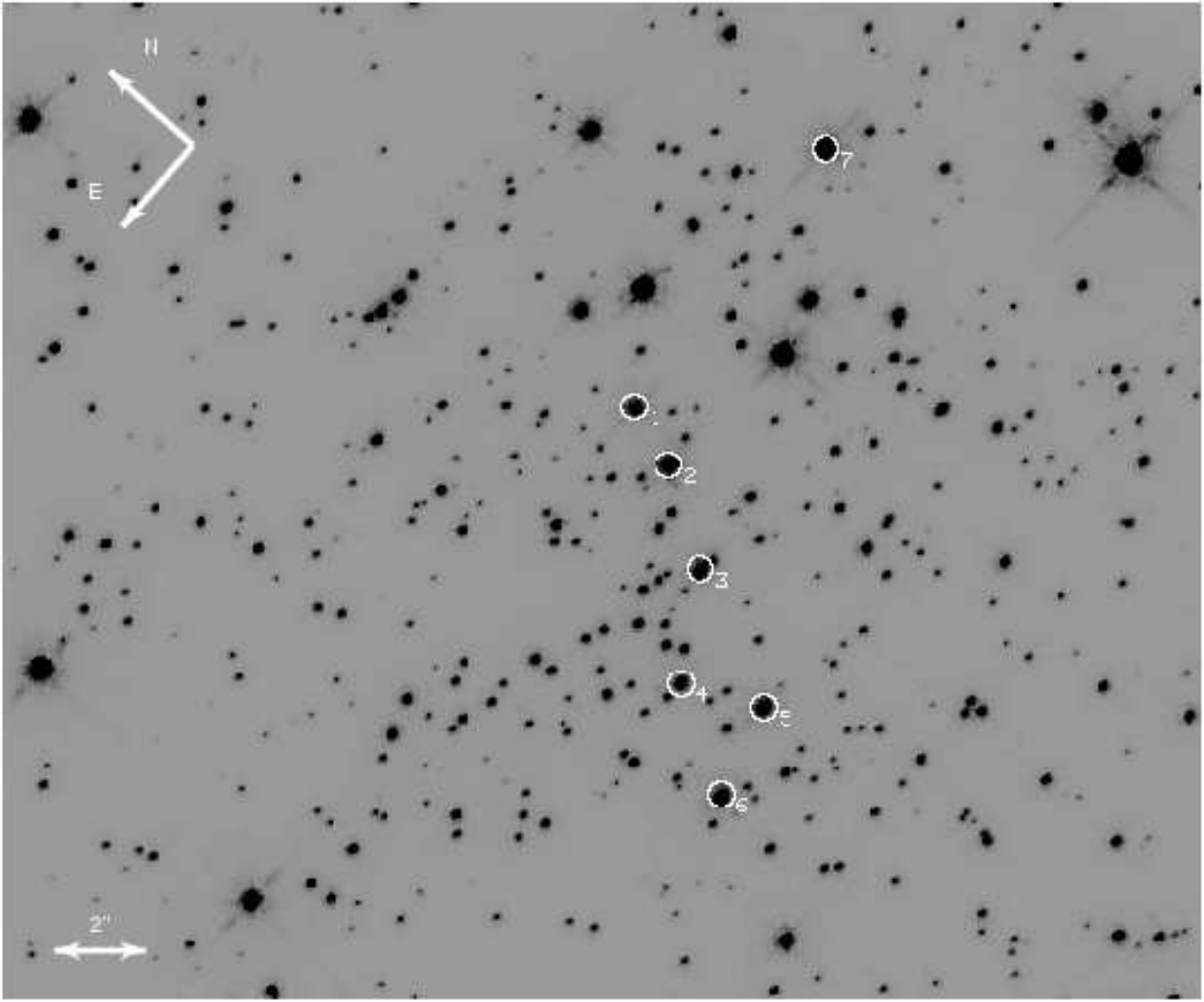}
\caption{A WFPC2/F555W image of NGC~6397 (u33r010kt). The analyzed objects are marked with
a circle (whose width is the same as the FOS 0.5-arcsec aperture width) and a numerical label 
corresponding to the labels in Table.~\ref{tab:results}.
Blends are discussed in Sections~\ref{sec:identification} and \ref{ssec:errvar}.
\label{fig:ngc6397_image}}
\end{figure}

In Table~\ref{tab:observations} we present a log of the observations.
HST/STIS image names are listed in Column 1 followed, in column 2, by
our own stellar identification names. RA \& Dec (column 3) and pixel positions
(column 4) were determined on the images displayed in Figs.~\ref{fig:47tuc_image}
to \ref{fig:ngc6397_image}. The slit positions (column 5) are the y-pixel positions of the
stellar spectra on the STIS images. In column~6 we list distances from the respective clusters'
centers. Finally, point-spread-function (PSF)
photometry carried out on Wide Field and Planetary Camera 2 (WFPC2) 
images is presented in columns 7 to 10 (described in detail
in Section~\ref{sec:photometry}).

\section{Spectra Extraction}
\label{sec:datareduction}

The entire GO-8226 and GO-6697 datasets were downloaded
from the HST archive at the Canadian Astronomy Data Center 
(CADC) in March 2001, with ``on-the-fly" calibration
and processed (Sepinsky et al. 2002) to
reduce the number of cosmic ray hits which were not eliminated by the
automatic pipeline calibration.

Despite the 0.043-arcsec per pixel spatial resolution of the STIS/HST, spectroscopy of 
GC stars is affected by severe blending, so that spectral extraction
is a real challenge and a source of
uncertainty. 

The size of the extracting aperture is a compromise between minimizing
contamination from other stars (width of the aperture $<$5 pixels) 
and insuring the full sampling of the spatial
PSF (width of the aperture between 8 and 12 pixels). 
The PSF shape is wavelength dependent,
with more and more stellar flux residing in the PSF's wings for
longer wavelengths.  Because of this, small apertures tend to extract
spectra which are systematically bluer (less flux falls within the aperture
at longer wavelengths). 
 
\begin{figure}
\vspace{18cm}
\includegraphics{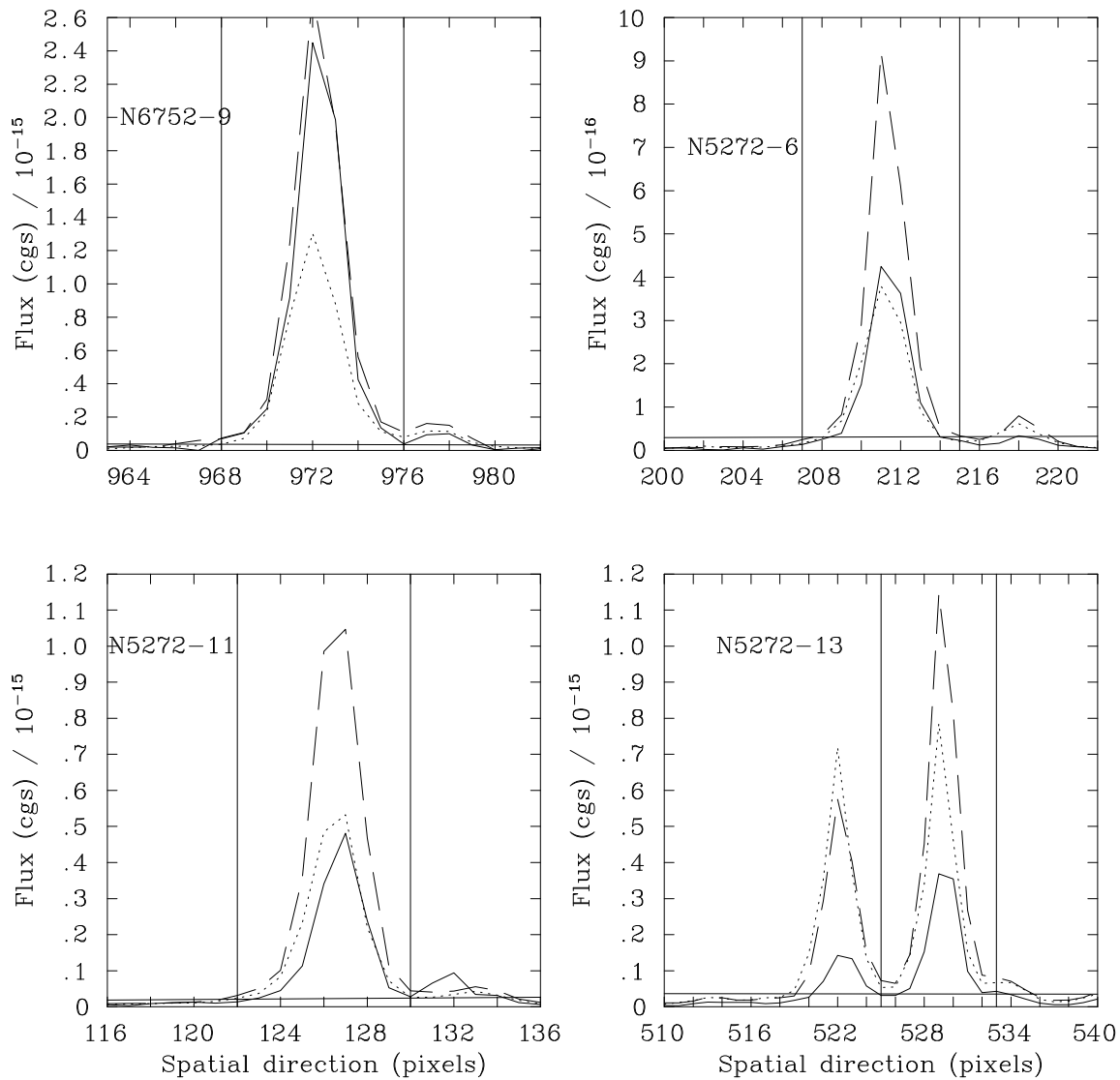}
\caption{Spatial cuts of four STIS spectral images to demonstrate 
extraction techniques. For each panel, cuts at 3300~\AA\ (solid line), 4400~\AA\ (dashed line)
and 5540~\AA\ (dotted line) were taken, corresponding to the central wavelength
of Johnson filters $U$, $B$ and $V$, respectively. The vertical lines show the limits of the
extraction apertures, while the horizontal line shows the background level used
in the extractions.
\label{fig:disk_spatialcut}}
\end{figure}

In Fig.~\ref{fig:disk_spatialcut} we show spatial cuts of the
STIS frames containing
four stars from two clusters taken at three wavelength positions,
corresponding to the central wavelengths of the Johnson $U$, $B$ and $V$ filter band passes.
The limits of the apertures used to extract the 4 stars, as well as the          
adopted background level are also shown.
In three out of four cases the star is blended with a very faint neighbor
which is not bright enough, nor close enough to contribute substantial flux.
In the fourth case, the neighbor is brighter, but again, not close enough
to constitute a problem. 

\begin{figure}
\vspace{11cm}
\includegraphics{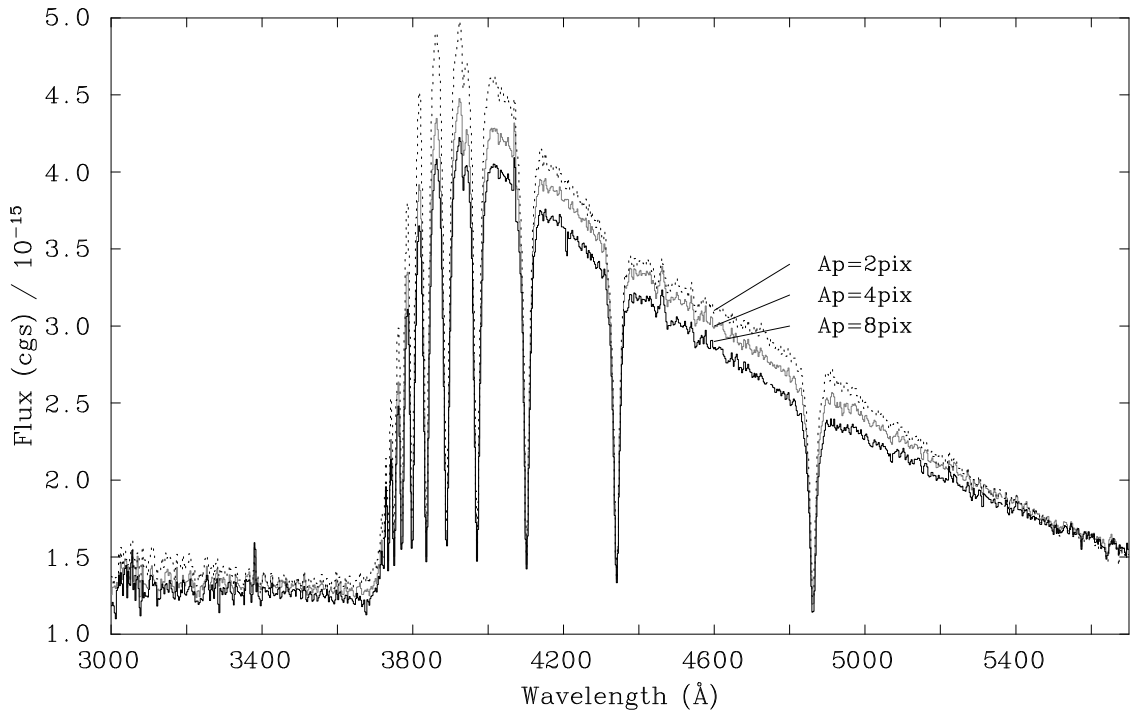}
\caption{Three extractions of the spectrum of star N5272-11 using
different size apertures (labeled). The smallest aperture excludes blends but has
an artificially bluer spectrum due to the wavelength-dependent nature of the PSF.
The 8-pixel aperture is optimal, as it contains virtually all of the PSF, suffering
minimum distortion and acceptable blending.
\label{fig:aperture_test}}
\end{figure}

In Fig.~\ref{fig:aperture_test} we present star N5272-11, extracted using three
aperture widths: 2 pixels, 4 pixels and 8 pixels. As already described, the
flux in the smaller two apertures is bluer and somewhat distorted. The large
8-pixel aperture allows flux from one faint neighbor to contribute. It was found that
apertures larger than 8 pixels increase contamination but the shape of the spectrum remains unchanged.
Hence, 8-pixel apertures are large enough to not incur the PSF problem, but small enough 
to not incur severe blending.

Overall, we maintain that we have carried out the best reduction and extraction that
can be carried out with these data. A better reduction and extraction demands
the use of software that can deconvolve blended spectra.
During the analysis of the data, great attention
has been put into checking the individual extractions before attributing an uncertainty to the 
derived parameters. We have carefully tracked the uncertainties
and propagated them into our results (Section~\ref{sec:uncertainties} and Table~\ref{tab:results}).

\section{Photometric observations}
\label{sec:photometry}

In order to determine spectroscopic masses one has to scale the stellar atmosphere model fluxes
to observed (de-reddened) stellar fluxes. If the spectra are photometric, one can rely on the spectroscopy alone. 
This is not the case 
for spectra acquired in crowded regions, or when there is a risk that the star is not
centered on the slit. For this reason we scaled our models
(Section~\ref{sec:stellarparameters}) not to the spectral fluxes but to stellar photometry
from archival WFPC2 images. We did not expect BSs to show significant variability, but some did.
This is in itself an important result of this work.
With hindsight, one should always
take photometric observations within a short time of the spectroscopic ones, since
parameters derived from spectra taken at a particular variability phase should not be
mixed with the brightness derived from photometry taken at a different phase.
Variable stars will be discussed in Sections~\ref{sec:identification} and \ref{sec:uncertainties}.

PSF photometry was carried out on a subsample of all the available archival HST/WFPC2 images of the four clusters.
A photometric reduction using all the available images
will be undertaken elsewhere. The data 
(Table~\ref{tab:photometry}) were obtained from the
MAST database with ``on-the-fly" calibrations. The photometric reductions were performed 
using the DAOphot/ALLFRAME software
(Stetson 1987). Briefly,
ALLFRAME determines the brightness of each star on each frame 
with point-spread-function fitting at the same time. This reduction process
increases the accuracy of the photometry by fixing the position of each star
across all the frames which increases the accuracy of de-blending the stars.

We calibrated the photometry to the VEGAmag system 
(WFPC2 data handbook\footnote{http://www.stsci.edu/instruments/wfpc2/Wfpc2\_hand/wfpc2\_handbook.html.}) in order to 
use the fewest corrections when
scaling the atmosphere models (Section~\ref{sec:stellarparameters}) to the observations.
The uncertainty of the
photometric magnitudes is estimated to be 0.02~mag for the $F555W$ and $F814W$ filters, 0.03~mag for
the $F439W$ filter and 0.04~mag for the $F336W$ one.

A calibration to any ``standard" system (e.g., Johnson)
includes many additional corrections because the WFPC2 filters are not
strictly ``standard". To convert the WFPC2 magnitudes listed in Table~\ref{tab:observations} to the 
Johnson system, one can use 
the transformations of Dolphin (2000a,b)\footnote{Also explained at the site: http://purcell.as.arizona.edu/wfpc2\_calib\/.}.

The observed STIS/VEGA $F336W$, $F439W$, $F555W$ and $F814W$
magnitudes obtained from averaging {\it all} images from Table~\ref{tab:photometry} 
are listed in Table~\ref{tab:observations} (columns 7 to 10). 
Stellar atmosphere models were scaled to the \V\
photometric magnitudes to obtain radii, luminosities and spectroscopic masses. 
For variable stars, it is not appropriate to obtain radii by scaling models to photometric magnitudes obtained
at a different time from the spectra. For variable stars we determine radii, luminosities and masses 
by scaling the models to the spectra.
In Section~\ref{sec:uncertainties} we will discuss in detail the uncertainties in this method.

\section{Stellar identification and classification}
\label{sec:identification}

The analyzed stars were identified on WFPC2 images 
by using the {\sc iraf} task {\it siapier}
to position the STIS slit on the image and match stars along the slit to spectra
in the STIS CCD images. For the FOS observations we relied on the coordinates provided in the image
header.
In Figs.~\ref{fig:47tuc_image} to \ref{fig:ngc6397_image} we present
portions of \V\ images containing the stars thus identified, which we label according to the
scheme presented in Table~\ref{tab:observations}.
RA (to the nearest tenth of second) and Dec (to the nearest tenth of arc-second),
as well as pixel coordinates (to the nearest pixel) for all analyzed stars were read off 
the WFPC2 images displayed using {\sc ds9} (Joye \& Mandel 2003)
and are also listed in Table~\ref{tab:observations}.

All stars in the instability 
strip are susceptible to stellar pulsations.
Variable HB stars (RR Lyrae) have by far the largest 
brightness amplitudes. They are typically in the temperature
range $\sim$6000--7300~K and have periods between 0.4 and 1~day. Variable pulsating
main sequence stars fall into the $\delta$~Scuti or SX~Phoenicis 
classes. These tend to pulsate with smaller amplitudes
($\sim$0.1~mag), although they can occasionally vary more
(0.4~mag) and they 
have periods between 0.04 and 0.2~days. The typical temperatures for main sequence pulsations are
marginally hotter than those of HB stars, and are in the range $\sim$7000-8200~K (from Allen's Astrophysical Quantities, 2000).
Stars can obviously also vary if they are in eclipsing binary systems.

To identify which ones of our stars are variable, we adopted four methods.
(i) Stars that vary by more than 0.2~mag over multi-epoch same-filter WFPC2 images 
(stars N104-8, N5272-8, 9, 12, 14, 15, N6397-5 and 6 in Table~\ref{tab:results}). This method 
does not detect all variables, since it depends on when the individual frames were acquired. 
(ii) Stars whose colors clearly vary when combining \U, \B\
and \V\ magnitudes from images taken at different dates (stars N104-1, 3, 6, N5272-8, 9, 12, 14, 15, 18, 20,
N6752-17, N6397-1, 2, 5, 6, and 7).
(iii) Stars whose low- and intermediate-resolution spectra have different 
flux levels (stars N104-8, N5272-13) and, finally,
(iv) stars whose spectral and photometric fluxes and colors 
are very different (stars N104-2, N5272-13, N6752-5, 9, N6397-1, and 2),
but the difference is inconsistent with other
causes (explained in detail in Section~\ref{ssec:errvar}).
Stars which satisfied any of these four criteria are considered variable and 
are marked ``v" in Table~\ref{tab:results}
and discussed further in Section~\ref{sec:uncertainties}.

The core of 47~Tuc was surveyed for variables by Edmonds et al. (1996).
None of our 47~Tuc stars matches the positions of the variables found
in that study, despite the fact that, based on the criteria listed above,
five of the 9 stars are variable.
(We note in passing that the positions quoted by Edmonds et al. [1996; their Table~2]
are with reference to star E in Guhathakurta et al. 1992, but the position of that star quoted in
that paper [their Table~3: 00$^h$21$^m$52$^s$.46 --72$^o$21$^{'}$31$^{''}$.68 [B1950.0] which maps to 
00$^h$24$^m$05$^s$.38 --72$^o$04$^{'}$54$^{''}$.30 [J2000.0]] does not correspond to
any star on the J2000.0 coordinate system of our CCD frame [e.g., j6ll01yiq]. This is not unexpected due to 
the positional uncertainty of the DSS1 system. We therefore re-downloaded their
dataset to find that there is a coordinate shift. Their star E corresponds to a bright star in J2000.0 position
00$^h$24$^m$05$^s$.36 --72$^o$04$^{'}$51$^{''}$.68.)   

For 47~Tuc and NGC~6397 we constructed \B -- \V\ vs. \V\ CMDs from images 
acquired on September 1, 1999 and March 6-7, 2001, respectively (Figs.\ref{fig:cmd_ngc104_bv} and
\ref{fig:cmd_ngc6397_bv}).
For M~3 we constructed a \V--\I\ vs. \V\ CMD from observations acquired on May 14, 1998 (Fig.\ref{fig:cmd_ngc5272_vi}).
For NGC~6752 we constructed a \U--\V\ vs. \V\ CMD from
observations acquired on March 22, 2001 (Fig.\ref{fig:cmd_ngc6752_uv}).
These ``same-epoch" CMDs were constructed to minimize the variability and it is on
them that we classified our stars as
BS, HB (EHB for extreme HB),
and turn-off (TO; ATO for ``above turn-off"; column 3 in Table~\ref{tab:results}). 
Star N104-3 has no clear classification based on its CMD position and
we mark it with a
``?" in Table~\ref{tab:results}.
To confirm the classifications we also
constructed CMDs with different color combinations using all the available photometry (Table~\ref{tab:photometry};
Figs.\ref{fig:cmd_ngc104_uv}, \ref{fig:cmd_ngc5272_uv}, 
\ref{fig:cmd_ngc6752_bv} and \ref{fig:cmd_ngc6397_uv}). 

Cluster membership could not be determined using the radial velocity of
the absorption lines of the stars in our sample, because we cannot derive
a sufficiently precise wavelength calibration from our spectra. The reason
for this is that the STIS spectra are taken through the wide 0.5-arcsec
slit and a star's wavelength is a function of its exact position in slit.
We can, however, assess that foreground contamination is close to zero.
From the total number of stars in the Guide Star Catalogue (version II;
170,000,000 to a limiting photographic J magnitude of 19.5 [
$\lambda_0=0.44~\mu m$]) we can estimate that there is $\sim$1 star
arcmin$^{-2}$ on average in the sky.
All our stars are two or more magnitudes brighter. Furthermore,
close to the high galactic latitudes characteristic of our four GCs
(28$^o$$<|l|<$72$^o$) there are often zero. 

Following the Referee's suggestion, we also used the calculations by Robin et al. (1996)
\footnote{A useful interface to their program can be found on-line at
http://www.obs-besancon.fr/modele/.} to estimate the number of foreground stars 
in the direction of the four GCs. Using the GCs'
galactic coordinates, a distance smaller than 10~kpc, and restricting the stellar
types to the apparent $V$ magnitude and color ranges of the stars in our
list. The simulated number of foreground stars is zero
for solid angles comparable to the angles subtended by the GC cores (typically
1~arcmin$^2$). Even enlarging the solid angle to 10~deg$^2$ the number of predicted stars
with the required characteristics was only about 10 for 47~Tuc and M~3 and zero for the
other two clusters.
We
therefore conclude that, although we cannot be certain that
every one of our stars belong to its cluster, it is
very unlikely that even 1 star per cluster is a foreground contaminant.

\section{Model Atmospheres and Stellar Parameters} 
\label{sec:stellarparameters}

All the selected spectra were de-reddened using $E(B-V)$ values
for their home clusters obtained from the literature ($E(B-V)$ = 0.032, 0.016, 0.056 and 0.18~mag for 
47~Tuc, M~3, NGC~6752 and NGC6397, respectively; see Table~\ref{tab:clusterproperties}) and fitted by synthetic
spectra convolved with Gaussian profiles to mimic the instrumental resolution (FWHM=5.46 and 0.56~\AA , for the
STIS G430L and G430M grating, respectively [2-pixel resolution], and 2.82 and 4.09~\AA\
for the G400H and G570H FOS gratings, respectively).

The model spectra were calculated with the Hubeny~\& Lanz (1995) spectrum 
synthesis code SYNSPEC\footnote{Hubeny \& Lanz, 2003, 
http://tlusty.gsfc.nasa.gov},
based on Local Thermodynamic Equilibrium (LTE) as well as on non-LTE (NLTE)
model atmospheres.

We started by calculating a grid of synthetic spectra for the parameter range,
5000 $\le T_{eff} \le$ 25\,000~K, $2.0 \le \log g \le 5.0$ (lowest gravities only for 
the coolest models), and metallicities, [Fe/H]=$-$0.83 (for 47~Tuc; VandenBerg 2000), 
$-$1.5 (for M~3 and NGC~6752), and $-$1.97 (for NGC~6397;
Harris [1996] quotes --1.57, --1.56 and --1.95, for M~3,
NGC~6752 and NGC~6397, respectively, similar to the values used),
using
LTE model atmospheres calculated with Kurucz ATLAS9 program (Kurucz 
1993). To achieve this,
we selected the ``ODFNEW'' grids of model atmospheres with
[Fe/H]=$-$1, $-$1.5, and $-$2 from the Kurucz
Web site\footnote{http://kurucz.harvard.edu/grids.html} and used them to
calculate detailed spectra
with the SYNSPEC code, scaling solar abundances for all
elements heavier than helium to the three selected metallicities. The 
helium and
$\alpha$-elements abundances were Y=He/H=0.24 (mass fraction), and 
[$\alpha$/H]=0.3
for all clusters, following the prescription of Bergbusch \& VandenBerg (2001). 
We cannot measure micro-turbulence
from our data, so we adopted a value of 2 km~s$^{-1}$ which is a good 
compromise from
previous studies of F and G stars (e.g., Edvardsson et al. 1993).

\begin{figure}
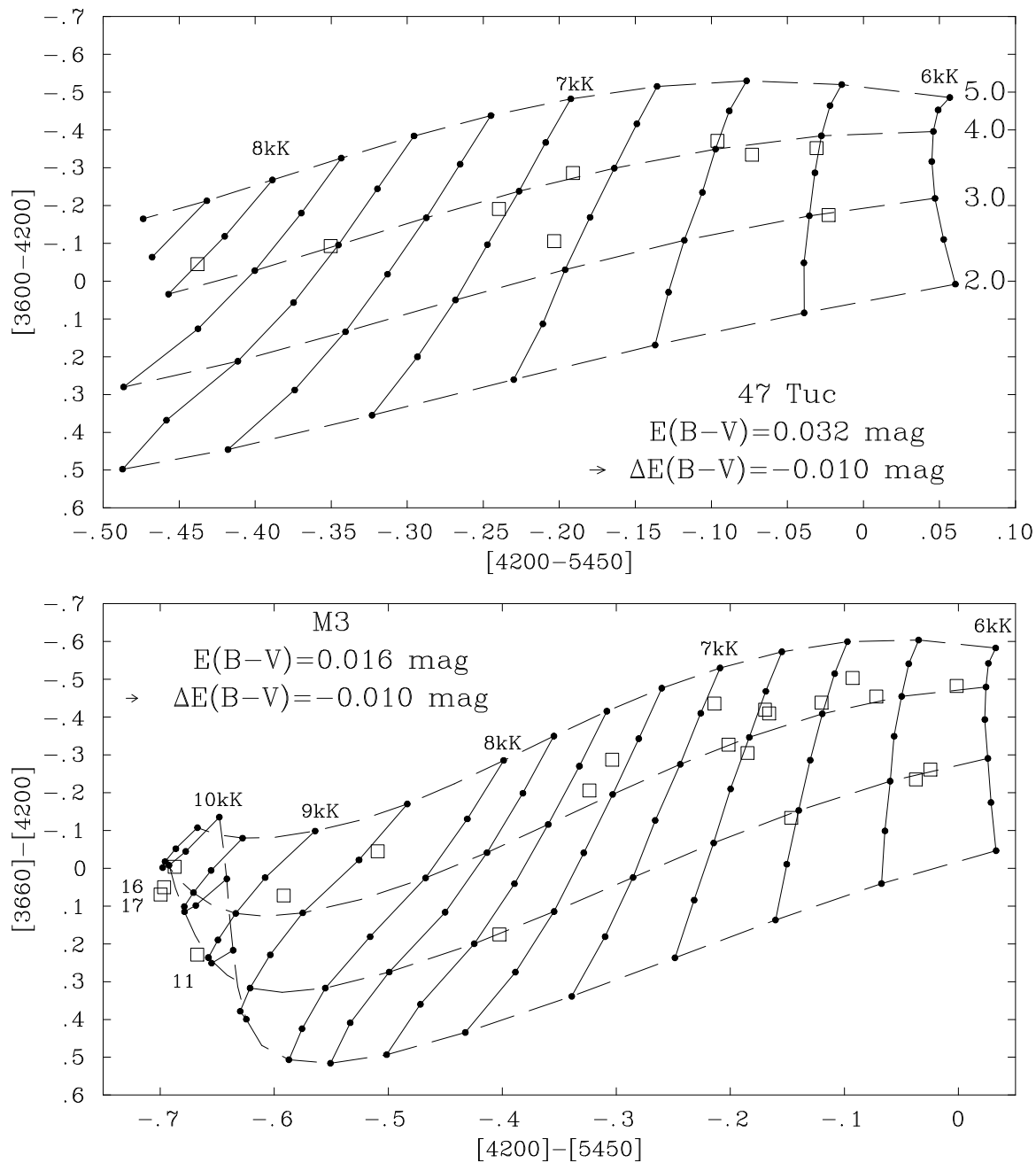

\vspace{17cm}
\includegraphics{f15a.eps}
\includegraphics{f15b.eps}
\caption{Monochromatic color-color diagram of the 47~Tuc (top) and M~3
(bottom) stars (large empty symbols) and
synthetic LTE models (small round filled symbols where solid lines join models of the 
same effective temperature and dashed lines join models of equal gravity). The models'
effective temperatures are labeled every 1000~K, while model's gravities are labeled on the right of the top plot every
1.0~dex. 
The adopted reddening is indicated, as is an arrow showing the shift in the data for a change in the
adopted reddening by -0.01~mag. The positions of stars NGC5272-11, 16 and 17 make them candidates for
circumstellar disks or high rotation (Section~\ref{sec:disk} and De Marco et al. [2004]).
\label{fig:col-col1}}
\end{figure}

\begin{figure}
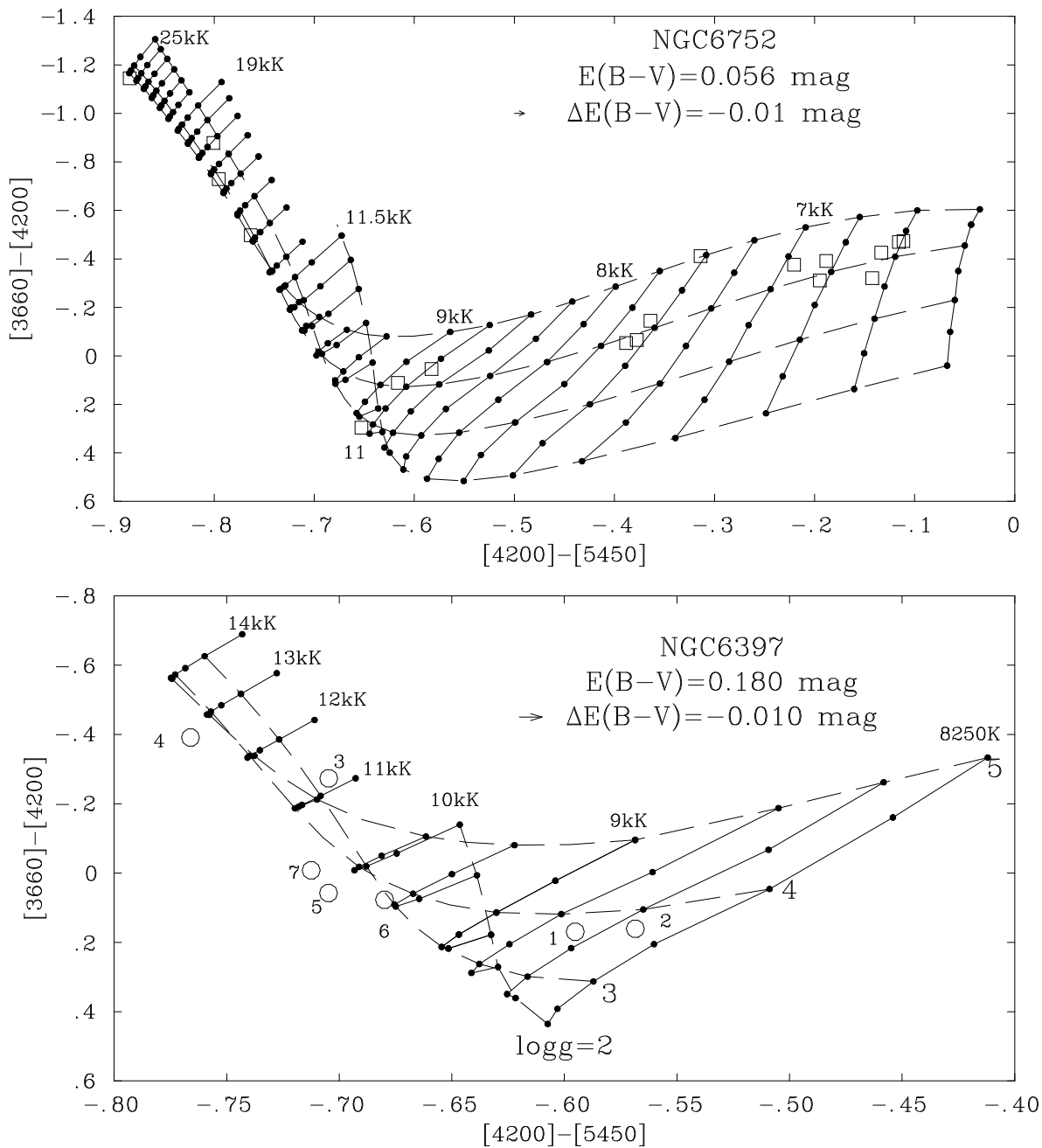

\vspace{17cm}
\includegraphics{f16a.eps}
\includegraphics{f16b.eps}
\caption{Monochromatic color-color diagram of the NGC~6752 (top) and NGC~6397 (bottom) stars (large empty symbols) and
synthetic LTE models (small round filled symbols where solid lines join models of the
same effective temperature and dashed lines join models of equal gravity). The models'
effective temperatures are labeled, while models' gravities are labeled in the bottom plot every \logg = 1.0.
The adopted reddening is indicated, as is an arrow showing the shift in the data for a change in the
adopted reddening by -0.01~mag. The positions of stars NGC6752-11 and NGC6397-4, 5, 6 and 7 make them candidates for
stars with circumstellar disks or high rotation (Section~\ref{sec:disk} and De Marco et al. [2004]).
\label{fig:col-col2}}
\end{figure}

In Figs.~\ref{fig:col-col1} and \ref{fig:col-col2}, we show mono-chromatic color-color 
diagrams including colors for all the LTE synthetic spectra
as well as the de-reddened stars (Table~\ref{tab:col-col}). We 
have
defined new, narrow-band magnitudes which we call [3600], [4200], and [5450], 
which
are constructed by averaging the theoretical and the observed fluxes in 
20-\AA\
wide spectral bands centered at 3600, 4200, and 5450\,\AA\  (these are 
the Johnson
bandpass centers, except for [4200] for which we used 4200~\AA\  
instead of
4400~\AA, thereby avoiding \ion{He}{1} $\lambda$4388, a line which is 
sensitive
to the adopted helium abundance in the hottest models), and converting 
them to
magnitudes using the same zero-points than Johnson UBV magnitudes 
(Bessel et al. 1998):

\[ [3600] = -2.5 \log f_{3600}-21.0-0.77 \]
\[ [4200] = -2.5 \log f_{4200}-21.0-0.12 \]
\[ [5450] = -2.5 \log f_{5450}-21.0 \]

Compared to the more familiar Johnson colors, these quasi-monochromatic 
colors
have the advantage that they are more sensitive to the continuum 
spectrum shape
from which we derive the stellar parameters, while the large breadth of 
the Johnson
bandpasses include many additional spectral lines that might be poorly 
matched
by these initial models (for instance, because of inadequate abundances 
of some
chemical species). 

We used these diagrams 
to derive
initial parameters for all stars. Most stars in our sample have colors 
consistent
with the models, that is, they are located within the model locii. 
There are, however,
a few stars whose colors do not match any of the models. These stars 
will be
discussed in detail in Section~\ref{sec:disk}.

For each star, we then recalculated a LTE ATLAS9 model atmosphere and 
the
detailed emergent spectrum using the stellar parameters derived from a comparison of
stellar and model colors. If the synthetic spectrum did not match the observations
to a high degree of accuracy, we calculated new models until agreement was reached
(often mismatches between model and observed colors were due to low SNR in
the faintest stars). The fitting process
was carried out by eye, using the Paschen continuum slope, the Balmer jump,
and the Balmer line wings from the higher resolution spectroscopy. Note that
their dependence on the basic parameters changes from hot to cool
stars (as seen in the color-color diagrams, Figs.~\ref{fig:col-col1} and \ref{fig:col-col2}). 
Accordingly, the accuracy of the derived parameters varies across the parameter space, and this is
reflected in the attributed uncertainties (see Table~\ref{tab:results} and Section~\ref{sec:uncertainties}).

Finally, we computed NLTE model atmospheres with the code TLUSTY, version
198 (Hubeny~\& Lanz 1995). TLUSTY assumes a static atmosphere, plane-parallel geometry,
hydrostatic and radiative equilibria. TLUSTY allows for explicit departures from LTE
for an extensive, user-selected set of species and levels. The present NLTE model atmospheres
explicitly incorporate hydrogen, helium, carbon, and iron. Table~\ref{tab:atom}
lists the detail of the atomic makeup of the model atmospheres. For Fe~{\sc i} and {\sc ii} we have 38 and 35
super-levels, respectively. Iron line blanketing is treated using opacity distribution functions
(Hubeny \& Lanz 1995). Additional sources of 
opacity include H$^-$ and H$_2^+$. The micro-turbulent velocity was maintained at 2~km~s$^{-1}$, and
convection was neglected. The abundance mix is chosen as for the LTE models.

For the hotter models, we found that the predicted LTE and NLTE 
model spectra are quite
similar, indicating that departures from LTE remain limited. 
The essential NLTE effects are small changes in the
predicted fluxes and in the $B-V$ colors, i.e., in the amplitude of the 
Balmer jump. For a calculation with T$_{eff}$=9000~K and $\log g$=3.3 the LTE model
is 3\% brighter at $V$ but 6\% brighter at $U$, making the Balmer jump slightly shallower.
This effect is alleviated for hotter calculations: for a T$_{eff}$=18\,000~K and $\log g$=4.3
the LTE atmosphere is 2\% brighter at $V$ and 5\% brighter at $U$. 
The different treatment of opacities is likely the cause of the different $V$ 
fluxes, while departures from LTE are more likely the cause of
the relative changes in the Balmer jump. These changes translate into slightly altered values of
the radii and masses obtained when using LTE and NLTE atmospheres and, while they
are smaller than the random uncertainty
(Section~\ref{sec:uncertainties}) the use of LTE, rather than NLTE, stellar atmosphere models 
would have introduced a {\it systematic}
source of error.

We should point out at this stage that
TLUSTY is a model atmosphere code originally designed to
model hot stellar atmospheres. We found that model convergence was often hard to
achieve for cool models ($T_{eff}$ $\le$ 8000~K) with the current version. 
The basic physical reason behind this difficulty is that hydrogen is partly ionized in these 
atmospheres, resulting in a strong, highly non-linear coupling between radiative 
equilibrium and the hydrogen equilibrium (H, H$^+$, H$_2^+$, and H$^-$). 
Spurious results might be derived from models that are not truly converged.
For instance, a poor conservation of the total flux would yield an
incorrect temperature stratification and result in an erroneous predicted
emergent spectrum. 

We exercised special care when converging
NLTE cool models, carrying out
additional iterations and changing the default limit between the
integral and the differential form of the radiative equilibrium equation
(see Hubeny~\& Lanz 1995) to ensure a satisfactory conservation of the total flux.

After obtaining $\log g$ and $T_{eff}$ from the spectral fits (columns 4 and 5 in Table~\ref{tab:results}), 
the stellar radii were obtained by scaling each model's flux ($H_\lambda$) to the de-reddened 
\V\ PSF photometric magnitudes or, for variable stars, to the spectroscopic fluxes at 5550~\AA\ ($f_\lambda$). 

From the scaling factor, the radius can be obtained using:

\begin{equation}
\left( \frac{R}{D} \right) ^2 =  \frac{1}{4 \pi} \frac{f_\lambda}{H_\lambda}
\end{equation}

where $f_\lambda$, $H_\lambda$, $R$ and $D$ are the observed (de-reddened) and theoretical stellar fluxes,
the stellar radius and the distance from Earth, respectively. 
From the radii and effective temperatures, 
the stellar luminosities ($L$) were obtained from the Boltzmann formula:
\begin{equation}
L = 4 \pi R^2 \sigma T_{\rm eff} ^4, 
\end{equation}

where $\sigma$ is the Stefan-Boltzmann constant, while from the radii and gravities,
we obtained the stellar spectroscopic masses:
\begin{equation}
M = g R^2 / G, 
\end{equation}
where $G$ is the gravitational constant. Radii, luminosities and spectroscopic masses
thus obtained are listed in columns 6 to 8 of Table~\ref{tab:results}.

While the metallicity of the stars is kept at
the adopted values, the strength of the theoretical He~{\sc i} lines is clearly too large to reproduce these lines 
in the high resolution spectra of stars with $T_{eff} >$ 12000~K and 
was therefore reduced to Y=He/H=0.014 (by mass). This behavior has been previously reported by, e.g.,
Behr et al. (1999). 

\begin{figure}
\vspace{20cm}
\includegraphics{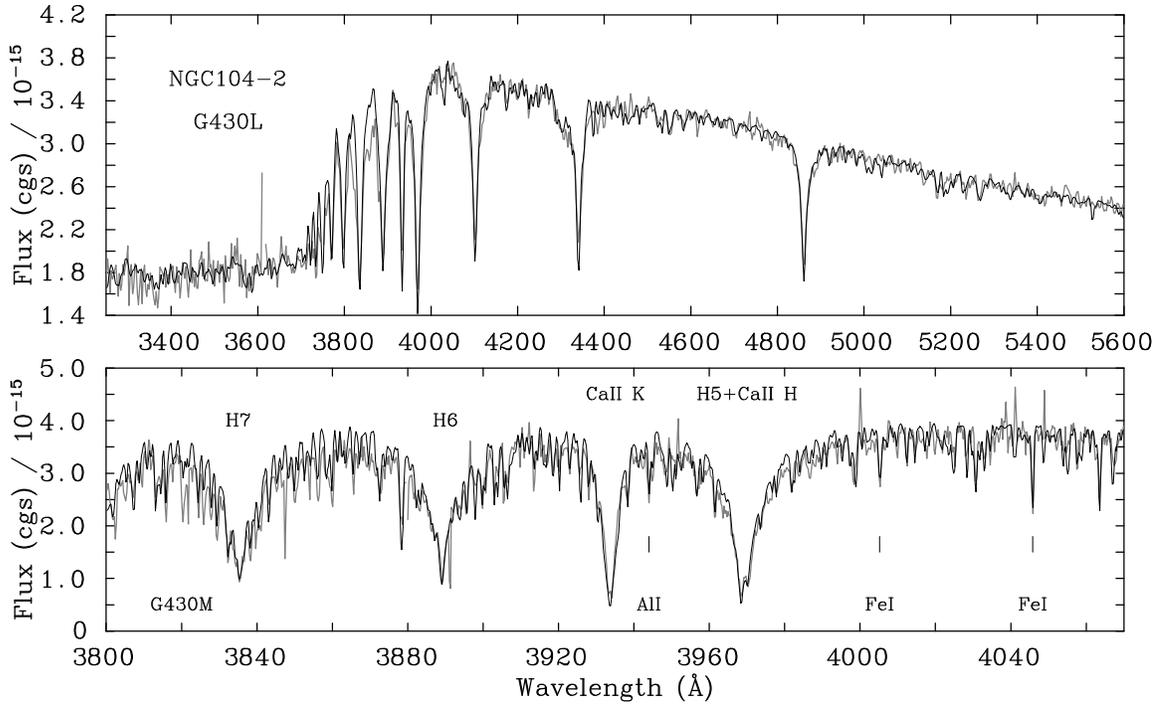}
\caption{A fit (thin black line) to the low resolution (top; thick gray line) 
and the intermediate resolution (bottom; thick gray line) 
spectra, of TO star NGC104-2 in 47~Tuc. The derived stellar parameters
are T$_{\rm eff}$ = (7040$\pm$100)~K and $log g$=3.85$\pm$0.15. 
\label{fig:ngc104_spectrum}}
\end{figure}

\begin{figure}
\vspace{20cm}
\includegraphics{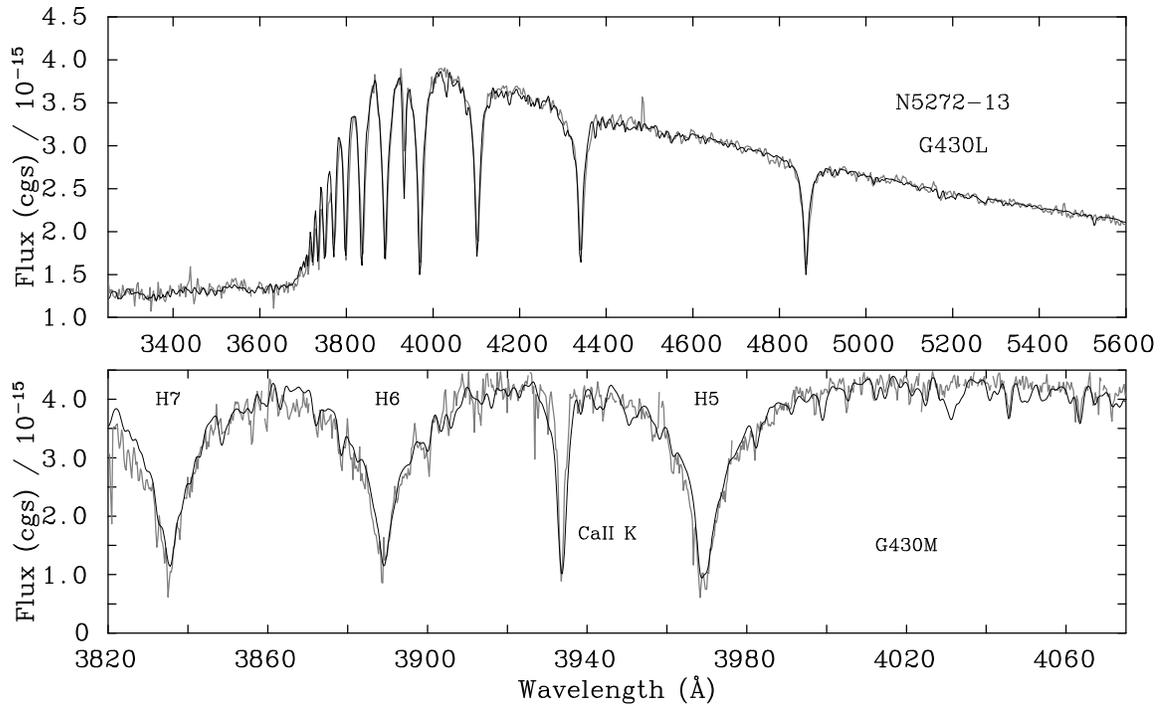}
\caption{A fit (thin black line) to the low resolution (top; thick gray line)
and the intermediate resolution (bottom; thick gray line)
spectra, of the BS NGC5272-13. The derived stellar parameters
are T$_{\rm eff}$ = (7050$\pm$100)~K and $log g$=2.6$\pm$0.15.
Model spectra 
are convolved with a Gaussian with FWHM=100~\kms\ and
Poisson noise so as to simulate the SNR of the data.
\label{fig:ngc5272_spectrum}}
\end{figure}

\begin{figure}
\vspace{20cm}
\includegraphics{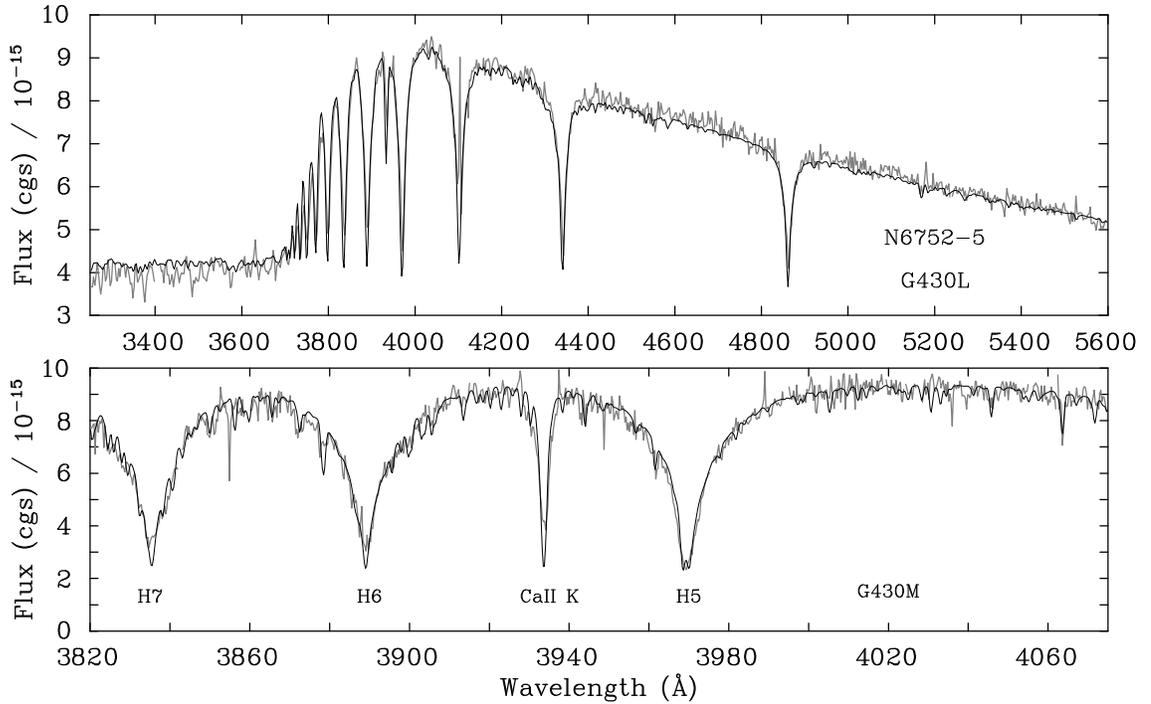}
\caption{A fit (thin black line) to the low resolution (top; thick gray line)
and the intermediate resolution (bottom; thick gray line)
spectra, of EHB star N6752-5. The derived stellar parameters
are T$_{\rm eff}$ = (7250$\pm$100)~K and $log g$=(3.70$\pm$0.10). 
Model spectra
are convolved with a Gaussian with FWHM=50~\kms, which improves the
fit to the metal lines. Choosing FWHM=100~\kms\ would further improve the fit
to the Ca~{\sc ii} H line.
\label{fig:ngc6752_spectrum}}
\end{figure}

\begin{figure}
\vspace{15cm}
\includegraphics{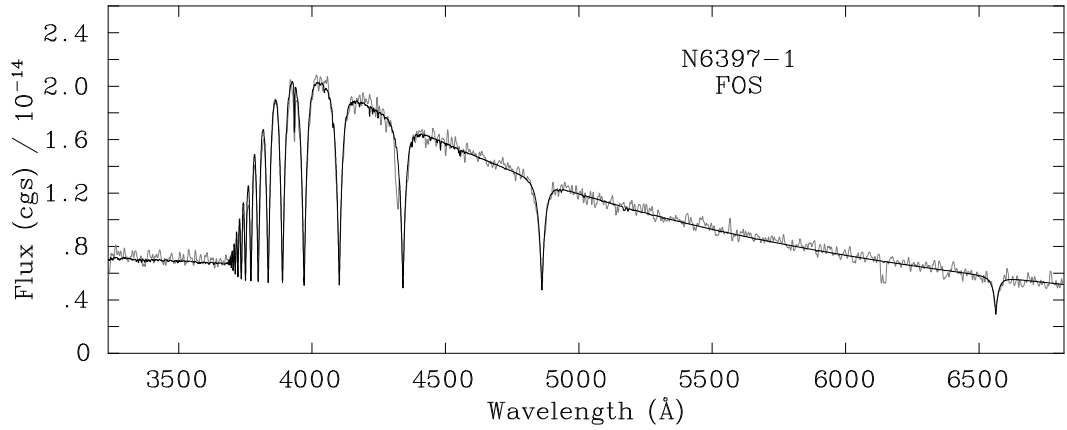}
\caption{Non-LTE model fits (thin black line) to the HST/FOS N6397-1 spectrum (thick gray line).
The derived parameters are \teff = (8600$\pm$100)~K and \logg = 3.7$\pm$0.1.
\label{fig:ngc6397_spectrum}}
\end{figure}

Except for the stars discussed in Sections~\ref{sec:disk} and \ref{sec:individualobjects},
model fits to all of the analyzed stars are excellent. Four such fits
(one per cluster) are presented in Figs.~\ref{fig:ngc104_spectrum}, \ref{fig:ngc5272_spectrum}, 
\ref{fig:ngc6752_spectrum} and \ref{fig:ngc6397_spectrum}. 
Fits to all other 
stars are available only electronically.

\section{Error Analysis}
\label{sec:uncertainties}

In order to assess the errors on the derived spectroscopic masses, we must 
carefully consider all of the sources of uncertainty. 
First, there are {\it random} errors on \logg\ and \teff, which derive from the sensitivity of
the synthetic spectra to changes in these two quantities as well as the SNR of the data.
Second, we have {\it random} errors on the factors to scale the models to the stellar 
photometry (or spectroscopy for variable stars). These
depend on the uncertainty in the photometry and the SNR of the spectroscopy. 
For variable stars, model fluxes are not scaled to the photometry, but
to the spectroscopy. These stars are therefore not corrected for 
{\it systematic} errors due to stellar blending and slit losses.
Stellar blending also adds an {\it additional}
source of systematic uncertainty to the derived effective temperatures.
In what follows we assess all sources of error and
propagate the uncertainties into our derived quantities.

\subsection{Errors in \logg\ and \teff}
\label{ssec:errorloggteff}

Errors on \logg\ and \teff\ are due to sensitivity of the synthetic spectra to these quantities as well as
the SNR of the data.
At low \teff, i.e., $<$7000~K, Balmer line widths as well as the Balmer jump are
not very sensitive to gravity and the value of $\log g$ cannot be obtained to an accuracy better than 
$\sim$0.10~dex, but more typically 0.15~dex (or worse in some cases; see Table~\ref{tab:results}).
It is important to emphasize that while there will be a minimum $\chi^2$ value for which the fit is
best, the formal error of the procedure is an underestimate of the true uncertainty.
The effective temperature is determined from the slope of the Paschen continuum (4200-5600~\AA ) and
its uncertainty is typically $\sim$100~K, with some higher values (Table~\ref{tab:results}).
For higher effective temperatures, it is possible to obtain more accurate values of gravity (0.10~dex typically), while
the temperature estimates remain similarly constrained. 
The sensitivity of \logg\ and \teff\ to abundance changes as large as [Fe/H]=0.2~dex is very low. We therefore conclude
that our adopting a constant cluster metallicity has negligible repercussion on the determination
of \logg\ and \teff.

The uncertainty on the ratio $f_\lambda$/$H_\lambda$ (Eq.~1) 
is a combination of the uncertainty on the
\V\ band photometry of 0.02~mag i(or the spectrophotometry for variable stars -- 0.15~mag)
and the uncertainty on the 
synthetic model fluxes. The latter is dominated by the uncertainty in the
temperature. For a 100~K change in \teff, the model flux changes by 7.5\% at 6000~K, 
and by 1.3\% at 13000~K (or $dH_V/H_V \sim  4 dT/T$  for \teff$<$6500~K,
$dH_V/H_V  \sim  2 dT/T$ for 6500$<$\teff$<$10\,000~K and $dH_V/H_V \sim dT/T$ for \teff$>$10\,000~K). 
For every star, we therefore derived the absolute error on $H_\lambda$ ($d H_\lambda$)
from the \teff\ value, the absolute error on \teff\ ($d$\teff) and the value of $H_{\lambda5540}$.

From the error on the $f_\lambda$/$H_\lambda$ ratio and a 10\% error on the distances
(listed in Table~\ref{tab:clusterproperties})
(this was chosen following the recent assessment by Gratton et al. 2003), we derive an error on
the radii by uncorrelated error propagation. To determine the error on the stellar luminosities,
however, we need to consider that the errors on $R$ and \teff\ are correlated, because, as explained above,
the uncertainty on \teff\ conditions the uncertainty on $H_\lambda$ and hence on $R$. We therefore
must use the error propagation formula for correlated errors:

\begin{equation}
\Delta L^2 = \left({\delta L \over \delta R}\right)^2 \Delta R^2 + 
             \left({\delta L \over \delta T}\right)^2 \Delta T^2 +
             2 r \Delta R \Delta T {\delta L \over \delta R} {\delta L \over \delta T}
\end{equation}

The cross-term is due to the fact that the errors on $R$ and \teff\ are not independent.
$r$ is negative since $R$ and \teff\ are inversely proportional. We set its value
at --0.5 which is intermediate between zero (no error correlation) and --1 (maximum correlation). 
The uncertainty on $R$ and $g$ propagates into the mass according to the formula
for uncorrelated uncertainty.

\subsection{Errors due to slit losses and stellar blending}
\label{ssec:errvar}

In the absence of stellar photometry, our determinations of radii, luminosities and masses would
be affected by spectral blending (which would make the spectra brighter and alter their color) and
slit losses (which would make the spectra fainter, but not alter their color). For non variable stars
we scaled the model spectra to photometric fluxes, thereby removing the uncertainty on the brightness,
though not on the color, due to blending and slit losses.
Variable stars,
which are not scaled to the photometry, are not corrected in this way.

To determine the impact of stellar blending and slit losses on our dataset,
all observed stellar spectra 
were convolved with the WFPC2 filter profiles 
(the {\sc iraf} routine {\it calcphot} was used with the Vega system; the
latest Vega magnitudes were used: {\it stsdas} Version 3.3) and the
spectro-photometric magnitudes thus obtained were compared with the photometric
values from Table~\ref{tab:observations}.
(Since the red end of the observed
spectra is slightly blue-ward of the red end of the F555W bandpass, we actually used the
synthetic spectra scaled to the observed spectra at 5540~\AA, to obtain the F555W spectrophotometric
magnitudes.)

The difference between the photometric and spectrophotometric \V\ magnitudes for all stars is
listed in Table~\ref{tab:results}, both as flux ratios (Column 9) and as magnitude differences (Column 10).
In columns 11 and 12 we list the differences between spectrophotometric and photometric colors,
$\delta$(\B--\V) and $\delta$(\U--\V), which we will use to assess additional uncertainty affecting 
our choice of temperatures. 

Magnitudes derived from
spectra can be brighter or dimmer than the corresponding photometry and the spectral colors can be bluer or redder
than the photometric ones. Brighter spectra with bluer, redder, or similar colors than the photometry can be the result
of spectral blending. Brighter spectra with the same colors as the photometry can be due to an eclipsing binary
with a faint companion. Dimmer spectra with bluer or redder colors than the photometry can be due to pulsational 
variability. Dimmer spectra with the same color, probably mean that the star was on the edge of the slit. 
In what follows we will take variable stars out of the discussion, as for these, spectroscopy and
photometry was not taken at the same time, and the the values of temperature and gravity
obtained from the spectra cannot be used together with the apparent brightness obtained from the photometry.
Variable stars will be treated later.

Although by scaling to the photometric fluxes we remove the source of uncertainty due to blending or slit losses
it is instructive to compare photometry and spectrophotometry.
In the absence of any systematic effect, F555W magnitude and color differences are expected to be distributed around
zero with a standard deviation of the order of the random errors.
The random error on the
magnitude differences is 0.15~mag,
determined by adding in quadrature the 0.02~mag error on the \V\
photometry and an average error of 0.15~mag on the spectrophotometry, due to the SNR of the spectra.
The random errors on the color differences are 0.22~mag for $\delta (\B-\V)$ and 0.25~mag for $\delta (\U-\V)$, and
were calculated by adding in quadrature
the uncertainties on the \V\ and \B\ (\U)
photometry and the expected average errors on the
equivalent spectrophotometric magnitudes of
15\% in \B\ and \V\ (20\% in \U) estimated from the SNR of the data.

\begin{figure}
\vspace{11cm}
\includegraphics{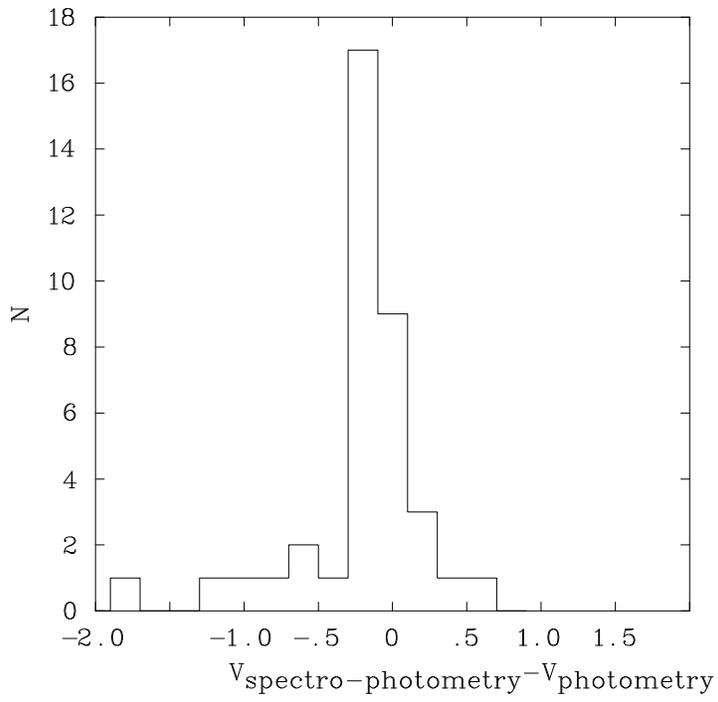}
\caption{A histogram of the differences between spectro-photometric and photometric magnitudes
for the non-variable stars
(Table~\ref{tab:results}, column 10).
\label{fig:scaling_hist}}
\end{figure}

In Fig.~\ref{fig:scaling_hist} we plot a histogram of all the \V\ magnitude
differences for the 34 non-variable stars
listed in Table~\ref{tab:results} (column 10). The mean and standard deviation 
are --0.18 and 0.29~mag, respectively (with a sigma clipping of 2).
The mean of the color differences (columns 11 and 12 in Table~\ref{tab:results})
are $<\delta (\B-\V)>$=0.02~mag and $<\delta (\U-\V)>$ = 0.14~mag.
If photometry and spectro-photometry were only affected by random errors we would
expect the mean of the histogram in Fig.~\ref{fig:scaling_hist} to be zero, with a standard 
deviation of the order of the random error expected
on the values (0.15~mag). The fact that the mean is less than zero means either that the predominant 
systematic uncertainty is stellar blending (which makes the spectra brighter) 
or that there is a systematic difference
in the methods used to convert flux values and \V\ bandpass counts to magnitudes. 
(The statistics are not very different if the comparisons are carried out on
STIS and FOS datasets separately, from which we conclude that the reason for the shift is not in 
the spectroscopic instruments, but rather in the calibrations carried out in the photometry and
spectrophotometry.)
The means of the color difference columns $\delta$(\B--\V) and $\delta$(\U--\V) are not the same
(0.02 and 0.14~mag, respectively). This suggests that there is an offset
in the transformations used to convert fluxes to magnitudes in the photometry and/or in the spectrophotometry,
which introduces a color term. This offset might therefore also be the reason for the fact
that the histogram in Fig.~\ref{fig:scaling_hist} does not peak at zero.  
The standard deviation of the magnitude differences (0.29~mag) is about twice as large as the random error 
(0.15~mag), reinforcing the fact that systematic as well as random uncertainties are at play.

To assess which stars are most affected by blending or slit losses,
we selected all stars with magnitude differences outside the range
--0.18$\pm$0.15~mag and whose color differences were outside the ranges 0.02$\pm$0.22~mag
(0.14$\pm$0.25~mag for $\delta$(\U--\V)). 
A total of 16 stars out of 34 non-variable stars have magnitude differences
outside the range. Of these, six stars have magnitude differences $<$(--0.18--0.15)~mag, indicative of blending
and we mark them with ``:".
Of these 6 stars, one (N104-5)
has a color difference ($\delta$(\B -- \V)) outside the range expected from random error alone, 
and we mark it with ``::" and raise its \teff\ uncertainty
to 300~K since, although this star's blended flux was corrected by scaling 
the model to the photometry,
its color was affected as well. Ten stars have magnitude differences $>$(--0.18+0.15)~mag,
indicative of slit losses. Although slit losses are not expected to alter the color of the star
we note that two of these ten stars (N5272-10 and N6752-10)
have color differences outside the range expected for random errors alone
and they too are given a larger temperature uncertainty and marked ``::" in Table~\ref{tab:results}. 

Six stars (N104-2, N5272-13, N6752-5, 9, N6397-1, and 2)
have magnitude differences consistent with random uncertainties, but {\it color} differences 
outside the range expected from random uncertainties alone. This is likely to be due to
variability, as discussed in Section~\ref{sec:identification} (variability detection criterion [iv]).

For the stars known to be variable (variability detection criteria [i-iv] in Section~\ref{sec:identification}), we 
calculated their radii, luminosities and spectroscopic masses scaling the
models to the spectra instead of the photometry, since, as explained, scaling to
a photometric flux determined at a different time from the spectroscopy is
inappropriate. However, since we concluded earlier in this section,
that there is a --0.18~mag offset between photometry and spectrophotometry,
(corresponding to a factor of 0.80),
{\it due to calibration}, scaling models to spectra will lead to masses
that are 20\% larger on average. We therefore account for this offset
in the derivation of the radii, luminosities and spectroscopic masses of the
variable stars, by multiplying these quantities by 0.80.

\subsection{NGC6397: comparison with Saffer et al.}

In the assessment of the uncertainty of a spectroscopic 
analysis such as ours it is also instructive to compare
results with those of other similar analyses. Here we compare our 
results for NGC~6397 with those obtained by Saffer et al. (2002).

The gravity values they obtained from an LTE analysis similar 
to the one reported here are either in
agreement (for two out of seven objects), 0.1~dex lower (for 
three objects) or 0.2~dex higher
(for the remaining two objects). Their effective temperature 
values are always within 5\%
of those determined here, except for NGC~6397-2, for which they 
are 12\% higher. Their photometry,
which, like in our study, is used to scale the FOS spectroscopy, 
is systematically brighter
than ours (by about 0.7~mag). 
The masses obtained here agree with those of Saffer et al. (2002) 
for N6397-4 and 7, while the others
are between 0.20 and 0.47~\msun\ lower, due to our dimmer photometric fluxes.
This reinforces what we will
state again in Section~\ref{sec:masses}: the spectroscopic masses are 
to be trusted only in a statistical sense.

\section{Stellar Masses}
\label{sec:masses}

In Table~\ref{tab:results} we list the spectroscopic masses derived in the present analysis,
along with their uncertainties.
In Table~\ref{tab:masses} we list (weighted) mean masses for the four clusters and the
three stellar types along with their variance. The weights are constructed using the masses
and their absolute errors: $w_i = (M_i/\sigma_{Mi})^2$, where $M_i$ are the mass values and $\sigma_{Mi}$ are their
absolute errors, calculated by averaging the positive and negative error bars listed in Table~\ref{tab:results}.
The variance ($\sqrt (\Sigma 1/w_i$) gives an idea of the error on the mean masses.

Considering the entire non-variable sample,
the BSs have masses which are marginally larger (1.04$\pm$0.17~\msun) than the HB masses (0.79$\pm$0.25~\msun), but are definitely larger than
the TO masses (0.58$\pm$0.22~\msun). Looking at the individual cluster results, it becomes clear that the reason
for the similar BS and HB mean masses is an anomaly in M~3 for which the mean (non-variable) BS mass 
(0.72$\pm$0.31~\msun) is smaller than
the mean (non-variable) HB mass (0.90$\pm$0.31~\msun). 
For NGC~6752, the only other cluster for which the comparison can
be done, the opposite is true, as expected. We later
show that it is likely that some of the M~3 HB stars are actually BSs and this 
might explain the high mean masses. 
From Table~\ref{tab:masses}
we can also see that the mean (non-variable)
spectroscopic mass for NGC~6397 is the largest (1.73$\pm$0.43~\msun), followed by
NGC~6752 (1.01$\pm$0.41~\msun), 47~Tuc (0.95$\pm$0.42~\msun) 
and M~3 (0.72$\pm$0.31~\msun). The mean \V\ brightness of our four BS samples (scaled to the distance of 47~Tuc)
are 15.80, 15.59, 16.00 and 16.01~mag for NGC~6397, NGC~6752, 47~Tuc and M~3, respectively. Masses
do scale with BS magnitude (see below) so the fact that the BSs in NGC~6397 (and not in NGC~6752) 
are by far the most massive might
have something to do with this cluster's high central density
(Table~\ref{tab:clusterproperties}). This might indicate that the GC central concentration plays a role
on the BS masses.

Including variable stars in the means improves the statistics, although the variable star
spectroscopic masses might be affected by worse uncertainty. 
The mean BS, HB and TO masses are not dissimilar, in particular the NGC~6397 mean BS mass
is still the highest (1.23~\msun) even if its mean scaled \V\ brightness (15.50~mag) is the same as that
of NGC~6752 (15.51~mag; the mean BS \V\ brightnesses of M~3 and 47~Tuc are both 15.90~mag and their 
mean masses are 1.03 and 0.99~\msun, respectively).

In Fig.~\ref{fig:teff_vs_mass}
we plot the spectroscopic masses for all 55 stars as a function of effective
temperature (left) and scaled \V\ brightness (right). The mean error on the masses is indicated on
the top-left panel. The latter plot gives a measure of the
relative brightness of the stars once all stars are scaled to the distance of 47~Tuc. 
As is expected, the BSs' masses correlate with their effective temperature and brightness although
the scatter is large. 
A Pearson $r$ test\footnote{Numerical Recipes F77, p. 630.} returns
a correlation factor of 0.28 (1.0 being perfect correlation) for effective temperature vs. mass
and --0.22, for brightness vs. mass (where the negative
sign results from large numbers expressing faint brightnesses). If
stars N6752-9 and N6397-4 are eliminated from the \teff--mass plot
the correlation index becomes 0.59, indicating a much better
correlation. These two stars have mass uncertainties of 0.39 and 0.24~\msun,
respectively, hence their low masses, which do not follow the trend,
are unlikely to be due to a particularly large error.

The TO star masses also correlate with 
\teff\ (0.17) and with \V\ (--0.30). 
The TO star spectroscopic masses sit at the low end of the BS mass range,
as expected.
Some TO stars have
masses larger than the turn-off mass, in particular for 47~Tuc.

HB star masses are not
expected to show a marked correlation with \teff, and they are not predicted to have masses
larger than the turn-off mass. Their masses are expected to be around 0.6~\msun (see
figure 9 in Moehler et al. 2003). Surprisingly, we find that 4 HB stars have masses
significantly larger than the turn-off mass (N5272-14, 17, N6752-5 and N6397-7). 
On the \logg - \teff\ plane only one of these (N6397-7) is
consistent with the HB locus, one is just below it (N5272-17), while two are well
below it (N5272-14 and N6752-5), 
in the region populated
by the BSs. 

Looking at the CMDs in Figs.~\ref{fig:cmd_ngc104_bv} to
\ref{fig:cmd_ngc6397_uv}, N5272-14 and N5272-17
have colors which are also consistent with being bright BSs, although on the \U--\V\ vs. \V\ plane
N5272-14 appears in the
RR-Lyrae region. N6752-5 and N6397-7 appear on the HB of their clusters, 
although, once again, they could also be bright BSs. 
Parts of the HB and BS loci, in 
fact, overlap. This is also the case for the \logg - \teff\ plane where 
evolutionary track for $\sim$2.5~\msun\ grazes the HB locus.

%
\section{Stellar Rotation}
\label{sec:rotation}


To measure stellar rotation spectroscopically requires high resolution spectroscopy of metal lines.
It is inappropriate to rely on fits to the troughs of the Balmer lines to measure stellar rotation, except
to impose broad limits. Small
changes in the model parameters 
result in significant differences in the troughs of the Balmer lines. In addition, the troughs of Balmer lines 
are affected by poor SNR more than the continuum. Relying on a
a handful of data points around these regions to determine stellar rotation will likely lead to erroneous results.

Although the resolution, SNR and spectral range of our spectroscopy are not ideal for
determining stellar rotation, we can impose some upper
limits and determine some values, using our STIS intermediate resolution
spectra and the following technique.
The best-fitting stellar atmosphere model spectrum (pre-convolved with the instrumental profile),
is convolved with 
Gaussian profiles of increasing FWHM and then with Poisson noise so as to reproduce the continuum SNR of the 
observations. Each of these model spectra is then compared to the data. For some stars a model is found for
which the fit is clearly worse. This will provide an upper limit to the stellar rotation above which
the model does not fit the data. For other stars, convolving models to broaden the lines 
beyond the instrumental effect improves the fit, and a best value of $v \sin i$ 
can therefore be determined.

Three types of spectral lines present in our spectra can be used: weak iron and other trace metal 
lines, the Ca~{\sc ii} K line (or H and K for those cooler stars where H$_\epsilon$ is weak compared to Ca~{\sc ii} H)
and, finally, the Balmer line wings. Weak iron and other metal lines can be used for spectra with low effective temperature
and reasonable SNR. However these lines are never resolved, hence only their troughs can be used. 
These are affected by noise as well as by the uncertainty on the abundance and temperature. Some spectra,
like those of N104-6 and N104-7 have models that reproduce well most of these metal lines. In such cases convolving the
model with a Gaussian with FWHM$>$100~km~s$^{-1}$ clearly deteriorates the fits.
In these cases, we can argue that $v \sin i$ is below this value. On the other hand, if metal lines
are not well fit anyway (whether with or without rotation convolution, e.g. N104-5), 
it is inappropriate to use them to determine a limit for the stellar rotation. 

The Ca~{\sc ii} K line (or H and K lines)
can also be used. 
This line is best at intermediate temperatures, since for high temperatures it is too weak and unresolved,
and at low temperatures it is very broad and similar to the Balmer lines;
for intermediate temperatures its wings and, to an extent even its trough,
can be used. This line, too, can be affected by abundance and temperature
uncertainties. The Balmer lines troughs, as already mentioned, are not reliable to measure rotation, 
but can still be used to impose broad limits. Finally, for stars hotter than about 10\,000~K, He~{\sc i}
lines can be used. These are ideal since they are always resolved and their shapes vary differently with abundance
and rotation.

In Fig.~\ref{fig:rotation} we present the intermediate resolution spectra of 
N5272-11,
one of the disk stars described in Section~\ref{sec:disk}. The high SNR of this spectrum and its weak, yet
resolved Ca~{\sc ii} K line makes the determination of the rotation upper limit more precise: (70$^{+40}_{-20}$)~km~s$^{-1}$. 
The wings of the line, which are less affected by small
temperature and abundance differences, tell us that the fit is already satisfactory with no rotation, however the
wings can also tolerate a rotation of 70~\kms and in fact the trough is better fit in this way. Higher values of the
rotation make the fit progressively worse until at 110~\kms\ both wings and trough no longer fit.
In Fig.~\ref{fig:disk} we present a similar plot for the EHB star N5272-17. This is the same plot
as in De Marco et al. (2004). It shows that the atmosphere model fits the Ca~{\sc ii} K line wings
only when convolved with a Gaussian with FWHM=200~\kms.

In Table~\ref{tab:rotation} we present values (and limits) of $v \sin i$ obtained with this method.
These results are the most extensive ever set of measurements and limits for BS
rotation velocities. Five BSs (N104-4, N5272-13 and 15, N6752-11 and 18) have reasonably
well measured $v \sin i$ (120$^{+100}_{-20}$, 100$\pm$20, 225$\pm$50, 50$\pm$20, 50$\pm$20~\kms, respectively).
The sample's median and average $v \sin i$ are 100 and 109~\kms, respectively. Five other BSs
(N104-7 and 8, N6752-4, 9 and 14) have only upper limits to their $ v \sin i$ (120, 120, 100, 25 and 50~\kms, respectively).
For a randomly oriented set of rotating stars, the average
value of $\sin i$ is 0.64. Thus the five BSs with measured $v \sin i$ suggest an average rotation velocity
of  $v = 160$~\kms. From values of radii
and masses of these stars, we can calculate their breakup speeds to be between 250 and 350~\kms (with the exception of star
N5272-15 which, with a very high mass of 3.74~\msun, has a breakup speed of $\sim$600~\kms. This star's mass is discussed in
Section~\ref{sec:individualobjects}).


Even taking the mean $v \sin i$ at face value, it is hard to tell whether this value is more in agreement with
BSs deriving from a collision or a binary merger. Both scenarios predict that the resulting star should
have a large angular momentum, and it should rotate very fast, especially when it starts to contract following
the post-formation expansion. Even if 
one scenario predicted a slower rotation rate than the other, one would still have
to contend with the various mechanisms that are likely to slow the initially-fast-rotating star down.
The often-cited paper by
Leonard and Livio (1995) predicts that collision mergers puff up enough to develop a convective zone that
would slow the star down, while binary mergers coalesce less violently and expand less, such that
no convective layer is expected and cannot therefore lose angular momentum
efficiently. However this is in contradiction with the recent
calculations of Sills et al. (2005), who predict that the product of an off-axis collision does not develop a
convective layer so the Leonard and Livio mechanism cannot operate. 
Sills et al. (2005) do however show that a disk, or even a magnetically locked wind can 
slow the star down, therefore providing an alternative slow-down mechanism for collision mergers. 

It appears that there is quite a bit of work still to be done in modeling BSs. On the other hand
these models need to be guided by observations and the only way to make headway observationally
is to carry out an analysis similar to the one carried out here on a very large dataset spanning
several GCs as well as open clusters and the field. In particular, field BSs are fundamental since 
they are highly unlikely to derive from collisions. Field BSs should therefore display characteristics 
which should make them stand aside from the rest of the BSs. Higher resolution and SNR spectra should be
obtained, however, since only then can rotation rates be determined in a meaningful way.

We should also note that all six $v \sin i$ values for HB stars (all of which are EHB stars),
are $\ge$50~\kms, including  one at 200$\pm$50~\kms. 
Values these high are not common for HB stars (e.g., Behr 2003). Field EHB stars are
called sub-dwarf O and B stars. Their rotation rates are by and large found to have small values
($v \sin i$$<$5~\kms; Napiwotzki 2001) as is the case for the EHB stars. Recently, Lanz et al. (2004)
and Ahmad et al. (2004)
have found values of the order of 100~\kms\ for sdOB stars. The latter study has determined that their
target sdB stars is actually a short-period 
double-lined spectroscopic binary, and that this is the reason for the broad absorption lines
and the large derived values of $v \sin i$. This, they argue, could also explain the high rotation rates found by
Lanz et al. (2004). It is therefore possible that this is the explanation for our rapidly-rotating EHB stars.
If this were the case a more careful look should be given to these stars as their parameters
would be affected by the companion.
Two of the six EHB stars with large rotation rates are among the
four objects which are suspected of being BSs. 
In Section~\ref{sec:disk} we return to the issue of rapidly-rotating EHB and HB stars.

\section{An edge-on disk around some hot stars?}
\label{sec:disk}

De Marco et al. (2004) showed that for 6 stars out of our sample (marked in Table~\ref{tab:results})
no reasonable fit
can be achieved. In all cases 
the model that fits the continuum redward of the Balmer jump (the Paschen continuum) has
a corresponding continuum blue-ward of the Balmer jump (the Balmer continuum) which is too bright.
The discrepancy
was evident at the 5$\sigma$ level and could not be reconciled with any change in model parameters.

On the color-color diagram of Figs.~\ref{fig:col-col1} and \ref{fig:col-col2}
the six exceptional stars fall leftward of the
region occupied by the models. For these stars the
[3600]-[4200] color (the slope of the Paschen continuum) is too blue (hot)
for the size of their Balmer jump, as parametrized by the [4200]-[5450] color. 
Here we identify two additional stars, NGC5272-11 and 16, where the same discrepancy, though
to a lesser extent, is identified.

De Marco et al. (2004) excluded a
purely instrumental problem as four of the stars in their group
were observed with the FOS while the others were observed with the STIS. 
Extraction problems that can affect the shape of the
spectrum such as background subtraction, tracing and the aperture size were investigated in great detail, but
were not responsible for the appearance of these spectra (see also Section~\ref{sec:datareduction}). 
An incorrect cluster reddening assumption does not solve the problem either.
Blending or binarity was also excluded as the cause of the discrepancy since
these effects would push stars to the upper side of the color-color diagram.
No two-temperature model, such as a disk or equatorial enhancement, can
account for the discrepancy. 

As a result of the impossibility of reconciling the data with the models, De Marco et al. (2004)
conjectured that a disk of partly ionized gas is responsible for the absorption of
Balmer continuum photons. Absorption models demonstrated that the column densities of material needed to
reconcile the data with the stellar models are quite modest, and that making reasonable assumptions for
the possible radii of these disks ($\sim$0.1~AU), their masses are quite low, of the order of 2$\times$10$^{-8}$~M$_\odot$. 
We refer the reader to that paper for further details and to our
conclusions in Section~\ref{sec:conclusions}.

A closer inspection revealed that two additional stars might have similar disks, although the hydrogen
column densities should be lower than for the six stars in De Marco et al. (2004). These two additional stars
are HB stars, as are two of the six stars in De Marco et al. (2004). In Section~\ref{sec:masses} we have already noted
that some HB stars, including two of the six disk stars of De Marco et al. (2004) are likely to be bright BSs. However,
the two additional HB stars with possible disks detected here appear to be bona fide HB stars (with low masses,
and locations on the CMD and the \logg--\teff\ plane consistent with HB stars). It is therefore unclear at this point
what the presence of disks might mean in the context of BS and other GC stars.

Recently, Porter and Townsend (2005) modeled the anomalous Balmer jump of N5272-17 and 
showed that they can reproduce the data with a rotating star whose shape is made oblate by
the centrifugal force. The rotation rates needed for the effect to explain the data depend on
whether the star is viewed pole- or equator-on, but there is a solution for the
values of $v \sin i$ determined for N5272-17 and N6752-11 (Table~\ref{tab:rotation}). Their Balmer line fits are not
of the highest quality, but they ascribe that to not having optimized the overall values
of temperature and gravity (they took the stellar parameters of De Marco et al. [2004] as their
starting values). 

There is however a remaining issue. In Fig.~\ref{fig:disk} we show the Ca~{\sc ii} K line region 
of our intermediate resolution spectrum of N5272-17. This is the same figure as on the upper right-hand
panel in Fig.~2 of De Marco et al. (2004). The Ca~{\sc ii} K line
shows three components, a broad, shallow one, interpreted as the stellar, rotationally-broadened component,
a narrow one at the same wavelength, due to the putative disk,
and a narrow one shifted by the GC's heliocentric velocity and interpreted as 
arising from the interstellar medium. If disks are not present, it might be hard to explain
the narrow stellar Ca~{\sc ii} K line component. On the other hand the SNR of the data
is not high enough to resolve the issue conclusively.

If Porter \& Townsend (2005) are right, and stellar rotation is solely responsible for the
anomalous Balmer jumps, then this confirms that
four HB stars in our sample are fast rotators. Two (N5272-17 and N6397-7) are, as we already explained,
likely to be bright BSs, but the remaining two, (N5272-11 and 16) are likely to be bona fide 
HB stars. Fast-rotating HB stars would be an interesting finding in itself, since rotation this high 
might indicate a binary
interaction in the star's past, leading us to wonder the extent to which
binarity influences the evolution of the horizontal branch.

\section{Individual Objects}
\label{sec:individualobjects}

In this Section we report the highlights of a subset of our objects which presented particular
challenges during the fitting process or which show other peculiarities.

{\it N104-3:} This star is variable and hard to classify without photometric observations taken
at the same time. Nonetheless it is likely to be a BS or a TO star, rather than an HB star. On the
basis of its spectrum alone it appears to be a BS. Its mass, obtained by scaling the model to the
spectrum is however only 0.42~\msun.

{\it N104-5:} In this high SNR spectrum, all of the metal lines are overestimated by the model. This is
also the case for the Ca~{\sc ii} K line which is broader in the model than in the observations. By convolving 
the model with a Gaussian profile with FWHM=80~km~s$^{-1}$ the overall fit to the metal lines is slightly improved. 
The Ca~{\sc ii} K line fit, however, remains poor. Rotation does not appear to be
responsible for the discrepancy between the intermediate resolution spectra and the model.
Rather, we conclude that there is a temperature inconsistency between the intermediate resolution
spectrum, where a higher model temperature (by about 300~K) would achieve a better fit, and the low resolution spectrum, where 
a higher temperature would misrepresent the Paschen continuum (gravity has little effect in this temperature regime). 
A lower abundance would help to fit better
the intermediate resolution spectrum, although the abundance change needed would be too large and inconsistent with
the cluster abundance. 

{\it N104-8:} The continuum level of the intermediate resolution spectrum for this BS 
is about 20\% brighter than for the
low resolution spectrum. No reason was found from the
data reduction to explain the discrepancy, since
the flux levels of the low and intermediate resolution spectra of stars on the same slit
are consistent. This might point
to an inherent photometric variability of N104-8. 
This star was marked as having a high probability of being a variable on the basis of the photometry
described in Section~\ref{sec:photometry} and \ref{ssec:errvar}. As a BS, this star might be a $\delta$~Scuti or
a SX~Phoenicis pulsating variable, or it might be an eclipsing binary.
We also found that the Balmer lines of the
low resolution spectrum are narrower than for the intermediate resolution one. 

{\it NGC104-9:} The spectrum of this BS suffers from the same problem as N104-5, although to a lesser extent. Here too we decided that
improving the fit to the metal lines by including rotation is not the correct interpretation of the
mismatch between data and model. 

{\it N5272-6:} This EHB star must be a slight blend of two stars, since it is impossible to fit perfectly
the Balmer continuum once the Paschen continuum is well fit. Since the gravity can be determined
by the Balmer line widths in its high-SNR intermediate resolution spectrum, we have retained this spectrum
in the sample and assigned a larger error to the temperature.

{\it N5272-13:} As is the case for NGC104-8, this BS's intermediate resolution spectrum is brighter 
than the low resolution spectrum by 7\%. 

{\it N5272-15:} This BS star has the largest spectroscopic mass encountered (3.74$\pm$2.26\msun). 
It was identified 
as a variable because its magnitude determined from individual images varied by just above 0.20~mag.
As a result we did not scale the stellar atmosphere model flux ($H_\lambda$) to the photometry,
but used instead the spectroscopic flux scaled by 0.8 as explained in
Section~\ref{sec:uncertainties}. One could suppose that the spectrum was
blended with that of another star, making $f_\lambda$ artificially bright, and the mass artificially
large. We note however, that even a blend with a star of equal brightness would only make the
mass twice as large. Such a blend is excluded on 
the basis of the imaging, where we see only one star close to N5272-15 which has only 
$\sim$25\% of its brightness and its flux was partly-to-mostly
excluded from the spectral extraction of N5272-15 by the extraction procedure. 
A blend with a very bright star is also highly improbable since the most likely
candidate would be a red giant and this would be obvious in the spectrum. The only blend that could feasibly have
augmented the stellar flux is one with another BS or an HB star almost coincident in position with N5272-15.
Alternatively this star is a main sequence foreground star. With a smaller
value of the distance $D$, its radius and spectroscopic mass would also be
smaller. As explained in Section~\ref{sec:identification} we do not expect
our sample of hot stars to be contaminated, but without radial velocity
measurements, the possibility cannot be ruled out.

{\it N6752-2:} It is likely that a small amount of contamination affects this HB star.
When its Paschen continuum is fitted and the gravity constrained by the intermediate resolution
spectrum, the Balmer continuum is a little too bright to be fitted by the model. This is the reason why
in the color-color diagram in Fig.~\ref{fig:col-col2} the star has the colors of a 17\,000-K object,
while the Balmer continuum is better fit with 18\,000 K. 
Since we do not know whether the contaminating star is bluer or redder, 
we adopt an intermediate temperature and increase the
uncertainty. 

{\it N6397-4:} This BS has a peculiarly low mass (0.55$\pm$0.24~\msun). Its high effective temperature
(\teff=13\,000~K) and gravity (\logg = 4.2) position it on the HB locus (Fig.~\ref{fig:loggvsteff_n6397}). Its position
on the CMD (Fig.~\ref{fig:cmd_ngc6397_bv}) is slightly blue-ward of the ZAMS. We therefore wonder whether this star could actually be
an EHB star.

\section{Comparison with stellar evolutionary models}
\label{sec:evolutionarytracks}

Stellar evolutionary models were calculated for a range of main sequence masses for each of the assumed
cluster metallicities (listed in Table 1). In
addition to the assumed metallicities, an alpha-element over-abundance
relative to iron of [$\alpha$/Fe]=+0.30 was assumed (Carney 1996).
The opacities and equation of state used in the calculations were those
described in VandenBerg et al. (2000). Evolutionary tracks
and isochrones were created using these models and the techniques described
in Bergbusch \& VandenBerg (1992).  For the three metal-poor
clusters (M~3, NGC~6752 and NGC~6397) an absolute age of 14~Gyr was
assumed, whereas an age 12~Gyr was assumed for the more metal-rich 47~Tuc:
these ages were accepted based on the agreement of the shape of the
isochrones and the shape of the cluster fiducial on the sub-giant branch,
turn-off, and upper main-sequence.  While assuming an age for a cluster
based on morphological considerations is problematic -- due to uncertainties
in abundances, opacities, and  color-temperature transformations -- the
exact cluster age is not critical to the results of this study.

In Fig.~\ref{fig:cmd_ngc104_bv}, \ref{fig:cmd_ngc5272_vi}, \ref{fig:cmd_ngc6752_bv} and \ref{fig:cmd_ngc6397_bv}, 
we over-plot theoretical isochrones,
zero-age main sequence (ZAMS) locii and evolutionary tracks on the \B--\V\ vs. \V, \V--\I\ vs. \V\ and \U--\V\ vs. \V\
CMDs of the four clusters. As previously observed,
the BSs tend to reside redward of the ZMAS. 
Ouellette \& Pritchet (1999) and Ouellette (2000)
simulated different synthetic BS populations using a range of assumptions for
their creation scenarios. By comparing the position of these synthetic populations to the real populations
in several cluster they deduced the most likely BS formation scenario. They
concluded that a 60:40 mix of BSs formed through collisions involving a turn-off star and a lower mass main sequence star 
and BSs formed by equal-mass collisions was
responsible for the BSs in 47~Tuc, while an 80:20 mix of the same two populations was responsible for the BSs
of NGC~6397. They also concluded that the BSs are not unevolved stars, in that their position
on the CMD is well above the position of the ZAMS. 
Sills et al. (2000) carried out similar simulations. They reported that the BS stars in 47~Tuc
must have formed in a burst that stopped several giga-years ago since in this GC bright BSs
are lacking.

\begin{figure}
\vspace{10cm}
\includegraphics{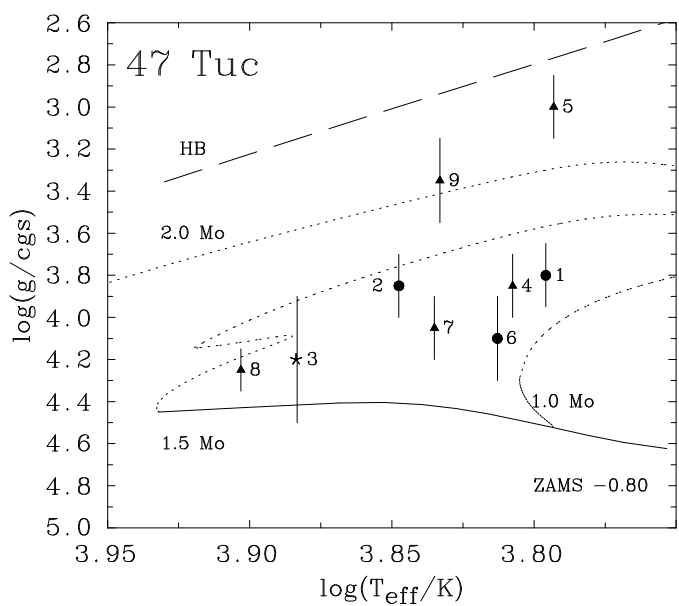}
\caption{A comparison of the stellar effective temperatures and gravities derived from our
model atmosphere fits to stars in 47~Tuc (TO star: solid circles; BSs: solid triangles; unclassified star: asterisk) 
with evolutionary calculations for metallicities appropriate for the
cluster. Gravity error bars 
are indicated by vertical segments, while temperature uncertainties are smaller than the width of the
symbols. The loci of the ZAMS (for [Fe/H]=--0.80) and HB are
indicated (solid and dashed lines, respectively). Evolutionary tracks for a sample of main sequence masses
(labeled) are also drawn (dotted lines).
\label{fig:loggvsteff_47Tuc}}
\end{figure}
\begin{figure}
\vspace{11cm}
\includegraphics{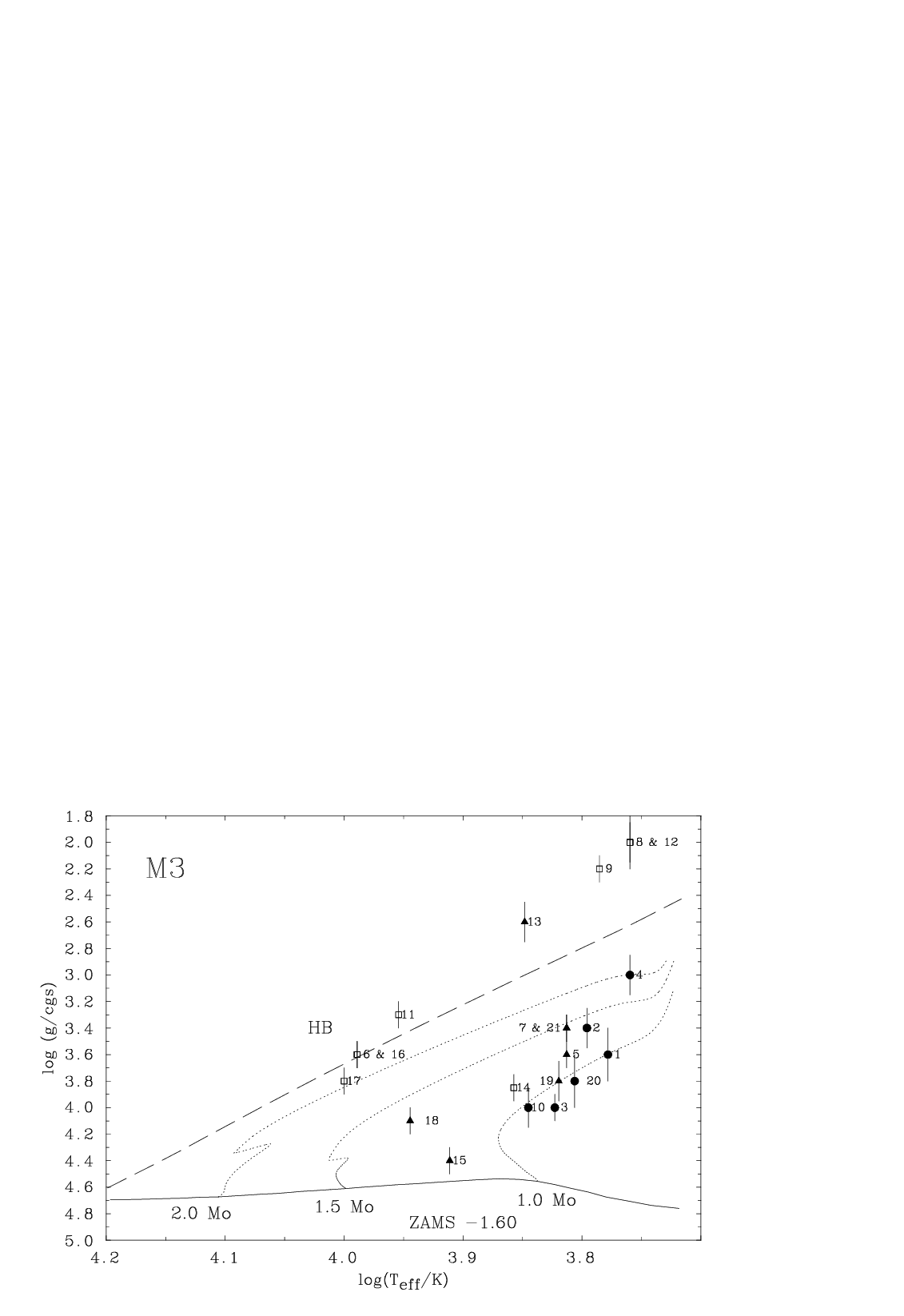}
\caption{A comparison of the stellar effective temperatures and gravities derived from our
model atmosphere fits to stars in M~3 (TO stars: solid circles; BSs: solid triangles; HB stars: squares)
with evolutionary calculations for metallicities appropriate for the
cluster. Gravity error bars
are indicated by vertical segments, while temperature uncertainties are smaller than the width of the
symbols. The loci of the ZAMS (for [Fe/H]=--1.60) and HB are
indicated (solid and dashed lines, respectively). Evolutionary tracks for a sample of main sequence masses
(labeled) are also drawn (dotted lines).
\label{fig:loggvsteff_m3}}
\end{figure}
\begin{figure}
\vspace{11cm}
\includegraphics{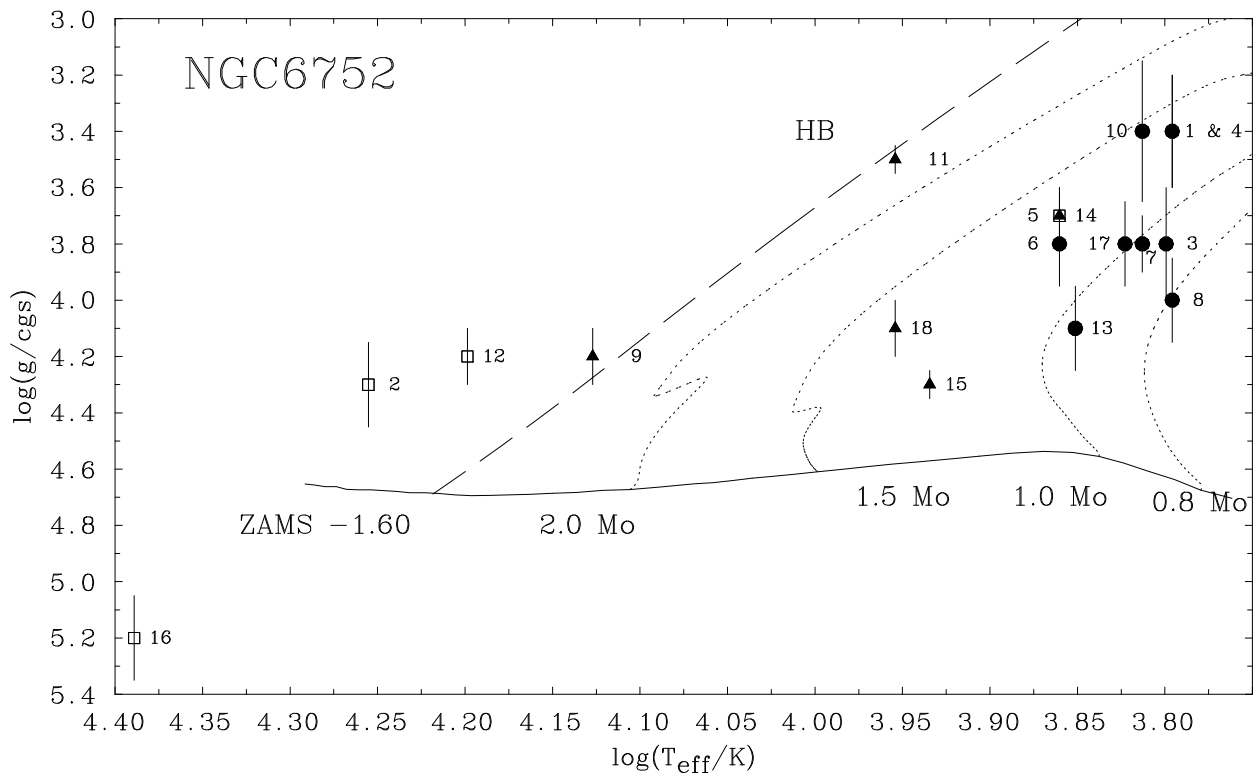}
\caption{A comparison of the stellar effective temperatures and gravities derived from our
model atmosphere fits to stars in NGC~6752 (TO stars: solid circles; BSs: solid triangles; HB stars: squares)
with evolutionary calculations for metallicities appropriate for the
cluster. Gravity error bars
are indicated by vertical segments, while temperature uncertainties are smaller than the width of the
symbols. The loci of the ZAMS (for [Fe/H]=--1.60) and HB are
indicated (solid and dashed lines, respectively). Evolutionary tracks for a sample of main sequence masses
(labeled) are also drawn (dotted lines).
\label{fig:loggvsteff_n6752}}
\end{figure}
\begin{figure}
\vspace{10cm}
\includegraphics{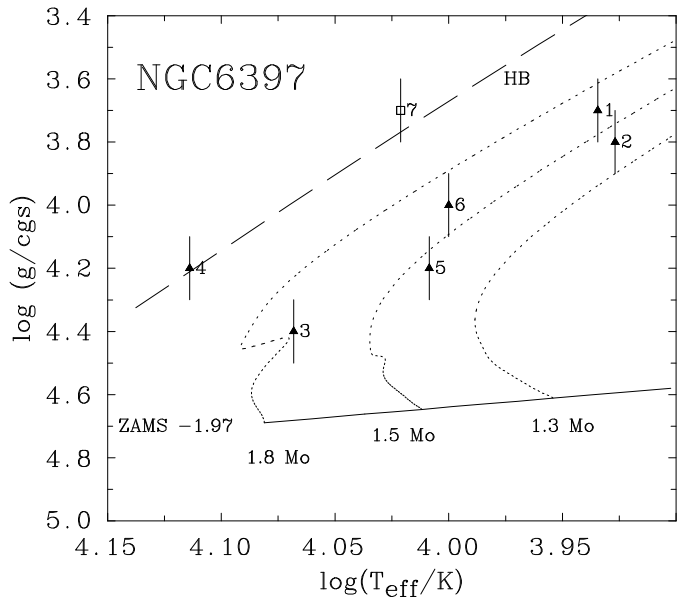}
\caption{A comparison of the stellar effective temperatures and gravities derived from our
model atmosphere fits to stars in NGC~6397 (BSs: solid triangles; HB stars: squares)
with evolutionary calculations for metallicities appropriate for the
cluster. Gravity error bars
are indicated by vertical segments, while temperature uncertainties are smaller than the width of the
symbols. The loci of the ZAMS (for [Fe/H]=--1.97) and HB are
indicated (solid and dashed lines, respectively). Evolutionary tracks for a sample of main sequence masses
(labeled) are also drawn (dotted lines).
\label{fig:loggvsteff_n6397}}
\end{figure}
\begin{figure}
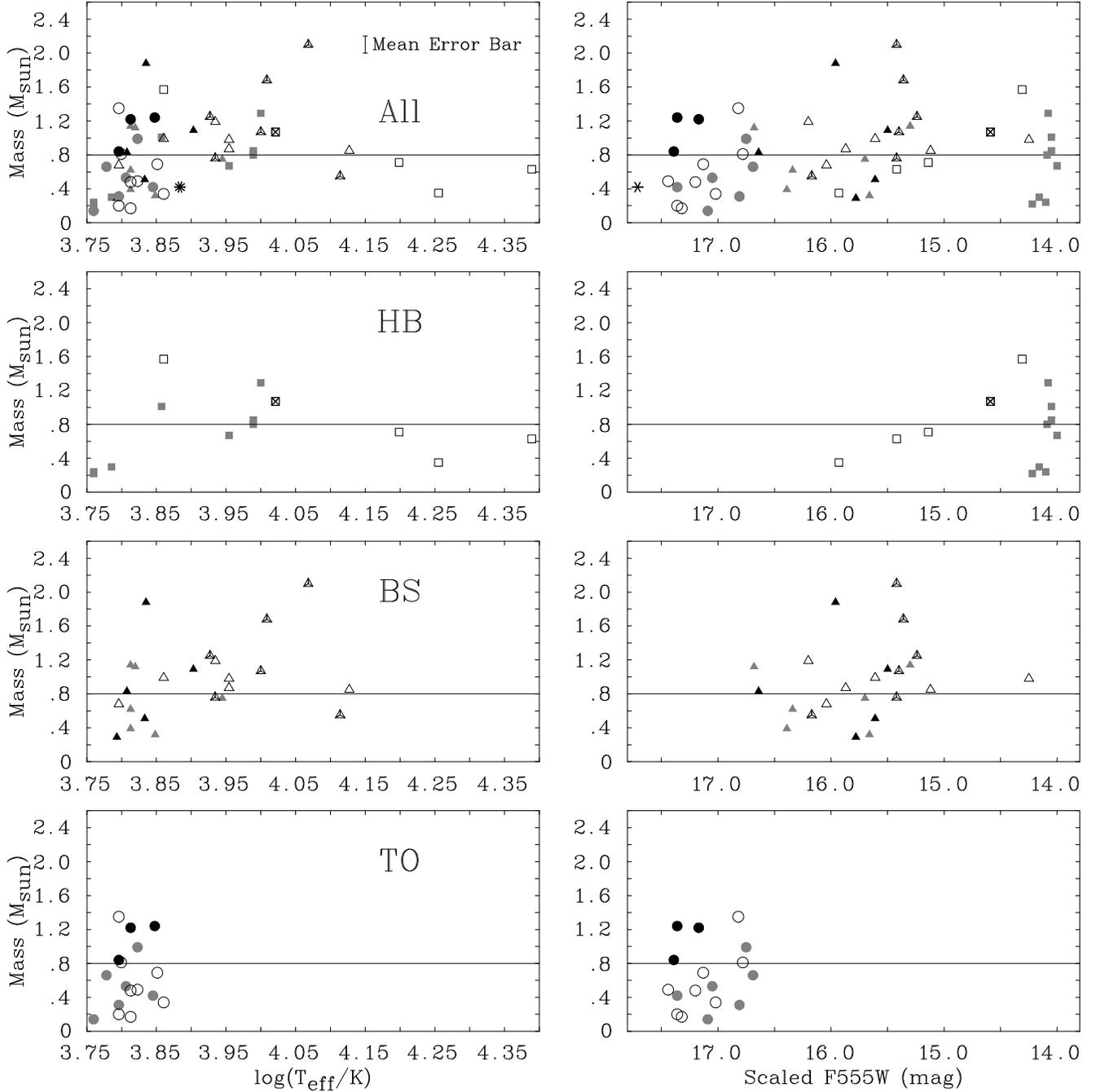

\vspace{21cm}
\includegraphics{f26a.eps}
\includegraphics{f26b.eps}
\caption{A plot of the spectroscopic masses as a function of effective temperature (left) 
and scaled photometric \V\ magnitudes (right). Triangles are 
the BSs, circles are the TO stars and squares are the HB stars. An asterisk symbol is used for the unclassified star
N104-3. Black filled symbols are for 47~Tuc, 
grey filled symbols are for M~3, un-filled symbols are for NGC~6752 and crossed un-filled 
symbols are for NGC~6397. The BS N5272-15, with a mass of (2.74$\pm$2.26)~\msun was eliminated 
from the plots for display reasons 
(but see discussion in Section~\ref{sec:masses}). 
\label{fig:teff_vs_mass}}
\end{figure}
\begin{figure}
\vspace{21cm}
\includegraphics{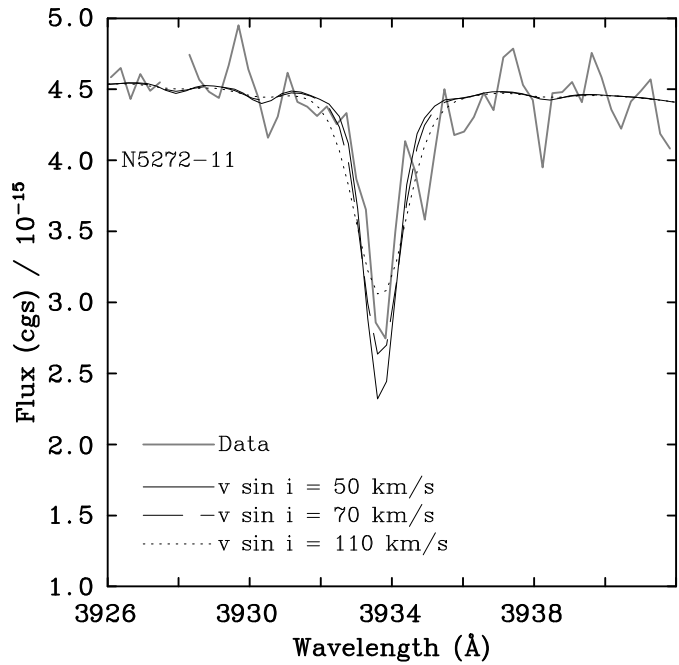}
\caption{Model fits to
N5272-11 intermediate resolution spectrum around the Ca~{\sc ii} K line. Fits with $v\sin i$=50,
70 and 110~\kms are presented (see legend) to demonstrate the deterioration of the spectral fit that
imposes limits on the value of $v \sin i$. 
\label{fig:rotation}}
\end{figure}
\begin{figure}
\vspace{21cm}
\includegraphics{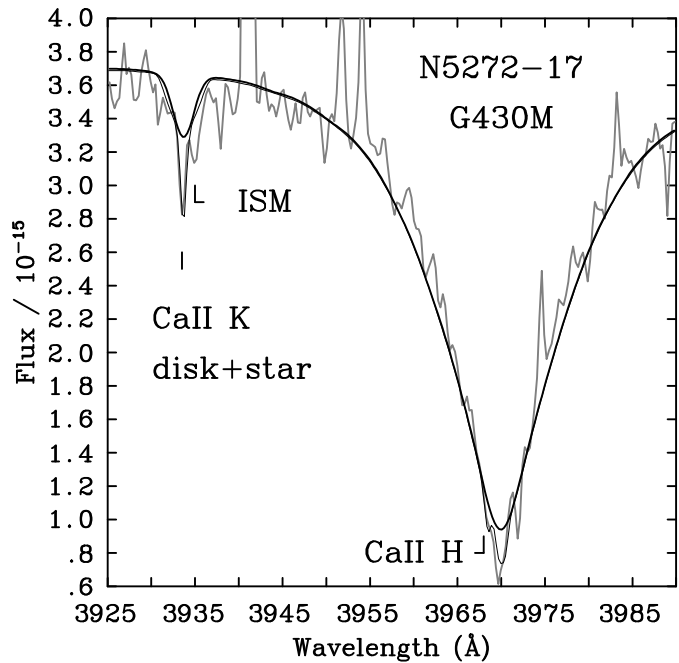}
\caption{Model fits (thick black line) to the EHB star
N5272-17 intermediate resolution spectrum (thick grey line) around the Ca~{\sc ii} K line. The stellar atmosphere model
(\teff=10\,000~K, \logg=3.8) is convolved with a Gaussian with FWHM=200~\kms. Additional absorption from 
a disk of partly-ionized hydrogen is modeled as well (thin black line). This is the same figure as the upper right-hand panel in
De Marco et al. (2004)
\label{fig:disk}}
\end{figure}

The ZAMS, theoretical evolutionary tracks for several initial masses, and
the theoretical locus of HB stars (from the calculations of VandenBerg et al. 2000),
are plotted in Figs.~\ref{fig:loggvsteff_47Tuc} to \ref{fig:loggvsteff_n6397}. 
As is the case for the \B--\V\ vs \V\ plane,
most of the BSs lie above the ZAMS, in a region of lower gravity.
This is unlikely to be due to high rotation rates since not even the largest rotation rates possible 
(the largest breakup speed for these stars is between $\sim$250 and 350~\kms) would lower the gravities of these stars
by the needed amount. In Table~\ref{tab:rotation} we were able to mostly exclude rotation 
rates that high.
The conclusion of Shara et al. (1997) and Saffer et al. (2001) that the
stars analyzed (which were re-analyzed in the present work) fall on the ZAMS
was due to their use of an incorrect ZAMS, located at gravities far smaller than the gravities
appropriate for the metallicities in their clusters. 

All of the BSs analyzed in this study have gravities lower than expected for main sequence stars,
in agreement with their positions on their clusters' CMDs, which is well above the main sequence.
We find in fact that most of the BSs reside {\it well above} not only the ZAMS but the 
{\it terminal-age} main sequence as well, and are in the region of the
Hertzsprung gap.
Stars formed by collision or merger binary scenarios are expected to be puffed-up
but should contract back to the main sequence on one or two thermal timescales,
of order 5$\times$10$^7$~yr. The main sequence lifetimes of these stars should be
much longer. This is a major puzzle at the heart of the BS phenomenon.


There is a mild correlation between the spectroscopic masses and the masses deduced from
a comparison with evolutionary tracks, in that the TO stars cluster near tracks of lower mass compared to the
BSs. 
Overall, the masses that would be derived from a comparison with the evolutionary tracks would be 
slightly larger than those determined from the spectroscopy.
Ouellette (2000)
found this to be the case also for 5 stars in NGC~6397 analyzed spectroscopically by Rodgers \& Roberts (1995), 
who found masses in the range 0.62-1.15~\msun . The masses obtained via a comparison with evolutionary
tracks are in the range 1.03-1.15~\msun , 
although the mass estimates from these two methods agreed
in the case of BSS-19 ($M=1.55$~\msun ) in 47~Tuc, which was analyzed spectroscopically by Shara et al. (1997).

Some HB stars in M~3 and NGC~6752 fall above the HB locus in the \logg\ and \teff\ plane
(Figs.~\ref{fig:loggvsteff_m3} and \ref{fig:loggvsteff_n6752}).
Moehler et al. (2003)
find that HB stars with temperatures above 12\,000 have gravities lower than predicted,
similarly to our HB stars in NGC~6752. Moehler et al. (2003) do not reach a satisfactory explanation for the
discrepancy, although they suggest it might have something to do with their using {\it scaled-solar}
model atmospheres (i.e., atmospheres calculated with solar abundances but whose level populations were scaled
to the cluster metallicity before the spectrum was computed). 
We, however, use model atmospheres calculated with the correct abundances, hence cannot
use their suggestion to explain our discrepancy.
At the high temperature end, the gravities of terminal-age HB stars are considerably lower
than those of zero-age HB stars (Dorman 1993), 
possibly presenting us with a solution for the low-gravities determined for our hot HB stars.

Our cool HB stars
N5272-8, 9 and 12 have gravities well below the model prediction. At these temperatures
zero-age and terminal-age locii occupy virtually the same position in the \logg-\teff\ plane,
which means that our determined low gravities are not readily explained. The fact that these three stars
are variable RR Lyrae stars does not readily present us with an explanation for their low
gravities.

As already discussed in Section~\ref{sec:masses},
HB stars N5272-14
and N6752-5 have gravities which are distinctly higher than the HB locus and more consistent with high mass
BSs. Their masses are well 
above the turn-off value (1.01 and 1.57~\msun, respectively) 
and their CMD positions, although not canonically 
that of BSs, are consistent with high mass BSs. 

\section{Summary}
\label{sec:conclusions}

We briefly summarize our results as follows:

\begin{enumerate}
\item BS spectroscopic masses are larger than the turn-off mass for their cluster
and are larger for the BSs than for either the TO or HB stars. 
Four out of 13 HB stars have masses significantly larger than the main sequence turn-off for their home clusters.
These stars could be high-luminosity BS. BS spectroscopic masses correlate both with their effective temperatures
and luminosities, although the scatter is large.
\item A comparison with evolutionary models leads to larger BS masses than the
spectroscopic masses obtained via a stellar atmosphere analysis.
\item {\it All} BSs have \logg\ values lower than expected for MS stars. Their positions on the CMD, as well
as \logg--\teff\ diagram is consistent with stars on the Hertzsprung gap. This is inconsistent with 
evolutionary timescales expected for stars out of thermodynamic equilibrium. We emphasize that this
is a major discrepancy with current theory that must be resolved by any successful theory of BS stars.
\item Five BSs have mean $v \sin i$ of 100~\kms, suggesting a mean value for the BS rotation rate of
$\sim$160~\kms. This is below the breakup speed for these stars, but rapidly rotating nonetheless.

\item De Marco et al. (2004) suggested that 4 BSs and 2 EHB stars 
have spectra consistent with
low-mass circumstellar disks. Two of the ``disk stars" discussed by De Marco et al. are HB stars
based on their CMD position. However, based on their high spectroscopic masses and position on the
\logg\ -- \teff\ they might actually be BSs. Two additional ``disk stars" are found here.
However these are both EHB stars based on their CMD locations, \logg\ -- \teff\
diagram locations as well as on the ground of their low masses. The fact that the presence of disks is suggested
around HB stars as well as BSs might point to a wider role for disks and binarity in GCs.
Recently, Porter and Townsend (2005) showed that the data are also consistent with a star whose
rotation rate makes it oblate. The rotation rates needed are of the order of those found in this study.
Although the shape of the Ca~{\sc ii} K line still argues for a disk being present, it is clear
at this time that more data are necessary to determine whether disks are present around these stars.

\item For some (mostly cool) HB stars we determine \logg\ which are lower than the HB locus. We
do not have a convincing explanation for this. We are confident, however that 
this is not a systematic effect that affects the whole sample.
\end{enumerate}

\acknowledgments

OD would like to acknowledge financial support from Janet Jeppson Asimov.
MMS acknowledges NASA support via HST grant GO-8226.
Data for this paper was obtained at the Canadian Astronomy Data Centre, 
which is operated by the Herzberg Institute of Astrophysics, National Research Council of Canada.
We thank Alison Sills for useful discussions and the anonymous referee for 
suggestions that improved the paper.




\clearpage

\clearpage

\clearpage

\clearpage

\clearpage

\clearpage

\clearpage

\clearpage



\clearpage 


\clearpage

\clearpage



\clearpage

\clearpage

\clearpage

\clearpage


\clearpage

\clearpage

\clearpage

\clearpage

\clearpage


\begin{deluxetable}{lllll}
\tabletypesize{\scriptsize}
\tablecaption{Cluster parameters. \label{tab:clusterproperties}}
\tablewidth{0pt}
\startdata
\hline
Cluster & 47~Tucanae& M~3  & NGC~6752  & NGC~6397\\
\hline
Alias                        & NGC~104                & NGC~5272               & --                     & -- \\
RA (J2000)                   &00$^h$ 24$^m$ 05.$^s$19 & 13$^h$ 42$^m$ 11.$^s$23& 19$^h$ 10$^m$ 51.$^s$78& 17$^h$ 40$^m$ 41.$^s$36 \\
Dec(J2000)                   & -72$^o$ 04$^\prime$49.$^{\prime\prime}$9 & +28$^o$ 22$^\prime$31.$^{\prime\prime}$6 & -59$^o$ 58$^\prime$54.$^{\prime\prime}$7 & -53$^o$ 40$^\prime$25.$^{\prime\prime}$3\\
$m-M$ (mag)\tablenotemark{a} & 13.37\tablenotemark{d} & 15.05\tablenotemark{d} & 13.25\tablenotemark{d} & 12.36\\
$E(B-V)$ (mag)               & 0.032\tablenotemark{d} & 0.016\tablenotemark{d} & 0.056\tablenotemark{d} & 0.18 \tablenotemark{e} \\
$[$Fe/H$]$                   & -0.83\tablenotemark{d} & -1.60\tablenotemark{d} & -1.54\tablenotemark{d} & -1.97\tablenotemark{f}\\ 
$M_V$ (mag)                  & -9.42                  & -8.93                  & -7.73                  & -6.63 \\
RV (\kms)                    & -18.7                  & -147.6                 & -27.9                  & 18.9 \\
c\tablenotemark{b}           & 2.03                   & 1.84                   & 2.50c                   & 2.50c  \\
r$_{core}$(arcmin)\tablenotemark{e}& 0.40             & 0.55                   & 0.17                    & 0.05   \\
$\rho_0$(\msun~pc$^{-3}$)\tablenotemark{c}   & 4.81   & 3.51                   & 4.91                    & 5.68   \\
Turn-off mass (\msun)        & 0.87                   & 0.78                   & 0.78                    & 0.77 \\
N(BS)                        & 61\tablenotemark{g}    & 137\tablenotemark{h}    & 28\tablenotemark{i}    & 21\tablenotemark{j}      \\
\enddata
\tablecomments{All values are from Harris 1996, unless otherwise specified.}
\tablenotetext{a}{Throughout this paper an uncertainty of 10\% was assumed for all distance values.}
\tablenotetext{b}{$c = \log (r_t / r_c)$, where $r_t$ is the half-mass radius and $r_c$ is the core radius. A `c' next to the value means the
cluster is core-collapsed.}
\tablenotetext{c}{Logarithm of the central luminosity in \lsun~pc$^{-3}$.}
\tablenotetext{d}{VandenBerg 2000}
\tablenotetext{e}{Gratton et al. 2003}
\tablenotetext{f}{Gratton et al. (2003) use --2.03, Harris (1996) has --1.95}
\tablenotetext{g}{Kaluzny et al. 1997}
\tablenotetext{h}{Ferraro et al. 1995}
\tablenotetext{i}{Sabbi et al. 2004}
\tablenotetext{j}{Sills et al. 2000}
\end{deluxetable}

\begin{deluxetable}{lcclcccccc}
\tabletypesize{\scriptsize}
\tablecaption{Observations. \label{tab:observations}}
\tablewidth{0pt}
\tablehead{
\colhead{STIS/FOS} & \colhead{Star} & 
\colhead{RA \&  Dec\tablenotemark{a}} &  \colhead{Pixel\tablenotemark{a}} & 
\colhead{Pixel\tablenotemark{b}} & r\tablenotemark{c} &
\colhead{\U\tablenotemark{d}} &
\colhead{\B\tablenotemark{d}} &
\colhead{\V\tablenotemark{d}} &
\colhead{\I\tablenotemark{d}} \\
Frame & ID &   (J2000.0) & Image   &  Slit & ('') & (mag)  & (mag) & (mag) & (mag) \\
(1)   & (2)&   (3)       & (4)     &  (5)  & (6)  & (7)    & (8)   & (9)   & (10) \\}
\startdata
\multicolumn{9}{c}{47~Tucanae (NGC~104)}\\
o5gx01010 & N104-1 & 00$^h$24$^m$06.$^s$78 --72$^o$04$^\prime$46.$^{\prime \prime}$54 &1993 2212  & 581 &6.64& 17.508  & 17.77&17.38& 17.08 \\
o5gx01010 & N104-2 & 00$^h$24$^m$06.$^s$77 --72$^o$04$^\prime$48.$^{\prime \prime}$52 &2032 2207  & 603 &5.44& 17.760  & 17.96&17.36& 16.76 \\
o5gx01010 & N104-3 & 00$^h$24$^m$06.$^s$74 --72$^o$04$^\prime$49.$^{\prime \prime}$62 &2054 2206  & 641 &5.37& 17.795  & 17.76&17.71& 17.81 \\
o5gx01010 & N104-4 & 00$^h$24$^m$06.$^s$68 --72$^o$04$^\prime$49.$^{\prime \prime}$90 &2061 2211  & 648 &4.88& 16.806  & 17.07&16.64& 16.23 \\
o5gx01010 & N104-5 & 00$^h$24$^m$06.$^s$65 --72$^o$04$^\prime$51.$^{\prime \prime}$27 &2088 2209  & 675 &4.72& 16.017  & 16.04&15.78& -- \\
o5gx02010 & N104-6 & 00$^h$24$^m$07.$^s$00 --72$^o$04$^\prime$32.$^{\prime \prime}$67 &1715 2235  & 784 &19.49& --      & 17.78&17.17& -- \\
o5gx03010 & N104-7 & 00$^h$24$^m$04.$^s$71 --72$^o$04$^\prime$52.$^{\prime \prime}$37 &2137 2382  & 646 &4.57& 16.219  & 16.39&15.96& -- \\
o5gx05010 & N104-8 & 00$^h$24$^m$06.$^s$54 --72$^o$04$^\prime$34.$^{\prime \prime}$54 &1759 2251  & 283 &16.75& --      & 15.84&15.50& -- \\
o5gx06010 & N104-9 & 00$^h$24$^m$05.$^s$81 --72$^o$04$^\prime$51.$^{\prime \prime}$30 &2100 2285  & 706 &1.00& 15.879  & 15.90&15.61& -- \\
\multicolumn{10}{c}{M~3 (NGC~5272)}\\
o5gx08010 & N5272-1 &  13$^h$42$^m$11.$^s$75 +28$^o$22$^\prime$46.$^{\prime \prime}$15 &740 916   & 348&16.78&18.55  &  --  &18.55  &  17.86 \\
o5gx08010 & N5272-2 &  13$^h$42$^m$11.$^s$57 +28$^o$22$^\prime$43.$^{\prime \prime}$13 &702 917   & 424&12.94&18.67  &  --  &18.64  &  18.00 \\
o5gx08010 & N5272-3 &  13$^h$42$^m$11.$^s$50 +28$^o$22$^\prime$41.$^{\prime \prime}$81 &686 918   & 456&10.42&18.61  &  --  &18.55  &  17.90 \\
o5gx08010 & N5272-4 &  13$^h$42$^m$11.$^s$47 +28$^o$22$^\prime$40.$^{\prime \prime}$80 &675 922   & 477&11.28&18.95  &  --  &18.95  &  18.34 \\
o5gx08010 & N5272-5 &  13$^h$42$^m$10.$^s$86 +28$^o$22$^\prime$29.$^{\prime \prime}$08 &533 933   & 758&4.32&18.20  &  --  &18.16  &  17.65 \\
o5gx08010 & N5272-6 &  13$^h$42$^m$12.$^s$04 +28$^o$22$^\prime$52.$^{\prime \prime}$10 &811 907   & 211&16.61&15.95  &  --  &16.21  &  16.08 \\
o5gx09010 & N5272-7 &  13$^h$42$^m$11.$^s$35 +28$^o$22$^\prime$28.$^{\prime \prime}$57 &571 987   & 691&4.31&17.16  &  --  &17.00  &  16.53 \\
o5gx09010 & N5272-8 &  13$^h$42$^m$11.$^s$55 +28$^o$22$^\prime$23.$^{\prime \prime}$34 &546 1040   & 806&10.02&16.08  &  --  &15.71  &  15.12 \\
o5gx10010 & N5272-9 &  13$^h$42$^m$11.$^s$64 +28$^o$22$^\prime$26.$^{\prime \prime}$33 &577 1030   & 423&8.65&16.02  &  --  &15.71  &  15.28 \\
o5gx10010 & N5272-10&  13$^h$42$^m$11.$^s$49 +28$^o$22$^\prime$25.$^{\prime \prime}$61 &560 1020   & 487&8.26&19.22  &  --  &19.23  &  18.67 \\
o5gx10010 & N5272-11&  13$^h$42$^m$12.$^s$87 +28$^o$22$^\prime$31.$^{\prime \prime}$47 &707 1108   & 126&21.08&15.86  &  --  &15.87  &  15.71 \\
o5gx10010 & N5272-12&  13$^h$42$^m$12.$^s$53 +28$^o$22$^\prime$30.$^{\prime \prime}$44 &684 1095   & 177&18.69&15.96  &  --  &15.59  &  14.84 \\
o5gx11010 & N5272-13&  13$^h$42$^m$12.$^s$31 +28$^o$22$^\prime$33.$^{\prime \prime}$42 &689 1054   & 529&16.21&17.52  &  --  &17.44  &  16.99 \\
o5gx11010 & N5272-14&  13$^h$42$^m$12.$^s$32 +28$^o$22$^\prime$28.$^{\prime \prime}$40 &651 1088   & 628&15.81&15.91  &  --  &15.64  &  15.33 \\
o5gx12010 & N5272-15&  13$^h$42$^m$10.$^s$80 +28$^o$22$^\prime$31.$^{\prime \prime}$53 &547 912   & 529&4.21&17.07  &  --  &17.08  &  17.02 \\
o5gx13010 & N5272-16&  13$^h$42$^m$10.$^s$80 +28$^o$22$^\prime$55.$^{\prime \prime}$73 &734 758   & 186&17.25&15.91  &  --  &16.15  &  15.98 \\
o5gx13010 & N5272-17&  13$^h$42$^m$10.$^s$57 +28$^o$22$^\prime$58.$^{\prime \prime}$72 &738 716   & 103&15.49&15.94  &  --  &16.14  &  15.996 \\
o5gx13010 & N5272-18&  13$^h$42$^m$12.$^s$09 +28$^o$22$^\prime$41.$^{\prime \prime}$43 &724 986   & 643&15.65&17.84  &  --  &17.27  &  16.41 \\
o5gx14010 & N5272-19&  13$^h$42$^m$10.$^s$51 +28$^o$22$^\prime$37.$^{\prime \prime}$05 &565 848   & 303&9.65&18.91  &  --  &18.94  &  18.32 \\
o5gx14010 & N5272-20&  13$^h$42$^m$11.$^s$25 +28$^o$22$^\prime$31.$^{\prime \prime}$10 &582 961   & 527&1.84&18.54  &  --  &18.57  &  18.08 \\
o5gx14010 & N5272-21&  13$^h$42$^m$11.$^s$63 +28$^o$22$^\prime$27.$^{\prime \prime}$86 &588 1021   & 650&8.04&18.25  &  --  &18.17  &  17.64 \\
\multicolumn{10}{c}{NGC~6752}\\
o5gx15010 & N6752-1&   19$^h$10$^m$51.$^s$89 --59$^o$58$^\prime$58.$^{\prime \prime}$35 &831 791  & 713&5.98&17.42 &17.65  &17.25  &  -- \\
o5gx15010 & N6752-2&   19$^h$10$^m$51.$^s$56 --59$^o$58$^\prime$55.$^{\prime \prime}$69 &868 796  & 785&9.73&14.61 &15.67  &15.82  &  -- \\
o5gx16010 & N6752-3&   19$^h$10$^m$51.$^s$35 --59$^o$58$^\prime$55.$^{\prime \prime}$75 &880 787  & 926&10.71&16.80 &17.09  &16.67  &  -- \\
o5gx16010 & N6752-4&   19$^h$10$^m$51.$^s$70 --59$^o$58$^\prime$58.$^{\prime \prime}$49 &842 782  & 850&6.88&16.09 &16.37  &15.93  &  -- \\
o5gx17010 & N6752-5&   19$^h$10$^m$49.$^s$93 --59$^o$58$^\prime$55.$^{\prime \prime}$33 &967 724  & 512&19.79&14.20 &--     &15.00  &  -- \\
o5gx17010 & N6752-6&   19$^h$10$^m$50.$^s$32 --59$^o$58$^\prime$57.$^{\prime \prime}$09 &932 728  & 554&16.31&17.13 &17.38  &16.91  &  -- \\
o5gx17010 & N6752-7&   19$^h$10$^m$52.$^s$15 --59$^o$59$^\prime$06.$^{\prime \prime}$92 &763 736  & 571&3.65&17.18 &17.54  &17.09  &  -- \\
o5gx17010 & N6752-8&   19$^h$10$^m$52.$^s$26 --59$^o$59$^\prime$07.$^{\prime \prime}$43 &753 737  & 904&3.93&16.87 &17.10  &16.71  &  -- \\
o5gx17010 & N6752-9&   19$^h$10$^m$52.$^s$50 --59$^o$59$^\prime$08.$^{\prime \prime}$48 &732 740  & 972&5.19&15.27 &15.26  &15.00  &  -- \\
o5gx18010 & N6752-10&  19$^h$10$^m$52.$^s$32 --59$^o$59$^\prime$08.$^{\prime \prime}$88 &741 729  & 544&5.36&17.07 &17.42  &17.02  &  -- \\
o5gx18010 & N6752-11&  19$^h$10$^m$51.$^s$98 --59$^o$59$^\prime$06.$^{\prime \prime}$99 &772 727  & 604&4.32&14.06 &14.92  &15.03  &  -- \\
o5gx18010 & N6752-12&  19$^h$10$^m$51.$^s$03 --59$^o$59$^\prime$02.$^{\prime \prime}$43 &857 719  & 772&9.78&14.14 &--      &14.10  &  -- \\
o5gx18010 & N6752-13&  19$^h$10$^m$50.$^s$95 --59$^o$59$^\prime$01.$^{\prime \prime}$97 &864 720  & 784&10.30&17.38 &17.63  &17.21  &  -- \\
o5gx19010 & N6752-14&  19$^h$10$^m$52.$^s$21 --59$^o$59$^\prime$04.$^{\prime \prime}$28 &776 759  & 414&1.16&15.61 &15.61  &15.50  &  -- \\
o5gx19010 & N6752-15&  19$^h$10$^m$52.$^s$91 --59$^o$59$^\prime$09.$^{\prime \prime}$09 &704 754  & 552&7.13&16.30 &16.35  &16.09  &  -- \\
o5gx20010 & N6752-16&  19$^h$10$^m$53.$^s$83 --59$^o$58$^\prime$59.$^{\prime \prime}$01 &713 877  & 741&12.11&15.31 &--      &16.87  &  -- \\
o5gx21010 & N6752-17&  19$^h$10$^m$53.$^s$17 --59$^o$58$^\prime$40.$^{\prime \prime}$55 &867 992  & 355&23.89&17.62 &17.14  &17.33  &  -- \\
o5gx21010 & N6752-18&  19$^h$10$^m$51.$^s$59 --59$^o$58$^\prime$41.$^{\prime \prime}$69 &953 908  & 591&22.53&15.86 &15.86  &15.75  &  -- \\
\multicolumn{10}{c}{NGC~6397}\\
y3fj010ft & N6397-1 &  17$^h$40$^m$41.$^s$69 --53$^o$40$^\prime$27.$^{\prime \prime}$1 & 548  456 & --&2.43& 14.89&14.83&14.67&14.32 \\
y3fj0109t & N6397-2 &  17$^h$40$^m$41.$^s$74 --53$^o$40$^\prime$28.$^{\prime \prime}$4 & 564  429 & --&1.39& 14.97&14.62&14.50&14.10 \\
y3fj010ht & N6397-3 &  17$^h$40$^m$41.$^s$88 --53$^o$40$^\prime$30.$^{\prime \prime}$4 & 579  380 & --&1.67& 14.34&14.67&14.62&14.44 \\
y3fj010ct & N6397-4 &  17$^h$40$^m$42.$^s$11 --53$^o$40$^\prime$31.$^{\prime \prime}$7 & 570  326 & --&3.57& 14.87&15.27&15.37&15.12 \\
y3fj010mt & N6397-5 &  17$^h$40$^m$42.$^s$02 --53$^o$40$^\prime$33.$^{\prime \prime}$4 & 609  315 & --&4.77& 14.70&14.61&14.57&14.26 \\
y3fj010pt & N6397-6 &  17$^h$40$^m$42.$^s$25 --53$^o$40$^\prime$34.$^{\prime \prime}$0 & 589  274 & --&6.12& 14.74&14.63&14.56&14.34 \\
y3fj0104t & N6397-7 &  17$^h$40$^m$40.$^s$92 --53$^o$40$^\prime$26.$^{\prime \prime}$5 & 638  577 & --&8.97& 14.90&13.79&13.85&13.54 \\
\enddata
\tablenotetext{a} {RA \& Dec and pixel positions of the stars analyzed were determined with the cursor on the images presented in
Figs.~\ref{fig:47tuc_image} to \ref{fig:ngc6397_image}. The uncertainties are 0.01$^s$, 0.01$^{''}$ and 1 pixel.}
\tablenotetext{b} {This is the y-pixel position of the spectrum on the CCD image (column 1).}
\tablenotetext{c}{Radial distance from cluster center.}
\tablenotetext{d} {Observed WFPC2 photometric magnitudes determined from PSF photometry.
The best-fitting synthetic spectra for non variable stars were scaled to these \V\ magnitudes in the way described in Sec.~\ref{sec:photometry}.}
\end{deluxetable}

\begin{deluxetable}{cccc}
\tabletypesize{\scriptsize}
\tablecaption{Imaging data \label{tab:photometry}}
\tablewidth{0pt}
\tablehead{
\colhead{Filter} & \colhead{Date of Observation} & \colhead{Exposure Time} & \colhead{\# Exposures} \\
      \colhead{} &   \colhead{(day/month/year)}  &  \colhead{(sec)}    & \colhead{} 
}
\startdata
\multicolumn{4}{c}{47 Tuc}\\
                    F336W  & 08/07/1999  & 160.0  &  1   \\
                    F336W  & 08/07/1999  & 200.0  &  6   \\
                    F336W  & 09/07/1999  & 300.0  &  1   \\
                    F336W  & 09/07/1999  & 400.0  &  6   \\
                    F336W  & 10/07/1999  & 600.0  &  1   \\
                    F336W  & 10/07/1999  & 700.0  &  6   \\
                    F336W  & 11/07/1999  & 900.0  &  7   \\
                    F336W  & 13/07/2001  & 400.0  &  6   \\
                    F336W  & 13/07/2001  & 1000.0  &  5   \\
                    F439W  & 01/09/1995  &  7.0  &  1   \\
                    F439W  & 01/09/1995  & 50.0  &  2   \\
                    F555W  & 01/09/1995  & 1.0  & 1    \\
                    F555W  & 01/09/1995  & 7.0  & 1    \\
                    F555W  & 28/10/1999  & 1.0  & 3    \\
                    F555W  & 28/10/1999  & 120.0  & 2    \\
                    F555W  & 03/07/1999  & 160.0  & 6    \\
                    F555W  & 28/10/1999  & 20.0  & 18    \\
                    F814W  & 03/07/1999  & 160.0  & 5    \\
\multicolumn{4}{c}{M3}\\
                    F336W  & 25/04/1995  & 70.0  & 2    \\
                    F336W  & 25/04/1995  & 600.0  & 16    \\
                    F336W  & 25/04/1995  & 800.0  & 4    \\
                    F555W  & 14/05/1998  & 3.0  & 2    \\
                    F555W  & 14/05/1998  & 60.0  & 21    \\
                    F555W  & 14/05/1998  & 100.0  & 4    \\
                    F555W  & 14/05/1998  & 400.0  & 6    \\
                    F555W  & 14/05/1998  & 500.0  & 2    \\
                    F814W  & 14/05/1998  & 3.0  & 2    \\
                    F814W  & 14/05/1998  & 100.0  & 12    \\
                    F814W  & 14/05/1998  & 140.0  & 4    \\
                    F814W  & 28/04/1999  & 100.0  & 4    \\
\multicolumn{4}{c}{NGC~6752}\\
                    F336W  & 22/03/2001  & 20.0  & 1    \\
                    F336W  & 22/03/2001  & 260.0 & 3 \\
                    F439W  & 17/08/1994  & 70.0 & 1 \\
                    F439W  & 17/08/1994  & 700.0 & 2 \\
                    F555W  & 22/03/2001  & 0.2 & 1 \\
                    F555W  & 22/03/2001  & 30.0 & 3 \\
\multicolumn{4}{c}{NGC~6397}\\
                    F336W  & 06/03/1996  & 10.0  & 2    \\
                    F336W  & 06/03/1996  & 80.0  & 2    \\
                    F336W  & 06/03/1996  & 400.0  & 28    \\
                    F336W  & 06/03/1996  & 500.0  & 6    \\
                    F336W  & 06/03/1996  & 700.0  & 1    \\
                    F439W  & 07/03/1996  & 10.0  & 2    \\
                    F439W  & 07/03/1996  & 80.0  & 2    \\
                    F439W  & 07/03/1996  & 400.0  & 16    \\
                    F439W  & 07/03/1996  & 500.0  & 4    \\
                    F555W  & 06/03/1996  & 1.0  & 1    \\
                    F555W  & 06/03/1996  & 8.0  & 1    \\
                    F555W  & 06/03/1996  & 40.0  & 6    \\
                    F555W  & 04/04/1999  & 1.0  & 2    \\
                    F555W  & 04/04/1999  & 8.0  & 2    \\
                    F555W  & 04/04/1999  & 40.0  & 24   \\
                    F814W  & 06/03/1996  & 1.0  & 1    \\
                    F814W  & 06/03/1996  & 8.0  & 1    \\
                    F814W  & 06/03/1996  & 40.0  & 2    \\
                    F814W  & 04/04/1996  & 1.0  & 2    \\
                    F814W  & 04/04/1996  & 8.0  & 2    \\
                    F814W  & 04/04/1996  & 40.0  & 24    \\
\enddata
\end{deluxetable}

\begin{deluxetable}{llllrrrllrrr}
\tabletypesize{\scriptsize}
\tablewidth{0pt}
\tablecaption{Results.}
\tablehead{
\colhead{ID}           & \colhead{Notes}          & \colhead{CMD}   & 
\colhead{log (g/ }     & \colhead{T$_{\rm eff}$}  & \colhead{R}     &
\colhead{L}            & \colhead{M(range)$^b$}   & $f^c$           & 
\colhead{$\Delta m^d$} & \colhead{$\delta(B-V)^e$}& \colhead{$\delta(U-V)^e$}\\
\colhead{}             & \colhead{}               & \colhead{Type}  & 
\colhead{cm~s$^{-2}$)} & \colhead{(K)}            & \colhead{(R$_\odot$)}  &
\colhead{(L$_\odot$)}  & \colhead{(M$_\odot$)}    & \colhead{}      &  
\colhead{(mag)}        & \colhead{(mag)}          & \colhead{(mag)} \\
\colhead{(1)}  & \colhead{(2)}    & \colhead{(3)}      &
\colhead{(4)}  & \colhead{(5)}    & \colhead{(6)}      & 
\colhead{(7)}  & \colhead{(8)}    & \colhead{(9)}      & 
\colhead{(10)} & \colhead{(11)}    & \colhead{(12)}         }
\startdata
\multicolumn{12}{c}{47 Tucanae (NGC104)}\\
1&v & TO  &  3.80 $\pm$  0.15 &   6250 $\pm$    150 &  1.9 $\pm$  0.2 &   4.9 $\pm$   1.3 & 0.84 (0.46 -  1.50) &  0.40 & -0.99 &  0.02 &  --    \\
2&v & TO  &  3.85 $\pm$  0.15 &   7040 $\pm$    100 &  2.2 $\pm$  0.3 &   10  $\pm$   3   & 1.24 (0.67 -  2.22) &  0.19 & -1.83 & -0.22 & -0.08  \\
3&v & ?   &  4.20 $\pm$  0.30 &   7650 $\pm$    250 &  0.9 $\pm$  0.2 &   2.2 $\pm$   1.0 & 0.42 (0.13 -  1.27) &  0.63 & -0.50 &  0.17 &  --    \\
4&  & BS  &  3.85 $\pm$  0.15 &   6420 $\pm$    100 &  1.8 $\pm$  0.2 &   4.8 $\pm$   0.9 & 0.83 (0.47 -  1.43) &  0.69 & -0.40 &  0.10 &  0.17  \\
5&::& BS  &  3.00 $\pm$  0.15 &   6210 $\pm$    300 &  2.8 $\pm$  0.4 &  11 $\pm$   3 & 0.29 (0.15 -  0.54) &  0.53 & -0.70 &  0.27 &  0.31  \\
6&v & TO  &  4.10 $\pm$  0.20 &   6500 $\pm$    100 &  1.6 $\pm$  0.2 &   4.2 $\pm$   1.2 & 1.22 (0.58 -  2.49) &  0.57 & -0.61 & -0.18 & --     \\
7&  & BS  &  4.05 $\pm$  0.15 &   6840 $\pm$    100 &  2.1 $\pm$  0.2 &   8.9 $\pm$   1.6 & 1.88 (1.07 -  3.21) &  1.18 &  0.18 & -0.01 &  0.03  \\
8&v & BS  &  4.25 $\pm$  0.10 &   8000 $\pm$    150 &  1.3 $\pm$  0.2 &  6.1 $\pm$ 1.8 & 1.09 (0.64 -  1.80) &  1.70 &  0.57 & -0.10 & --     \\
9&  & BS  &  3.35 $\pm$  0.20 &   6810 $\pm$    100 &  2.5 $\pm$  0.3 &  12 $\pm$   2 & 0.51 (0.26 -  0.99) &  0.80 & -0.25 &  0.10 &  0.12  \\
\multicolumn{12}{c}{M~3 (NGC5272)}\\
1&  & TO  &  3.60 $\pm$  0.20 &   6000 $\pm$    100 &  2.1 $\pm$  0.2 &   5.2 $\pm$   1.0 & 0.66 (0.33 -  1.28) &  0.86 & -0.17 &  -- &  0.20  \\
2&  & TO  &  3.40 $\pm$  0.15 &   6250 $\pm$    100 &  1.8 $\pm$  0.2 &   4.5 $\pm$   0.8 & 0.31 (0.17 -  0.53) &  0.99 & -0.01 &  -- &  0.00  \\
3&::& TO  &  4.00 $\pm$  0.10 &   6650 $\pm$    300 &  1.7 $\pm$  0.2 &   4.7 $\pm$   0.8 & 0.99 (0.64 -  1.52) &  1.20 &  0.19 &  -- & -0.29  \\
4&  & TO  &  3.00 $\pm$  0.15 &   5750 $\pm$    300 &  1.9 $\pm$  0.3 &   3.6 $\pm$   0.9 & 0.14 (0.07 -  0.25) &  0.34 & -1.17 &  -- &  0.30  \\
5&  & BS  &  3.60 $\pm$  0.10 &   6500 $\pm$    100 &  2.1 $\pm$  0.2 &   6.8 $\pm$   1.2 & 0.62 (0.40 -  0.95) &  0.97 & -0.03 &  -- &  0.15  \\
6&  & EHB &  3.60 $\pm$  0.10 &   9750 $\pm$    700 &  2.4 $\pm$  0.2 &  44 $\pm$  11 & 0.80 (0.51 -  1.24) &  0.90 & -0.11 &  -- &  0.20  \\
7&  & BS  &  3.40 $\pm$  0.10 &   6500 $\pm$    100 &  3.5 $\pm$  0.4 &  20 $\pm$   4 & 1.14 (0.73 -  1.74) &  1.05 &  0.05 &  -- &  0.05  \\
8&v & HB  &  2.00 $\pm$  0.15 &   5750 $\pm$    100 &  7.8 $\pm$  0.9 &  59 $\pm$  14 & 0.22 (0.12 -  0.39) &  0.89 & -0.12 &  -- &  --    \\
9&v & HB  &  2.20 $\pm$  0.10 &   6100 $\pm$    150 &  7.1 $\pm$  0.8 &  63 $\pm$  16 & 0.30 (0.18 -  0.47) &  0.82 & -0.21 &  -- &  --    \\
10&:& TO  &  4.00 $\pm$  0.15 &   7000 $\pm$    100 &  1.1 $\pm$  0.1 &   2.4 $\pm$   0.4 & 0.42 (0.24 -  0.71) &  0.42 & -0.94 &  -- &  0.08  \\
11\tablenotemark{f}&& EHB &  3.30 $\pm$  0.10 &   9000 $\pm$    100 &  3.0 $\pm$  0.3 &  54 $\pm$  10 & 0.67 (0.43 -  1.02) &  0.88 & -0.14 &  -- &  0.20  \\
12&v   & HB  &  2.00 $\pm$  0.20 &   5750 $\pm$    100 &  8.1 $\pm$  0.9 &  63 $\pm$  15 & 0.24 (0.12 -  0.47) &  0.95 & -0.06 &  -- &  --    \\
13&v   & BS  &  2.60 $\pm$  0.15 &   7050 $\pm$    100 &  4.7 $\pm$  0.6 &  49 $\pm$  12 & 0.32 (0.18 -  0.57) &  0.19 & -1.82 &  -- &  0.40  \\
14&v   & HB  &  3.85 $\pm$  0.10 &   7200 $\pm$    100 &  2.0 $\pm$  0.3 & 9.3 $\pm$ 2.5 & 1.01 (0.60 -  1.63) &  5.18 &  1.79 &  -- & --     \\
15&v   & BS  &  4.40 $\pm$  0.10 &   8150 $\pm$    100 &  2.0 $\pm$  0.4 &  16 $\pm$   6 & 3.74 (2.01 -  6.54) &  0.78 & -0.27 &  -- &  --    \\
16\tablenotemark{f} && EHB &  3.60 $\pm$  0.10 &   9750 $\pm$    100 &  2.4 $\pm$  0.2 &  47 $\pm$   9 & 0.85 (0.55 -  1.29) &  0.81 & -0.23 &  -- &  0.21  \\
17\tablenotemark{f} && EHB &  3.80 $\pm$  0.10 &  10000 $\pm$    200. &  2.4 $\pm$  0.2 &  49 $\pm$   9 & 1.29 (0.83 -  1.96) &  0.89 & -0.13 &  -- &  0.22  \\
18&v   & BS  &  4.10 $\pm$  0.10 &   8800 $\pm$    300 &  1.3 $\pm$  0.2 & 8.7 $\pm$ 3.0 & 0.75 (0.42 -  1.27) &  1.12 &  0.12 &  -- &  --    \\
19&:   & TO  &  3.80 $\pm$  0.20 &   6400 $\pm$    100 &  1.5 $\pm$  0.2 &   3.4 $\pm$   0.6 & 0.53 (0.27 -  1.02) &  0.62 & -0.51 &  -- &  0.07  \\
20&v   & BS  &  3.80 $\pm$  0.15 &   6600 $\pm$    300 &  2.2 $\pm$  0.3 &  8.2  $\pm$  2.2  & 1.12 (0.60 -  2.02) &  0.48 & -0.81 &  -- &  --    \\
21&    & BS  &  3.40 $\pm$  0.10 &   6500 $\pm$    100 &  2.1 $\pm$  0.2 &   6.7 $\pm$   1.2 & 0.39 (0.25 -  0.59) &  1.05 &  0.05 &  -- &  0.04  \\
\multicolumn{12}{c}{NGC~6752}\\
1&   & TO  &  3.40 $\pm$  0.20 &   6250 $\pm$    100 &  1.5 $\pm$  0.2 &   3.0 $\pm$   0.6 & 0.20 (0.10 -  0.39) &  0.89 & -0.12 &  0.10 &  0.09  \\
2&   & EHB &  4.30 $\pm$  0.15 &  18000 $\pm$    500 &  0.7 $\pm$  0.1 &  44 $\pm$   8 & 0.35 (0.20 -  0.60) &  0.89 & -0.13 &  0.09 &  0.11  \\
3&   & ATO &  3.80 $\pm$  0.20 &   6300 $\pm$    100 &  1.9 $\pm$  0.2 &   4.9 $\pm$   0.9 & 0.81 (0.41 -  1.56) &  0.86 & -0.16 &  0.05 &  0.06  \\
4&   & BS  &  3.40 $\pm$  0.20 &   6250 $\pm$    100 &  2.7 $\pm$  0.3 &  10 $\pm$   2 & 0.68 (0.34 -  1.31) &  0.89 & -0.12 &  0.06 &  0.13  \\
5& v & EHB &  3.70 $\pm$  0.10 &   7250 $\pm$    100 &  2.9 $\pm$  0.3 &  21 $\pm$   5 & 1.57 (0.99 -  2.45) &  0.80 & -0.24 &  -- &  1.12  \\
6&   & TO  &  3.80 $\pm$  0.15 &   7250 $\pm$    100 &  1.2 $\pm$  0.1 &   3.6 $\pm$   0.7 & 0.34 (0.20 -  0.59) &  0.81 & -0.23 & -0.16 &  0.07  \\
7&   & TO  &  3.80 $\pm$  0.10 &   6500 $\pm$    150 &  1.4 $\pm$  0.1 &   3.3 $\pm$   0.6 & 0.48 (0.31 -  0.74) &  0.99 & -0.01 &  0.02 &  0.11  \\
8&   & ATO &  4.00 $\pm$  0.15 &   6250 $\pm$    100 &  1.9 $\pm$  0.2 &   5.0 $\pm$   0.9 & 1.35 (0.77 -  2.33) &  1.35 &  0.32 &  0.12 & -0.03  \\
9& v & BS  &  4.20 $\pm$  0.10 &  13400 $\pm$    150 &  1.2 $\pm$  0.1 &  42 $\pm$  10 & 0.85 (0.53 -  1.32) &  0.90 & -0.11 & -0.29 & -0.90  \\
10&:: & TO  &  3.40 $\pm$  0.25 &   6500 $\pm$    300 &  1.4 $\pm$  0.1 &   3.0 $\pm$   0.5 & 0.17 (0.08 -  0.37) &  1.02 &  0.02 & -0.18 & -0.22  \\
11\tablenotemark{f}& & BS  &  3.50 $\pm$  0.05 &   9000 $\pm$    150 &  2.9 $\pm$  0.3 &  49 $\pm$   9 & 0.98 (0.71 -  1.33) &  0.76 & -0.29 &  -- &  0.25  \\
12&  & EHB &  4.20 $\pm$  0.10 &  15800 $\pm$    100 &  1.1 $\pm$  0.1 &  68 $\pm$  13 & 0.71 (0.46 -  1.09) &  0.97 & -0.04 &  0.03 &  0.06  \\
13&  & TO  &  4.10 $\pm$  0.15 &   7100 $\pm$    100 &  1.2 $\pm$  0.1 &   3.4 $\pm$   0.6 & 0.69 (0.40 -  1.18) &  1.12 &  0.12 &  0.11 &  0.39  \\
14&  & BS  &  3.70 $\pm$  0.10 &   7250 $\pm$    100 &  2.3 $\pm$  0.2 &  13 $\pm$   2 & 0.99 (0.63 -  1.50) &  1.76 &  0.61 &  0.18 &  0.18  \\
15&  & BS  &  4.30 $\pm$  0.05 &   8600 $\pm$    100 &  1.3 $\pm$  0.1 &   7.9 $\pm$   1.4 & 1.19 (0.86 -  1.61) &  0.58 & -0.59 & -0.09 & -0.01  \\
16&  & EHB &  5.20 $\pm$  0.15 &  24500 $\pm$    500 &  0.3 $\pm$  0.0 &  35 $\pm$   6 & 0.63 (0.36 -  1.07) &  0.97 & -0.04 &  -- &  0.11  \\
17&v & TO &  3.80 $\pm$  0.15 &   6650 $\pm$    100 &  1.5 $\pm$  0.2 &   3.7 $\pm$   0.9 & 0.49 (0.27 -  0.86) &  0.58 & -0.60 &  --   & -0.09  \\
18&  & BS  &  4.10 $\pm$  0.10 &   9000 $\pm$    100 &  1.4 $\pm$  0.1 &  11 $\pm$   2 & 0.87 (0.56 -  1.33) &  0.97 & -0.04 &  0.01 &  0.09  \\
\multicolumn{12}{c}{NGC~6397}\\
1&v  & BS  &  3.70 $\pm$  0.10 &   8600 $\pm$    100 &  2.0 $\pm$  0.2 &  20 $\pm$   5 & 0.76 (0.48 -  1.18) &  0.89 & -0.13 &  0.10 &  0.47  \\
2&v  & BS  &  3.80 $\pm$  0.10 &   8450 $\pm$    100 &  2.3 $\pm$  0.3 &  25 $\pm$   6 & 1.25 (0.78 -  1.95) &  0.85 & -0.17 &  0.11 &  0.46  \\
3&   & BS  &  4.40 $\pm$  0.10 &  11700 $\pm$    200 &  1.5 $\pm$  0.2 &  38 $\pm$   7 & 2.10 (1.35 -  3.20) &  0.88 & -0.14 &  0.08 &  0.10  \\
4\tablenotemark{f}&  & BS  &  4.20 $\pm$  0.10 &  13000 $\pm$    300 &  1.0 $\pm$  0.1 &  24 $\pm$   4 & 0.55 (0.35 -  0.84) &  0.82 & -0.21 &  0.13 &  0.07  \\
5\tablenotemark{f}&v  & BS  &  4.20 $\pm$  0.10 &  10200 $\pm$    300 &  1.7 $\pm$  0.2 &  28 $\pm$   7 & 1.68 (1.05 -  2.61) &  0.87 & -0.15 & -0.06 &  --    \\
6\tablenotemark{f}&v  & BS  &  4.00 $\pm$  0.10 &  10000 $\pm$    100 &  1.7 $\pm$  0.2 &  26 $\pm$   6 & 1.07 (0.67 -  1.67) &  0.88 & -0.14 &  0.02 &  --    \\
7\tablenotemark{f}&v   & HB  &  3.70 $\pm$  0.10 &  10500 $\pm$    200 &  2.4 $\pm$  0.3 &  63 $\pm$  15 & 1.07 (0.67 -  1.66) &  0.83 & -0.20 &  0.10 &  --    \\
\enddata
\tablenotetext{a}{``v": stars that are suspected of variability (Section~\ref{sec:photometry}). 
Radii, luminosities and masses of these stars
have been determined by scaling the models to the spectroscopy $\times$0.80, not the photometry 
as is the case for the rest of the sample (see Section~\ref{ssec:errvar}).
``:" - stars suspected of {\it low-level} blending based on the \V\ magnitude photometry/spectrophotometry comparison (Section~\ref{ssec:errvar}).
``::" - stars suspected of moderate blending or where the magnitude {\it and} color 
photometry/spectrophotometry comparison is outside the random error range; these stars
have been given a larger temperature error (Section~\ref{ssec:errvar}).
}
\tablenotetext{b}{An error range rather than a formal error is presented due to the error asymmetry.}
\tablenotetext{c}{Scaling factors are derived from a ratio of the spectrophotometric (SP) and photometric (P; Table~\ref{tab:observations}) 
values: $10^{(F555W_{SP}-F555W_P)/2.5}$.}
\tablenotetext{d}{$\Delta m = F555W_{SP}-F555W_P$.}
\tablenotetext{e}{$ \delta (B-V) = (F439W_{SP}-F555W_{SP}) - (F439W_P-F555W_P)$; $ \delta (U-V) = (F336W_{SP}-F555W_{SP}) - (F336W_P-F555W_P)$.}
\tablenotetext{f}{This star's parameters were derived from the Paschen continuum and Balmer lines, 
but for these values the data Balmer continuum is
fainter than the model's. Stars with this characteristic are discussed in Section~\ref{sec:disk}}
\label{tab:results}
\end{deluxetable}

\begin{deluxetable}{lccc}
\tabletypesize{\scriptsize}
\tablecaption{Monochromatic spectrophotometric colors. \label{tab:col-col}}
\tablewidth{0pt}
\tablehead{\colhead{Star} & \colhead{[3600]$^a$} & \colhead{[3600]$^a$-[4200]$^b$} & \colhead{[4200]$^a$-[5450]$^c$} \\
}
\startdata
\multicolumn{4}{c}{47~Tuc}\\
1	 &   15.99 &   -0.35 &   -0.03 \\
2	 &   15.07 &   -0.19 &   -0.24 \\
3	 &   16.69 &   -0.09 &   -0.35 \\
4	 &   15.80 &   -0.33 &   -0.07 \\
5	 &   14.84 &   -0.17 &   -0.02 \\
6	 &   16.09 &   -0.37 &   -0.10 \\
7	 &   15.64 &   -0.29 &   -0.19 \\
8	 &   15.58 &   -0.05 &   -0.44 \\
9	 &   15.00 &   -0.11 &   -0.20 \\
\multicolumn{4}{c}{M~3}\\
1	 &   17.88 &   -0.45 &   -0.07 \\
2	 &   18.05 &   -0.50 &   -0.09 \\
3	 &   18.10 &   -0.44 &   -0.21 \\
4	 &   17.32 &   -0.48 &    0.00 \\
5	 &   17.64 &   -0.33 &   -0.20 \\
6	 &   15.46 &    0.00 &   -0.69 \\
7	 &   16.60 &   -0.31 &   -0.18 \\
8	 &   15.35 &   -0.26 &   -0.02 \\
9	 &   15.25 &   -0.13 &   -0.15 \\
10	 &   17.72 &   -0.29 &   -0.30 \\
11	 &   15.33 &    0.23 &   -0.67 \\
12	 &   15.28 &   -0.24 &   -0.04 \\
13	 &   15.40 &    0.17 &   -0.40 \\
14	 &   16.91 &   -0.21 &   -0.32 \\
15	 &   16.28 &   -0.05 &   -0.51 \\
16	 &   15.31 &    0.05 &   -0.70 \\
17	 &   15.43 &    0.07 &   -0.70 \\
18	 &   17.12 &    0.07 &   -0.59 \\
19	 &   17.87 &   -0.44 &   -0.12 \\
20	 &   17.18 &   -0.41 &   -0.17 \\
21	 &   17.65 &   -0.42 &   -0.17 \\
\multicolumn{4}{c}{NGC~6752}\\
1	 &   16.42 &   -0.47 &   -0.12 \\
2	 &   13.94 &   -0.88 &   -0.80 \\
3	 &   15.26 &   -0.32 &   -0.14 \\
4	 &   15.83 &   -0.43 &   -0.13 \\
5	 &   14.20 &   -0.07 &   -0.38 \\
6	 &   16.07 &   -0.14 &   -0.36 \\
7	 &   16.42 &   -0.38 &   -0.22 \\
8	 &   16.35 &   -0.47 &   -0.11 \\
9	 &   13.57 &   -0.50 &   -0.76 \\
10	 &   16.30 &   -0.41 &   -0.31 \\
11	 &   13.40 &   -0.73 &   -0.80 \\
12	 &   13.36 &    0.30 &   -0.65 \\
13	 &   16.52 &   -0.39 &   -0.19 \\
14	 &   15.58 &   -0.05 &   -0.39 \\
15	 &   14.88 &    0.05 &   -0.58 \\
16	 &   14.76 &   -1.14 &   -0.88 \\
17	 &   15.10 &    0.11 &   -0.62 \\
18	 &   16.08 &   -0.31 &   -0.19 \\
\multicolumn{4}{c}{NGC~6397}\\
1	 &   13.61 &    0.17 &   -0.60 \\
2	 &   13.41 &    0.16 &   -0.57 \\
3	 &   13.05 &   -0.27 &   -0.70 \\
4	 &   13.56 &   -0.39 &   -0.77 \\
5	 &   13.32 &    0.06 &   -0.70 \\
6	 &   13.40 &    0.08 &   -0.68 \\
7	 &   12.43 &   -0.01 &   -0.71 \\
\enddata
\tablenotetext{a} {Average of 5 flux points (20\AA) centered at 3600~\AA\ and converted to magnitude using $-2.5 \log f_{3600}-21.0-0.77$ 
from Bessel et al. (1998).}
\tablenotetext{b} {Average of 5 flux points (20\AA) centered at 4200~\AA\ and converted to magnitude using $-2.5 \log f_{4200}-21.0-0.12$
from Bessel et al. (1998).}
\tablenotetext{c} {Average of 5 flux points (20\AA) centered at 5450~\AA\ and converted to magnitude using $-2.5 \log f_{5450}-21.0$
from Bessel et al. (1998).}
\end{deluxetable}

\clearpage

\begin{deluxetable}{lcccc}
\tabletypesize{\scriptsize}
\tablecaption{The number of explicit levels in the model atom used in the NLTE stellar atmosphere calculations.\label{tab:atom}}
\tablewidth{0pt}
\tablehead{\colhead{} & \colhead{I} & \colhead{II} & \colhead{III} & \colhead{IV}\\
}
\startdata
H  & 16& 1  & --   &--\\
He & 24& 14 & 1    &--\\
C  & 36& 39 & 1    &--\\
Fe$_{\rm cool}$\tablenotemark{a} & 38& 35 & 1   & -- \\
Fe$_{\rm hot}$\tablenotemark{a} & 0& 26 & 50    & 1 \\
\enddata
\tablenotetext{a}{Models cooler than $\sim$9750~K have a different iron atom than those hotter
than 9750~K. The atomic makeup had
to be changed due to convergence problems when running hotter models with neutral iron.}
\end{deluxetable}
\clearpage

\begin{deluxetable}{lcccccc}
\tabletypesize{\scriptsize}
\tablecaption{Spectroscopic masses.\label{tab:masses}}
\tablewidth{0pt}
\tablehead{
\colhead{Sample} &
\colhead{BS}     &
\colhead{BS}     &
\colhead{TO}     & 
\colhead{TO}     & 
\colhead{HB}     &
\colhead{HB}     \\
\colhead{ }      & 
\colhead{Mass\tablenotemark{a}}      & 
\colhead{Variance\tablenotemark{b}}      & 
\colhead{Mass\tablenotemark{a}}      & 
\colhead{Variance\tablenotemark{b}}      & 
\colhead{Mass\tablenotemark{a}}      & 
\colhead{Variance\tablenotemark{b}}      \\
\colhead{}       &
\colhead{(\msun)}     &
\colhead{}     &
\colhead{(\msun)}     &
\colhead{}     &
\colhead{(\msun)}     &
\colhead{}     
}
\startdata
&\multicolumn{6}{c}{Non-variable stars\tablenotemark{c}}\\
47~Tuc   & 0.95 & 0.42(4) & --   & --      &--    & --         \\
M~3      & 0.72 & 0.31(3) & 0.56 & 0.56(6) & 0.90 & 0.31(4) \\
NGC~6752 & 1.01 & 0.41(5) & 0.61 & 0.61(7) & 0.59 & 0.41(3)   \\
NGC~6397 & 1.73 & 0.43(2) & --   &  --     & --   & --               \\
All      & 1.04 & 0.17(14)& 0.58 &0.22(13) & 0.79 &0.25(7)  \\
&\multicolumn{6}{c}{All stars}\\
47~Tuc   & 0.99 & 0.37(5) & 1.09 & 0.52(3) &--    & --    \\
M~3      & 1.03 & 0.24(7) & 0.56 & 0.56(6) & 0.74 & 0.24(8)   \\
NGC~6752 & 0.99 & 0.35(6) & 0.59 & 0.59(8) & 0.87 & 0.35(4)   \\
NGC~6397 & 1.27 & 0.26(6) & --   &  --     & 1.07 & 0.65(1)   \\
All      & 1.07 & 0.14(24)& 0.65 &0.20(17) & 0.81 &0.19(13)   \\
\hline
\enddata
\tablenotetext{a}{Weighted mean, where the weights are $w_i = (M_i/\sigma_Mi)^2$ ($M_i$ are the mass values and $\sigma_Mi$ are their absolute
errors from Table~\ref{tab:results}, where positive and negative error bars were averaged. }
\tablenotetext{b}{Variance=$1/\Sigma w_i$.}
\tablenotetext{c}{Mean masses obtained excluding possibly variable stars, which are marked ``v" in Table~\ref{tab:results} and discussed in
Section~\ref{ssec:errvar}.}
\end{deluxetable}
\clearpage

\begin{deluxetable}{llcccc}
\tabletypesize{\scriptsize}
\tablecaption{Limits on and values of $v \sin i$.\label{tab:rotation}}
\tablewidth{0pt}
\tablehead{ 
Name & CMD & \multicolumn{4}{c}{$v \sin i$ (\kms)} \\
                 & Type&Metal & Ca~{\sc ii} K&Balmer&He~{\sc i}
}
\startdata
\multicolumn{6}{c}{47 Tuc (NGC~104)}\\
N104-1  &TO & $<$140             &--&--&-- \\
N104-2  &TO & $<$70              &--&--&-- \\
N104-4  &BS & 120$^{-20}_{+100}$ &--&--&-- \\
N104-6  &TO & $<$120             &--&--&-- \\
N104-7  &BS & $<$120             &--&--&-- \\
N104-8  &BS & $<$50              &--&--&-- \\
\multicolumn{6}{c}{M~3 (NGC~5272)}\\
N5272-6 &EHB&  --                &  100$\pm$50 & --       & --      \\
N5272-11&EHB&  --                &  70$\pm$30  & --       &--\\
N5272-13&BS &  100$\pm$20        &  $<$150     & $<$200   &--\\
N5272-15&BS &  --                &  225$\pm$50 & $<$300   &--\\
N5272-16&EHB&  --                &  100$\pm$30 & --       &--  \\
N5272-17&EHB&  --                &  200$\pm$50 & --       &--  \\
\multicolumn{6}{c}{NGC~6752}\\
N6752-1 &TO &   --               & $<$200     & --        & --      \\
N6752-2 &EHB&   --               & --         & $<$100    & $<$100  \\
N6752-3 &ATO&   $<$75            & $<$150     & $<$150    & --      \\
N6752-4 &BS &   $<$100           & $<$100     & $<$200    & --      \\
N6752-5 &EHB&   $<$50            & 150$\pm$30 & $<$300    & --      \\
N6752-6 &TO &   --               & --         & $<$100    & --      \\
N6752-8 &ATO&   --               & $<$100     & $<$250    & --      \\
N6752-9 &BS &   --               & $<$25      & $<$200    & --      \\
N6752-11&BS &   $<$50            & 50$\pm$20  & $<$200    & $<$50   \\
N6752-12&EHB&   --               & --         & $<$200    & 50$\pm$20 \\
N6752-13&TO &  --                & --         & $<$50     & --      \\
N6752-14&BS &  --                & $<$50      & --        & --      \\
N6752-16&EHB&  --                & --         & --        & $<$100  \\
N6752-17&TO &  --                &  $<$100    & --        & --      \\
N6752-18&BS &  --                &  50$\pm$20 & $<$200    & --      \\
\enddata
\tablecomments{Stellar classification: TO (ATO): turn-off (above turn-off ) star. BS: blue straggler star. HB (EHB):
horizontal branch (extreme horizontal branch) star.}
\end{deluxetable}
\clearpage

\end{document}